\documentclass[amssymb,aps]{revtex4-2}
\emergencystretch=\maxdimen
\usepackage{graphicx}
\usepackage{dcolumn}
\usepackage{amsmath}
\usepackage[latin1]{inputenc}
\usepackage{amssymb}
\usepackage[colorlinks=true, citecolor=blue, urlcolor = blue, linkcolor= red, bookmarks=true]{hyperref}
\usepackage{float}
\usepackage{amsfonts}
\usepackage{hyperref}
\usepackage{subfigure}
\usepackage{pgfplots}
\usepackage{booktabs}
\usepackage{mathrsfs}
\usepackage{multirow}
\pgfplotsset{compat=1.16}

\usepackage{wrapfig}
\usepackage{adjustbox}
\makeatletter
\def\btt#1{\texttt{\@backslashchar#1}}
\DeclareRobustCommand\bblash{\btt{\@backslashchar}} \makeatother

\begin{document}

\title[]{Parameters estimation and strong gravitational lensing of nonsingular Kerr--Sen black holes}

\author{Sushant~G.~Ghosh$^{a,\;b}$} \email{sghosh2@jmi.ac.in, sgghosh@gmail.com}
\author{Rahul Kumar$^{a}$ } \email{rahul.phy3@gmail.com}
\author{Shafqat Ul Islam$^{a}$ } \email{Shafphy@gmail.com}

\affiliation{$^{a}$ Centre for Theoretical Physics, 
	Jamia Millia Islamia, New Delhi 110025, India}
\affiliation{$^{b}$ Astrophysics and Cosmology Research Unit, 
	School of Mathematics, Statistics and Computer Science, 
	University of KwaZulu-Natal, Private Bag 54001, Durban 4000, South Africa}
\begin{abstract}
The recent time witnessed a surge of interest in strong gravitational lensing by black holes is due to the Event Horizon Telescope (EHT) results, which suggest comparing the black hole lensing in both general relativity and heterotic string theory. That may help us to assess the phenomenological differences between these models. Motivated by this, we consider gravitational lensing by the nonsingular Kerr--Sen black holes, which encompass Kerr black holes as a particular case, to calculate the light deflection coefficients $p$ and $q$ in strong-field limits, while the former increases with increasing parameters $k$ and charge $b$, later decrease. We also find a decrease in the light deflection angle $\alpha_D$, angular position $\theta_{\infty}$ decreases more slowly and impact parameter for photon orbits $u_{m}$ more quickly, but angular separation $s$ increases more rapidly with parameters $b$ and $k$. We compare our results with those for Kerr black holes, and also the formalism is applied to discuss the astrophysical consequences in the case of the supermassive black holes NGC 4649,  NGC 1332, Sgr A* and M87*.  In turn, we also investigate the shadows of the nonsingular Kerr--Sen black holes and show that they are smaller and more distorted than the corresponding Kerr black holes and nonsingular Kerr black holes shadows. The inferred circularity deviation $\Delta C\leq 0.10$, for the M87* black hole shadow, put constraints on the nonsingular Kerr--Sen black hole parameters ($a, k$) and ($a, b$).  The maximum shadow angular diameter for $b=0.30M$ and $k=0.30M$ are, respectively, $\theta_d=35.3461\,\mu$as and $\theta_d=35.3355\,\mu$as. We also estimate the parameters associated with nonsingular Kerr--Sen black holes using the shadow observables. 
\end{abstract}

\maketitle
\thispagestyle{empty}
\setcounter{page}{1}

\section{Introduction}
The Kerr-Newman black hole solution \cite{Newman:1965my}, characterized by the only three parameters mass, angular momentum and charge, is the unique electro-vacuum solution of Einstein field equations and leads to the \textit{no-hair} theorem. However, in the presence of other fundamental or exotic fields, hairy black hole solutions are also possible, which as expected show a significant departure from their general relativity counterparts in the strong-field regimes. One of the important exact rotating axisymmetric black hole solutions in the Einstein-Maxwell-axion-dilaton gravity was obtained by Sen \cite{Sen:1992ua}, commonly known as Kerr--Sen black hole. This solution belongs to the effective theory of the heterotic string theory and has been widely investigated including its uniqueness \cite{Rogatko:2010hf} and hidden conformal symmetries \cite{Ghezelbash:2012qn}. This solution has some similarities in physical properties to those for the Einstein-Maxwell theory, but still has some prominent differences in several salient aspects.  The Kerr--Sen solution, besides the metric tensor and $U(1)$ vector field, is governed by two additional nongravitational fields, namely, an anti-symmetric third-order tensor field and a dilaton scalar field. Therefore, in the vacuum, the solution of string theory in weak-field limits matches with those in the general relativity. Similar to the Kerr-Newman black holes, the weak cosmic censorship holds in Kerr--Sen black holes for considering up to second-order perturbations \cite{Jiang:2019vww}; however, this can be violated by neglecting the radiative and self-force effects \cite{Siahaan:2015ljs}. The Kerr--Sen and Kerr--Newman black holes have been compared extensively in the varieties of issues, namely, the photon captured regions \cite{Hioki:2008zw}, black hole shadows \cite{Xavier:2020egv,Younsi:2016azx}, evaporation process \cite{Koga:1995bs}, gyromagnetic ratios \cite{Horne:1992zy}, the instability of bound state charged massive scalar fields \cite{Furuhashi:2004jk,Siahaan:2015xna} and the CFT$_2$ holographic dual for the scattering process \cite{Ghezelbash:2014aqa}. Recently, Jayawiguna \cite{Jayawiguna:2020qok} used  the Hassan-Sen transformation to  obtained a nonsingular Kerr--Sen black holes from the rotating nonsingular black hole \cite{Ghosh:2014pba}. The resulting metric is regular at the center, as all the curvature invariant are finite at $r=0$, and in contrary to the generic regular black holes it has a Minkowski flat core.

Gravitational lensing by black holes is one of the most powerful astrophysical tools for investigations of the strong-field features of gravity. Photons propagating in the curved spacetime get deviated from their original path that they follow in the absence of curvature, this phenomenon is known as the gravitational bending of light. Darwin \cite{Darwin} pioneered the study of strong gravitational lensing by compact astrophysical objects with a photon sphere, such as black holes. Later, Virbhadra \cite{Virbhadra:1999nm} obtained the gravitational lens equation in the strong-field limits and also analyzed the gravitational lensing by a Schwarzschild black hole. Bozza \textit{et al.} \cite{Bozza:2001xd,Bozza:2002zj} proposed an alternate formalism to  analytically investigate the strong-field deflection limit for the general static spherically symmetric black holes. Since then, gravitational deflection of light by black holes has received great attention due to the amazing advancement of current observational facilities  \cite{Will,Bhadra:2003zs,Bozza:2007gt,Wei:2011nj,Eiroa:2010wm,Sadeghi}. The analytical and numerical computations of the positions, magnifications and time delays of the relativistic images have been done in weak and strong deflection limits \cite{Ono:2017pie,Virbhadra:2007kw,Man,Chen:2009eu,Sarkar:2006ry,Javed:2019qyg,Shaikh:2019itn}. The gravitational lensing has far-reaching impacts on our understanding of black hole spacetimes as the real part of the black hole quasinormal modes are related with the deflection angle in the strong-field limits and the shadow radius \cite{Stefanov:2010xz, Cardoso:2008bp,Hod:2009td}. Photon geodesics and shadows of the Kerr--Sen black holes have also been investigated by several authors \cite{Lan:2018lyj,Xavier:2020egv,Dastan:2016bfy,Narang:2020bgo,Blaga:2001wt,Uniyal:2017yll,Liu:2018vea}.

The last decade witnessed a spectacular progress in the astrophysical black hole observations. The Event Horizon Telescope (EHT) collaboration unraveled the first-ever image of the supermassive black hole M87* at the center of the nearby galaxy Messier 87 \cite{Akiyama:2019cqa,Akiyama:2019brx,Akiyama:2019sww,Akiyama:2019bqs,Akiyama:2019fyp,Akiyama:2019eap}. The M87* shadow results offer a fascinating probe of the strong gravitational fields and place the theories of gravity on the testing grounds. Although the primary studies suggested that the observed shadow is consistent with the expected image of the Kerr black hole as predicted by the general relativity \cite{Akiyama:2019cqa}. However, the current uncertainty in the measurement of spin angular momentum and the deviation parameters does not wholly exclude the possibilities of the non-Kerr black holes in general relativity as well as in modified theories of gravity \cite{Bambi1,Medeiros:2019cde}. Moreover, considering the parametric non-Kerr metrics, stringent constraints are placed on the second post-Newtonian metric coefficient using the M87* shadow angular size \cite{Psaltis:2020lvx}. Thus, the non-Kerr black holes which predict significant departure from the Kerr black hole could be ruled out \cite{Psaltis:2020lvx}.

With the existing observational facilities, it is now possible to measure the theoretically predicted potential deviation in the strong-field limits in the context of gravitational lensing and shadow. Motivated by this, in this paper, we aim to investigate the strong gravitational lensing of light and the shadows of the nonsingular Kerr--Sen black holes. The impact of the axionic field parameters and the free parameter (causing the regularity at the center) on the lensing and shadow morphology is also calculated in the context of supermassive black holes. Furthermore, we use the M87* shadow results to check whether these black holes can be viable candidates for the astrophysical black holes.

The rest of the paper is organized as follows. In the Sect.~\ref{Sec2}, we briefly review the nonsingular Kerr--Sen black hole. Formalism for gravitational deflection of light in the strong-field limit is setup in Sect.~\ref{Sec3}. Whereas strong-lensing observables, numerical estimations of deflection angle, and lensing by supermassive black holes are discussed in Sect.~\ref{Sec4}. Black holes shadows and estimation of the parameter using shadows observables is topic of Sect.~\ref{Sec5}. Finally, we summarize our main findings in  Sect.~\ref{Sec6}.\\

\section{Nonsingular Kerr--Sen black hole}\label{Sec2}
We begin with a brief introduction of the Kerr--Sen black holes. Applying a solution generating technique to the Kerr solution, Sen  \cite{Sen:1992ua} obtained a charged rotating black holes solution to the low-energy limit of the heterotic string theory field equations, which is famously known as the Kerr--Sen black holes. Thus, the Kerr--Sen metric describes an asymptotically flat charged rotating black hole spacetime which arises due to modification of general relativity from the low energy heterotic string theory effective action that is given by    
\begin{equation}
\label{aksi}
S= \int d^4x \sqrt{-\tilde{g}} ~e^{-\Phi} \left[\mathcal{R} + (\nabla\Phi)^2 -\frac{1}{8} F^2-\frac{1}{12} H^2\right],
\end{equation}
where $ \tilde{g} $ is a determinant of metric tensor $ g_{\mu\nu} $, $ \mathcal{R} $ is a Ricci scalar, $F=F_{\mu\nu}F^{\mu\nu}$ with $F_{\mu\nu}$ being the $U(1)$ Maxwell field strength tensor, $ \Phi $ is a scalar dilaton field, and $H = H_{\mu \nu \rho} H^{\mu \nu \rho}$ is the field strength for the axion field
\begin{equation}\label{action2}
H_{\kappa\mu\nu} =\partial_{\kappa}B_{\mu\nu}+ \partial_{\nu}B_{\kappa\mu}+\partial_{\mu}B_{\nu\kappa}-\frac{1}{4}\left(A_{\kappa}F_{\mu\nu}+A_{\nu}F_{\kappa\mu}+A_{\mu}F_{\nu\kappa}\right).
\end{equation}
It is a 3-form tensor field which contain the antisymmetric 2-form tensor field where the last term in Eq.~(\ref{action2}) is the gauge Chern-Simons term. On using the conformal transformation 
\begin{equation}
\label{conformal}
ds_{E}^2 = e^{-\Phi} \tilde{ds}^2,
\end{equation}
one obtains  the action in Einstein frame
\begin{equation}
S= \int d^4 x \sqrt{-g} \left[R(g)-\frac{1}{2}(\nabla\Phi)^2 -\frac{e^{-\Phi}}{8}F^2-\frac{e^{-2\Phi}}{12}H^2\right],\label{EinAction}
\end{equation}

which encompasses the Einstein-Hilbert action as a special case in the absence of dilaton, vector, and axion fields. In Boyer-Lindquist coordinate, the Kerr--Sen metric reads as 
\begin{equation}\label{kerrsen}
ds^2_{E} = -\left(1-\frac{2M r}{\rho^2_{KS}}\right) dt^2 -\frac{4aMr \sin^2\theta}{\rho^2_{KS}} dt\,d\phi + \rho^2_{KS} \left(\frac{dr^2}{\Delta_{KS}}+d\theta^2\right) + \left[r^2 +a^2 +2br+ \frac{2M r a^2 \sin^2\theta}{\rho^2_{KS}}\right]\sin^2\theta~ d\phi^2,
\end{equation}
where $ \rho^2_{KS} = r^2+2br +a^2\cos^2\theta $, $ \Delta_{KS} = r^2+ 2br -2Mr +a^2$, and $ b=Q^2/2M $. Parameters $M$ and $a$, respectively, can be identified as the Arnowitt-Deser-Misner mass and the rotation parameter $a=J/M$, with $J$ as the black hole angular momentum. Here, $Q$ is the $U(1)$ gauge field charge. $A_{\mu}$, $B_{\mu\nu}$ and the dilaton field $\Phi$, respectively, are given by
\begin{align}
A_{\mu}dx^{\mu}=& \frac{Q r}{\rho^2_{KS}} \left(dt^2-a \sin^2\theta d\phi^2 \right),\\
B_{t\phi}=&\frac{2bra \sin^2\theta}{\rho^2_{KS}},
\end{align}
and
\begin{equation}
e^{-2\Phi}= \frac{\rho^2_{KS}}{r^2 +a^2 \cos^2\theta}.
\end{equation}

In the limit, $b \to 0$, the metric (\ref{kerrsen}) reduces to the Kerr metric. The horizons of Kerr--Sen black hole are determined by the real positive roots of $\Delta_{KS}=0$, as
\begin{equation}\label{ksh}
r_{\pm} = (M-b) \pm \sqrt{(M-b)^2 - a^2},
\end{equation}
where $r_{\pm}$, respectively, denote the radii of outer (event) and inner (Cauchy) horizons. The regularity of the horizons requires $a \leq (M-b)$ and the extremal case leading to degenerate horizons is obtained for the $a=(M-b)$. When $b$ and $a$ are bounded, respectively, we obtain the corresponding ranges for $ a $ and $ b $ as $ 0 \leq a \leq M $, $ 0 \leq b \leq M $.

Here, we are interested in the recently reported nonsingular counterpart of Kerr--Sen black hole metric \cite{Jayawiguna:2020qok} that is governed by an additional parameter $k$ arising from the nonlinear electrodynamics \cite{Ghosh:2014pba} and measures the potential difference from the Kerr--Sen solution. The metric in Boyer-Lindquist coordinates reads 
\begin{eqnarray}\label{metric1}
ds^2_{E} &=& -\left(1-\frac{2Mre^{-k/r}}{\Sigma}\right) dt^2 - \frac{4aMre^{-k/r}\sin^2\theta}{\Sigma}dt d\phi+ \Sigma \left[\frac{dr^2}{\Delta}+d\theta^2\right] \nonumber \\ && + \left[r^2 +a^2 + 2bre^{-k/r}+ \frac{2Mra^2 e^{-k/r} \sin^2\theta}{\Sigma}\right]\sin^2\theta ~d\phi^2,
\end{eqnarray}
where 
\begin{eqnarray}
 \Sigma=r(r+2be^{-k/r}) + a^2\cos^2\theta,~~~~~~\Delta = r(r  +2be^{-k/r}) +a^2 -2Mre^{-k/r}, 
\end{eqnarray}
 and
\begin{eqnarray}
A_{\mu} dx^{\mu}= \frac{Qr e^{-k/r}}{\Sigma}(dt-a\sin^2\theta d\phi),~~~~B_{t\phi} = \frac{2 b r a e^{-k/r}\sin^2\theta}{\Sigma},~~~ \textrm{and}~~~ e^{-2\Phi} = \frac{\Sigma}{\rho^2}.
\end{eqnarray}
The metric (\ref{metric1}) describes the nonsingular Kerr--Sen black holes, which is a solution of appropriately modified field equations but certainly different from that resulting from action Eq.~(\ref{EinAction}). The metric (\ref{metric1}) encompasses the Kerr--Sen black hole \cite{Sen:1992ua}  as special case of ${k}=0$ and reverts back to the Kerr black holes \cite{Kerr:1963ud} when $k=0=b$. The rotating nonsingular black hole \cite{Ghosh:2014pba} is obtained when $b=0$ in the metric~(\ref{metric1}).  
To proceed further, we introduce the dimensionless variables as  
\begin{eqnarray}
x \to \frac{r}{2M},\; a \to \frac{a}{2M}, \;t \to \frac{t}{2M}, \; \tilde{k}=\frac{k}{2M},\; \tilde{b}=\frac{b}{2M},
\end{eqnarray}
accordingly the metric (\ref{metric1}) can be rewritten as
\begin{equation}\label{metric2}
ds^2 =  - \left( 1- \frac{xe^{-\frac{\tilde{k}}{x}}}{\Sigma} \right) dt^2 +
\frac{\Sigma}{\Delta}dx^2 + \Sigma d \theta^2 - \frac{2axe^{-\frac{\tilde{k}}{x}}
}{\Sigma  } \sin^2 \theta dt \; d\phi + \left[x^2+ a^2 + 2 \tilde{b} x e^{-\frac{\tilde{k}}{x}} +
 \frac{x a^2 e^{-\frac{\tilde{k}}{x}}}{\Sigma} \sin^2 \theta
\right] \sin^2 \theta d\phi^2,
\end{equation} 
where
\begin{eqnarray}
\Sigma=x^2+a^2\cos^2{\theta} + 2 \tilde{b} x e^{-\frac{\tilde{k}}{x}},\;\;\;\;\Delta =x^2+ a^2+2 \tilde{b} x e^{-\frac{\tilde{k}}{x}}-xe^{-\frac{\tilde{k}}{x}}.
\end{eqnarray}
Now $a$ is the specific angular momentum and $0<\tilde{k}<1$.
 \begin{figure}
    \centering
    \includegraphics[scale=0.7]{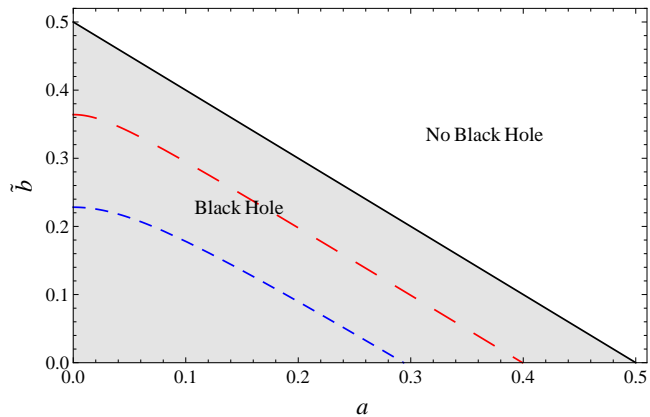}
    \caption{The parameter space $(a, \tilde{b})$ for the existence of the nonsingular Kerr-Sen  black hole horizons. The black solid line is for $\tilde{k}=0.0$, red dashed line is for $\tilde{k}=0.10$, and blue dashed line is for $\tilde{k}=0.20$. The colored lines correspond to the extremal black hole delineating the boundary for the existence of black hole to the no black hole case. The black solid line ($\tilde{k}=0$) is for the Kerr-Sen black hole. }
    \label{NoBH}
\end{figure} 
The nonsingular Kerr--Sen spacetime admits two linearly independent killing vectors, 
$\xi^{\mu}_{(t)}=\delta^{\mu}_t $ and $\xi^{\mu}_{(\phi)}=\delta^{\mu}_{\phi}$ associated with the time translation and rotational invariance of metric (\ref{metric2}) \cite{Chandrasekhar:1992}. The horizon is a null hypersurface boundary of black hole that comprises of outer directed null geodesics unable to reach the future null infinity, the  horizons are zeroes of $g^{rr}=\Delta=0$, i.e.,

\begin{equation}
\label{horizoneq}
x^2+a^2+ 2 \tilde{b} x e^{-\frac{\tilde{k}}{x}}-x e^{-\frac{\tilde{k}}{x}} = 0.
\end{equation}
\begin{figure}[t]
	\begin{centering}
		\begin{tabular}{c c}
		    \includegraphics[scale=0.77]{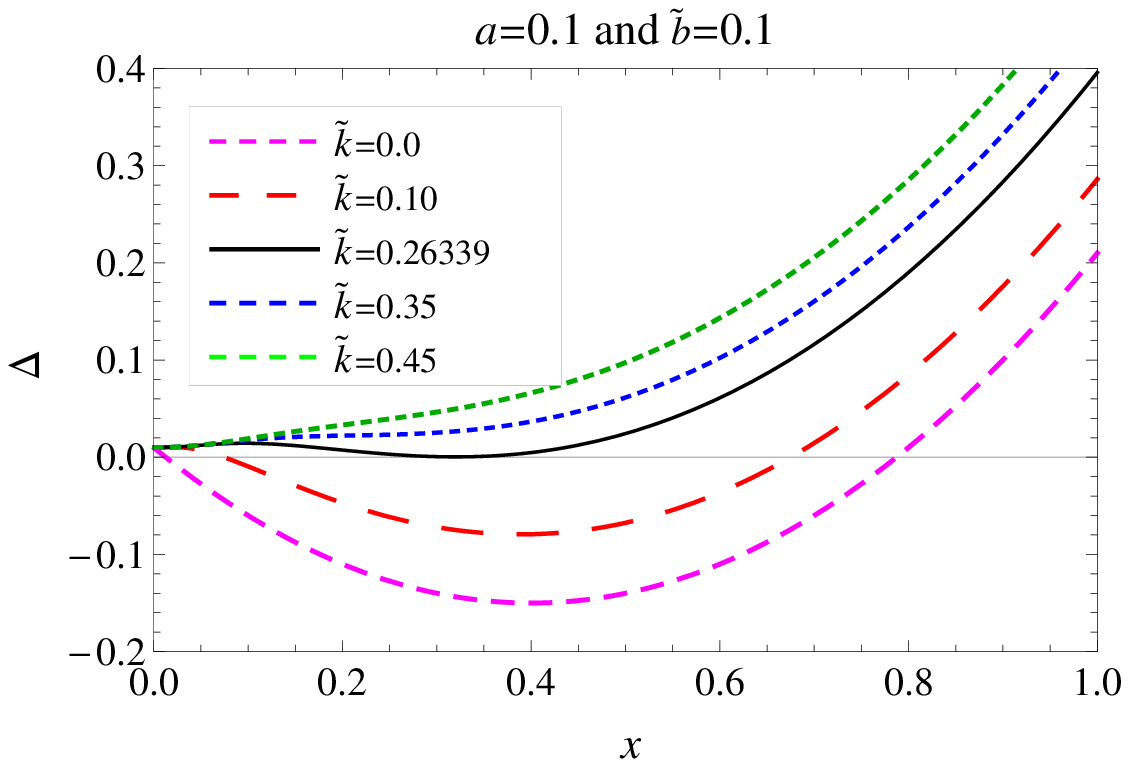}\hspace*{-0.9cm}&
		    \includegraphics[scale=0.77]{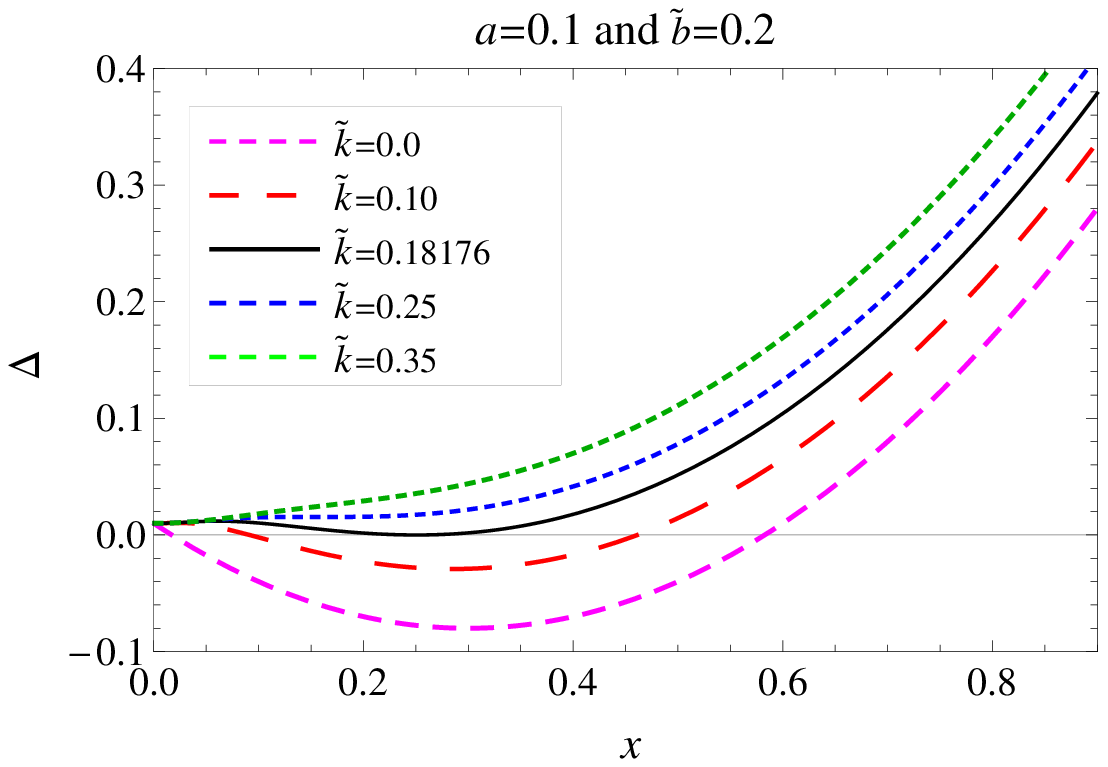}\\
		    \includegraphics[scale=0.78]{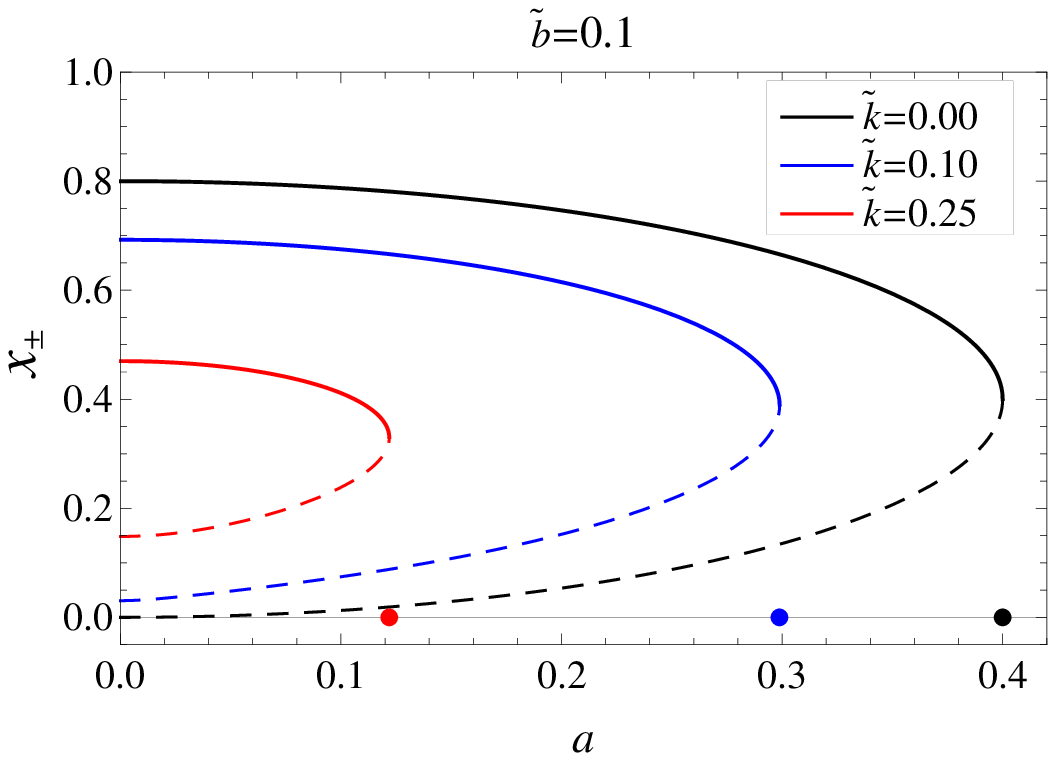}\hspace*{-0.9cm}&
		    \includegraphics[scale=0.78]{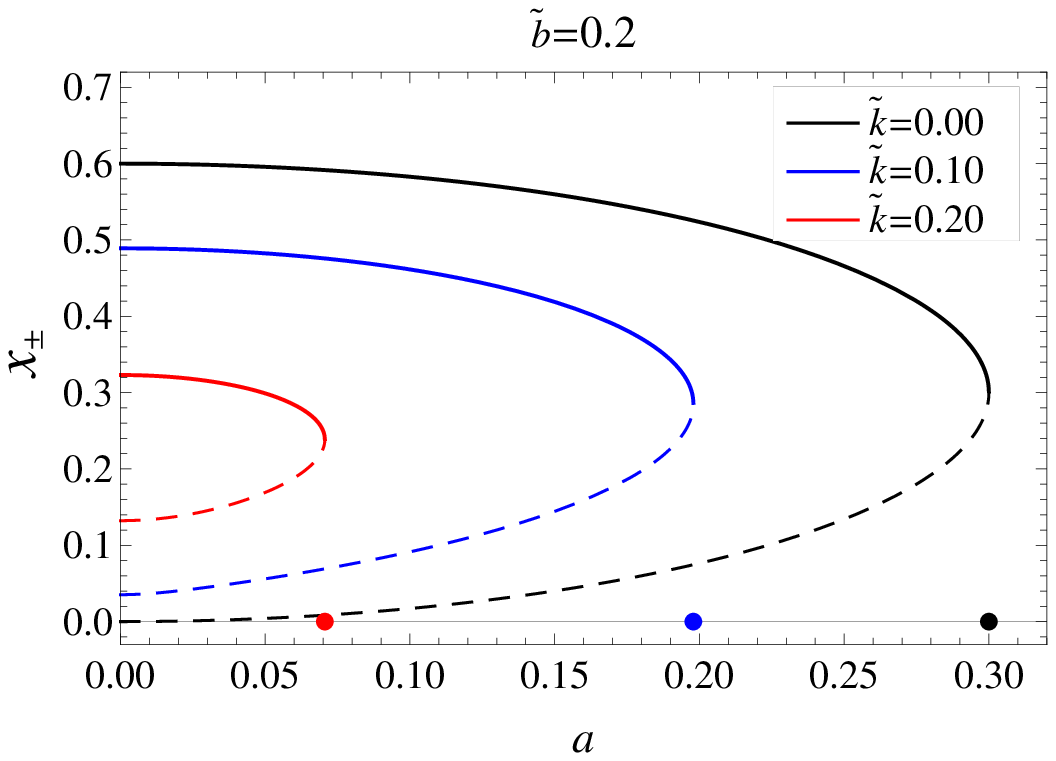}
		 \end{tabular}
	\end{centering}
	\caption{Plot showing (upper panel) the behaviour of horizon ($\Delta$ vs $x$), with varying black hole parameter $a, \tilde{b}$ and $\tilde{k}$. The black solid line corresponds to the extremal black holes. (Lower panel)  the variation of event horizon $x_{+}$ (solid lines) and Cauchy horizon $x_{-}$ (dashed lines)  with spin $a$. Points on the horizontal axis correspond to the extremal values of $a$.}\label{plot1}		
\end{figure}  
The numerical analysis of the above transcendental equation suggest that one can found non-zero values of $a$, $\tilde{b}$ and $\tilde{k}$ for which Eq.~(\ref{horizoneq}) admits two positive roots ($x_{\pm}$) corresponding to Cauchy $(x_{-})$ and event horizons $(x_{+})$. Two distinct roots of Eq.~(\ref{horizoneq}) correspond to the nonextremal black holes with Cauchy and event horizons, while no horizon exists when Eq.~(\ref{horizoneq}) has no real solutions. It turns out that for a given value of $a$ and $\tilde{b}$, one can found the critical values $\tilde{k}=\tilde{k_c}$, such that Eq.~(\ref{horizoneq}) has one double zero at $ x_-=x_+\equiv x_c$ corresponding to an extremal black hole. Equation~(\ref{horizoneq}) has no zeroes for $\tilde{k} > \tilde{k_c}$ and has two distinct roots for $\tilde{k} <\tilde{k_c}$  (cf. Fig.~\ref{plot1}). These two cases correspond, respectively to no black holes and black hole with event and Cauchy horizons as depicted in Fig.~\ref{NoBH}.  

For the nonsingular Kerr--Sen spacetime, there also exists the static limit surface (SLS) that can be determined by  $ g_{tt}(x)=0 $
\begin{equation}
\label{rsl}
x^2 + a^2\cos^2\theta + 2\tilde{b}x e^{-\tilde{k}/x} -x e^{-\tilde{k}/x} = 0.
\end{equation}
The SLS gets its name from the fact that no observer can stay static inside this region and they are bound to co-rotate around the black hole. We employ the numerical analysis to investigate the natural characteristic of $ g_{tt}$. The SLS radius becomes smaller with the increasing parameter $k$. 

\begin{figure}
	\begin{centering}
	        \includegraphics[scale=0.88]{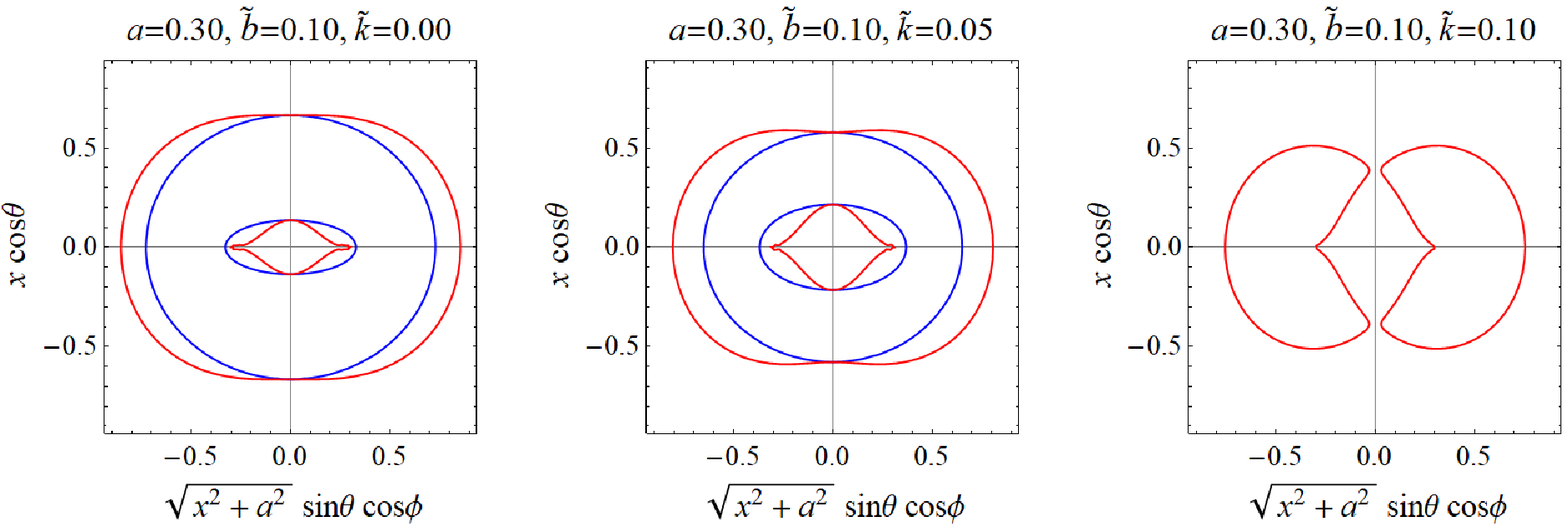}\\
		    \includegraphics[scale=0.88]{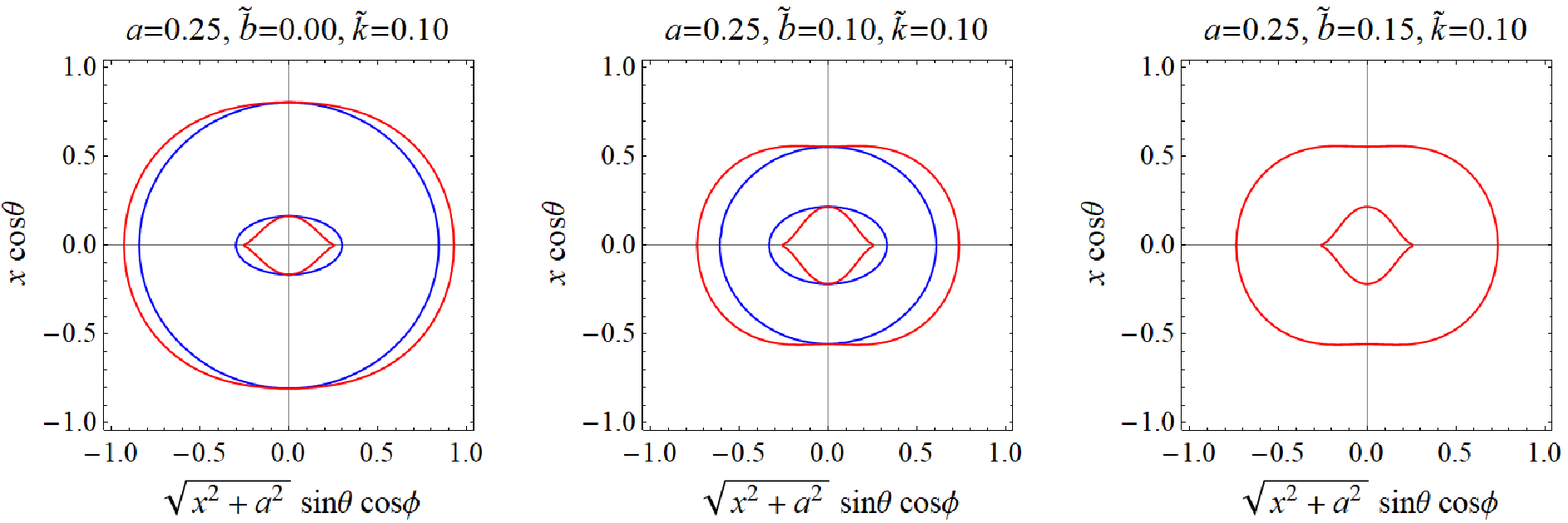}
	\end{centering}
	\caption{The cross-section of event horizon (outer blue line), SLS (outer red line) and ergosphere (region between event horizon and SLS) for different values of parameters.}\label{Ergo}		
\end{figure}

The region between the SLS and the event horizon is called the ergosphere that is depicted in Fig.~\ref{Ergo}. Penrose \cite{Penrose:1969pc} stated that energy can be extracted from black hole's ergosphere. The shape of this region  strongly depends on free parameter $\tilde{k}$, charge $\tilde{b}$, and rotational parameter $a$ (cf. Fig.~\ref{Ergo}). The shared feature with Ref.~\cite{Ghosh:2014pba} is that the ergosphere grows as free parameter $ \tilde{k} $ gets larger.

\section{Light deflection angle}\label{Sec3}
We shall focus on the strong deflection limit of light by the nonsingular Kerr--Sen  black holes in the equatorial plane ($\theta=\pi/2$), i.e., we consider  both the observer and the source in the equatorial  plane ($\theta = \pi/2$) and the trajectory of the photons is limited to this plane only. The metric (\ref{metric2}), now reduces to the form
\begin{eqnarray}\label{NSR}
\mathrm{ds^2}=-A(x)dt^2+B(x) dx^2 +C(x)d\phi^2-D(x)dt\,d\phi,
\end{eqnarray}
where
\begin{eqnarray}
A(x)&=& 1-\frac{x e^{-\frac{\tilde{k}}{x}}}{x^2+ 2 \tilde{b} x e^{-\frac{\tilde{k}}{x}}},\;\;\;\;\;
B(x)=\frac{x^2 + 2 \tilde{b} x e^{-\frac{\tilde{k}}{x}}}{\Delta}\nonumber\\
C(x)&=& x^2+a^2+x^2+ 2 \tilde{b} x e^{-\frac{\tilde{k}}{x}}+\frac{x a^2 e^{-\frac{\tilde{k}}{x}}}{x^2+ 2 \tilde{b} x e^{-\frac{\tilde{k}}{x}}},\;\;\;
D(x)=\frac{2 a x e^{-\frac{\tilde{k}}{x}}}{x^2+ 2 \tilde{b} x e^{-\frac{\tilde{k}}{x}}}
\end{eqnarray}
and $ \Delta= x^2+a^2+ 2 \tilde{b} x e^{-\frac{\tilde{k}}{x}}-x e^{-\frac{\tilde{k}}{x}}$.

Photons geodesic equations can be derived  using the  Lagrangian
\begin{eqnarray}
\mathcal{L}=g_{\mu\nu}\dot{x}^\mu\dot{x}^\nu,
\end{eqnarray}
where an over dot denotes the partial derivative with respect to the affine parameter of geodesics. Due to isometries of rotating nonsingular Kerr--Sen spacetime along the $t$ and $\phi$ coordinates, photons possess two obvious constants of the motion -- energy $E$ and component of angular momentum $L$ as follows
\begin{align}
2E&=\frac{\mathcal{\partial{L}}}{\partial \dot{t}}=g_{tt}\dot{t}+g_{t\phi}\dot{\phi},\\
-2L&=\frac{\mathcal{\partial{L}}}{\partial \dot{\phi}}=g_{t\phi}\dot{t}+g_{\phi\phi}\dot{\phi}.
\end{align}
For simplicity we suitably choose the affine parameters to set the energy $E$ to be unity. The above two equations together with the null radial  geodesics equation in the equatorial plane give the following  set of  first order geodesic equations 
\begin{eqnarray}
\dot{t} &=& \frac{4C-2 L D}{4AC + D^2},\label{tdot}\\
\dot{\theta} &=& 0,\\
\dot{\phi} &=& \frac{2D+4 A L}{4AC + D^2},\\
\dot{x} &=& \pm 2 \sqrt{\frac{C-DL-AL^2}{B(4AC + D^2)}}.\label{xdot} 
\end{eqnarray}
The photons path becomes a straight line at infinity as the spacetime is asymptotically flat. Equations~(\ref{tdot}-\ref{xdot}) completely describe the motion of photons around the nonsingular Kerr-Sen black hole at $\theta=\pi/2$. The effective potential $V_{\text{eff}}$ for radial motion in metric (\ref{NSR}) is given by
\begin{eqnarray}
V_{\text{eff}} &=& \frac{4(C-DL-AL^2)}{B(4AC + D^2)}.
\end{eqnarray}
Apart from characterizing the different types of possible orbits, the effective potential is also regarded as an important quantity to determine the boundary of the shadow. At this point, we should mention that the impact parameter in the equatorial plane coincides with the angular momentum $L$. In the asymptotic limit, a light ray from the infinity approaches the black hole and reaches the minimum distance $x_0$ only to leave again towards infinity. At the distance of minimum approach $x_0$, effective potential vanishes. Hence solving $V_{\text{eff}} =0 $  for $L$ gives 
\begin{eqnarray}\label{angmom}
L&=&u=\frac{-D_0+\sqrt{D_0^2+4A_0C_0}}{2A_0},\nonumber\\
&=&\frac{a - \left(2\tilde{b}+x_0 e^{\frac{\tilde{k}}{x_0}} \right) \sqrt{a^2 + x_0 e^{-\frac{\tilde{k}}{x_0}}(2\tilde{b}+x_0e^{\frac{\tilde{k}}{x_0}}-1 )}}{1 - x_0 e^{\frac{\tilde{k}}{x_0}}-2\tilde{b}},
\end{eqnarray}
where $u$ is the impact parameter and all the metric functions with the subscript $0$ are evaluated at $x=x_0$. Therefore, the impact parameter $u$ is determined once $x_0$ is found or vice versa. The closest approach distance $x_0$ is different than the impact parameter $u$ as long as the deflection angle is not vanishing. We choose the positive sign before the square bracket describing only the counterclockwise winding of light rays.
The light bending angle in a general rotating stationary spacetime described by the line element (\ref{NSR}), for a closest distance approach $x_0$ is given by
\begin{eqnarray}\label{bending1}
\alpha_{D}(x_0)=I(x_0)-\pi,
\end{eqnarray} 
where
\begin{eqnarray}\label{bending2}
I(x_0) = 2 \int_{x_0}^{\infty}\frac{d\phi}{dx} dx
= 2\int_{x_0}^{\infty}P_1(x,x_0)P_2(x,x_0)dx,
\end{eqnarray}
\begin{eqnarray}\label{bending3}
P_1(x,x_0)&=&\frac{\sqrt{B}\left(2A_0AL+A_0D\right)}{\sqrt{CA_0}\sqrt{4AC+D^2}},\nonumber\\
P_2(x,x_0)&=&\frac{1}{\sqrt{A_0-A\frac{C_0}{C}+\frac{L}{C}\left(AD_0-A_0D\right)}}.
\end{eqnarray}

\begin{figure}[t]
	\begin{centering}
		\begin{tabular}{c c}
		    \includegraphics[scale=0.77]{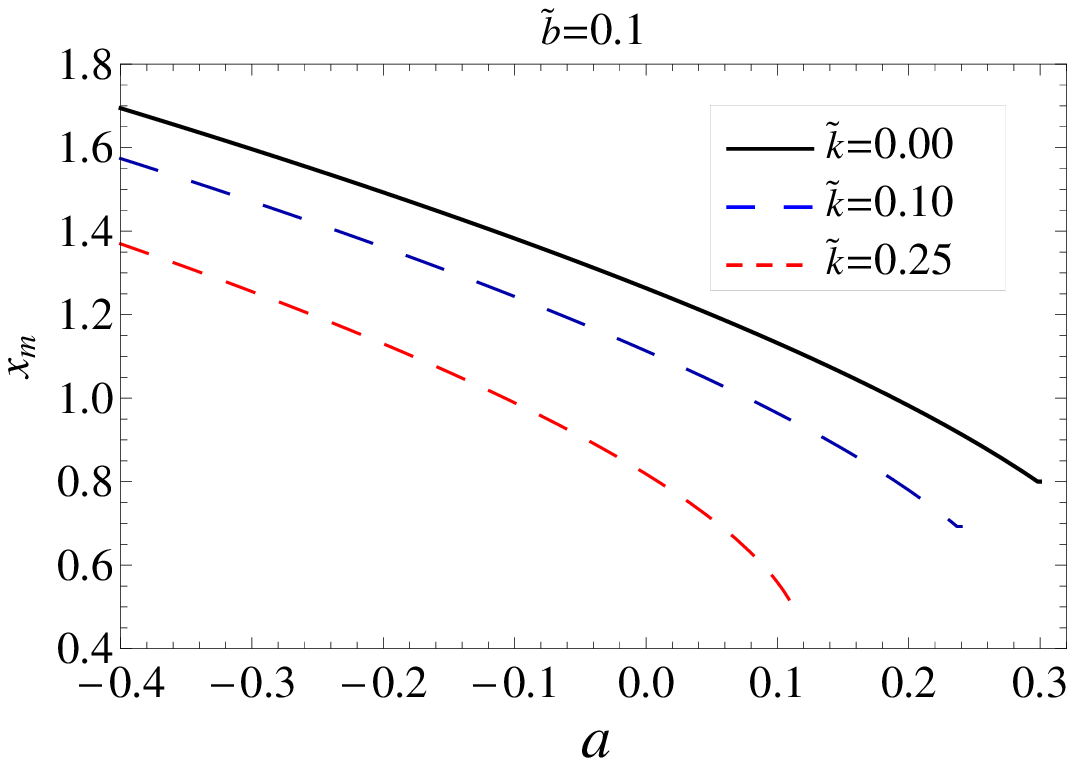}\hspace*{-0.9cm}&
		    \includegraphics[scale=0.77]{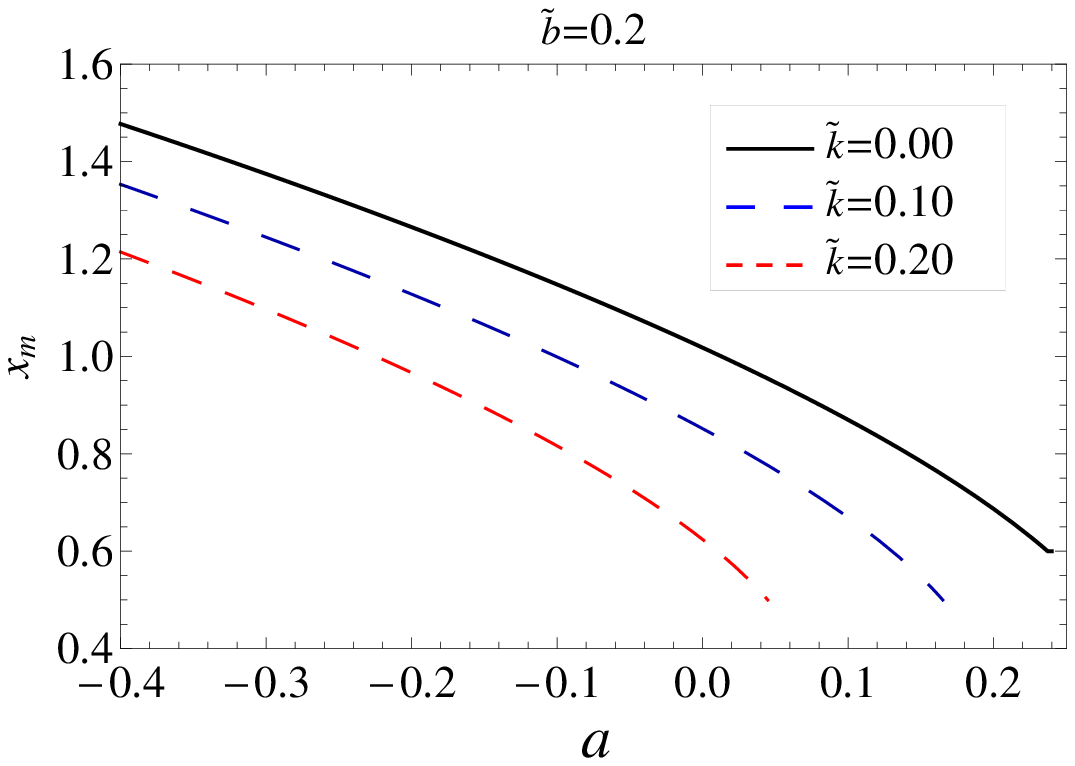}\\
		    \includegraphics[scale=0.77]{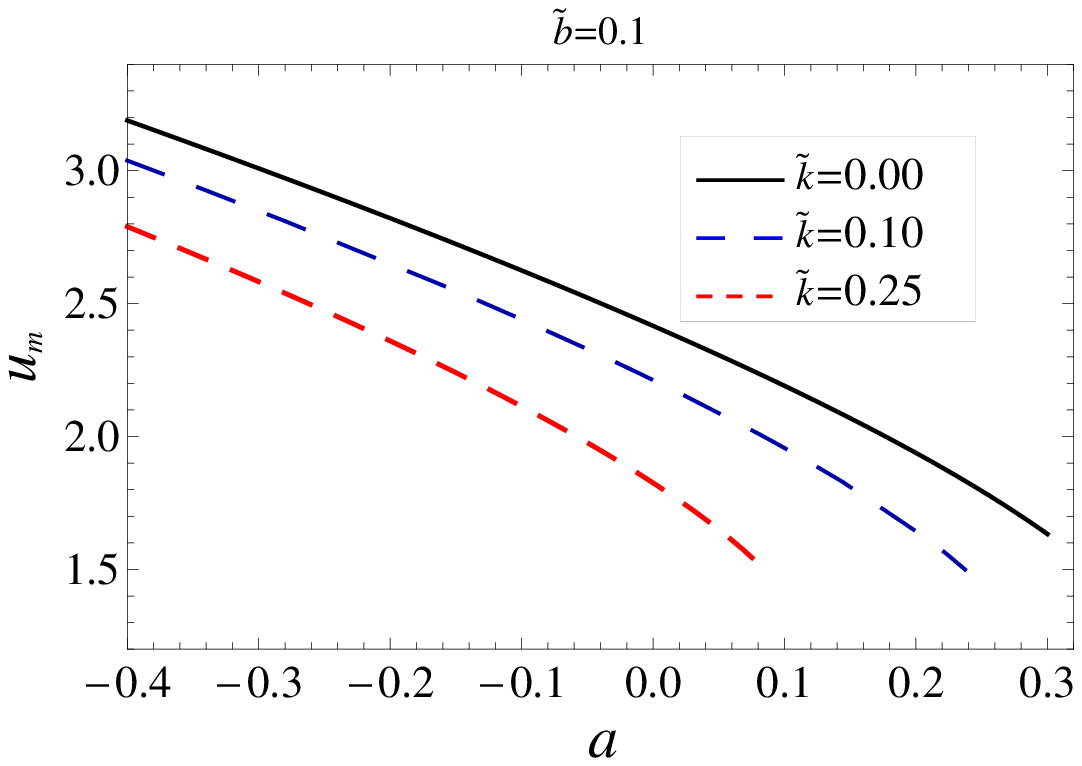}\hspace*{-0.9cm}&
		    \includegraphics[scale=0.77]{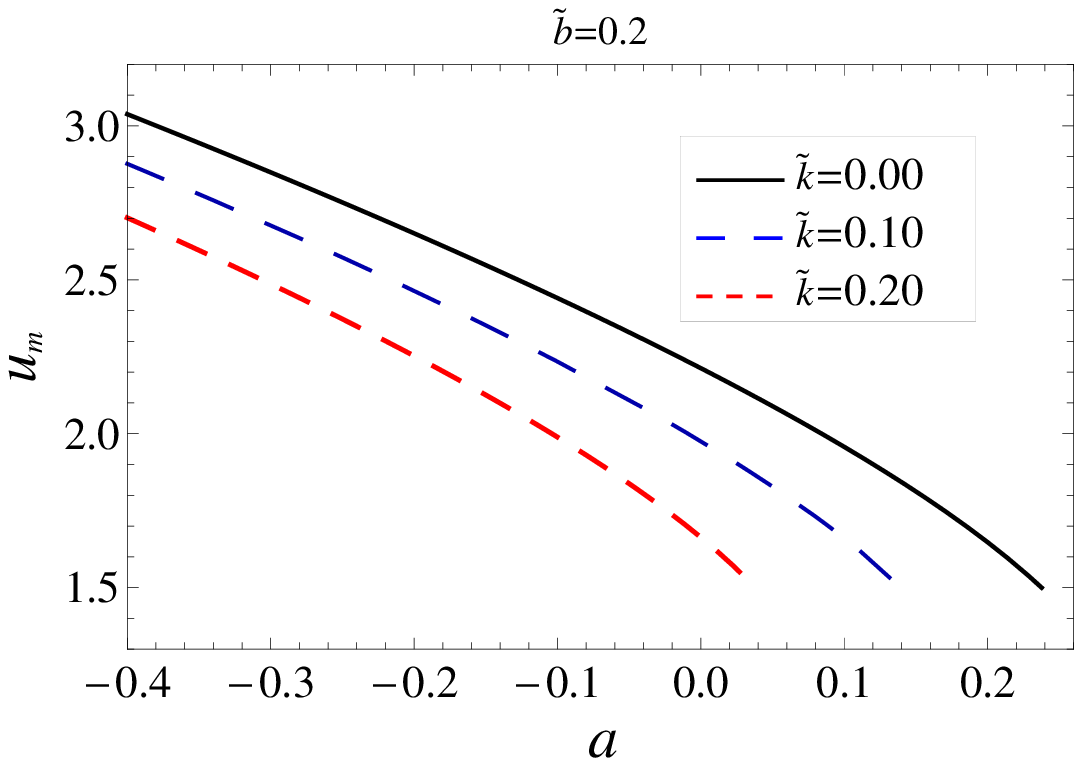}
		 \end{tabular}
	\end{centering}
	\caption{Plot showing (upper panel) the behaviour of unstable circular photon orbit radii versus the black hole spin $a$. (Lower panel)  the variation of impact parameter with spin $a$ for different values of $\tilde{b}$ and $\tilde{k}$.}\label{plot3}	
\end{figure}  

Here, $I(x_0)$ is the total azimuthal shift in the light geodesics direction, which in the absence of black hole is a straight line and thus $I(x_0)=\pi$. However, the light rays bent in the presence of gravitational field and $I(x_0)\neq \pi$. The  angle grows as $x_0$ decreases and at certain radius $x_0 = x_m$ the deflection angle diverges. $x_m$ is the unstable photon orbit radius. The integral (\ref{bending2}) cannot be solved explicitly, therefore, we follow the method developed by Bozza \cite{Bozza:2002zj} in which the integral is expanded near the photon orbit and hence provides an analytical representation of strong gravitational lensing. We define new variable in order to separate the divergent part and the regular part of the $I(x_0)$ as  \cite{Bozza:2002zj}
\begin{eqnarray}
z=1 - \frac{x_0}{x},
\end{eqnarray}
then the  integral (\ref{bending2}) reads 
\begin{eqnarray}\label{integral}
I(x_0)=\int_{0}^{1} R(z,x_0)f(z,x_0)dz,
\end{eqnarray}
where
\begin{eqnarray}
R(z,x_0)=\frac{2x^2}{x_0}P_1(x,x_0),
\end{eqnarray}
\begin{eqnarray}
f(z,x_0)=P_2(x,x_0).
\end{eqnarray}

The function $R(z,x_0)$ is regular everywhere for all  values of $z\;\text{and}\;x_0$, whereas $f(z,x_0)$ diverges as $z\to 0$. The order of divergence in $f(z,x_0)$ can be determined if one does the Taylor series expansion of the argument of the square root in $f(z,x_0)$ to the second order in $z$, i.e.,
\begin{eqnarray}
f(z,x_0)\sim f_0(z,x_0)=\frac{1}{\sqrt{\alpha z+\gamma z^2}},
\end{eqnarray}
where
\begin{eqnarray}\label{alpha}
\alpha&=&\frac{x_0}{C_0}\left[\left(C_0^\prime A_0 -A_0^\prime C_0\right)+L\left(A_0^\prime D_0-A_0 D_0^\prime \right)\right]\\
\gamma &=& \frac{x_0}{2 C_0^2 } \left[ 2 C_0 (A_0 C'_0 - A'_0 C_0 ) + 2 x_0 C'_0 (C_0 A'_0  - A_0 C'_0) - x_0 C_0 (C_0 A''_0 - A_0  C''_0 )\right] + \nonumber\\&& L \left[\frac{x_0^2 C'_0 (A_0 D'_0 - D_0 A'_0)}{C_0{^2}} + \frac{(x_0^2/2) (D_0 A''_0 - A_0 D''_0) + x_0 (D_0 A'_0 - A_0 D'_0))}{C_0}\right];
\end{eqnarray}

The radius of circular photon orbit, $x_{m}$, is the largest real positive root of the Eq.~(\ref{alpha}), i.e.,
\begin{eqnarray}\label{xm}
8 \tilde{b}^3 (\tilde{k} + x)- 8 \tilde{b}^2\left(\tilde{k} + x-x(\tilde{k}+2x)e^{\tilde{k}/x}\right) + 2 \tilde{b} \left( \tilde{k} + x +  x^2 (\tilde{k} + 5 x) e^{2 \tilde{k}/x}-  x (\tilde{k} + 7 x)e^{\tilde{k}/x} \right)
 +e^{\tilde{k}/x} \nonumber\\ \left( 2 a^2 e^{\tilde{k}/x} (\tilde{k} - x) + 2 a e^{\tilde{k}/x} (-\tilde{k} + x) \sqrt{a^2 + x^2+ (-1 + 2 \tilde{b})x e^{-\tilde{k}/x}}+x (-1 +x e^{\tilde{k}/x} (\tilde{k} + x (-3 + 2x e^{\tilde{k}/x} \right) = 0
\end{eqnarray}

\begin{figure*}[t]
	\begin{centering}
		\begin{tabular}{c c}
		    \includegraphics[scale=0.77]{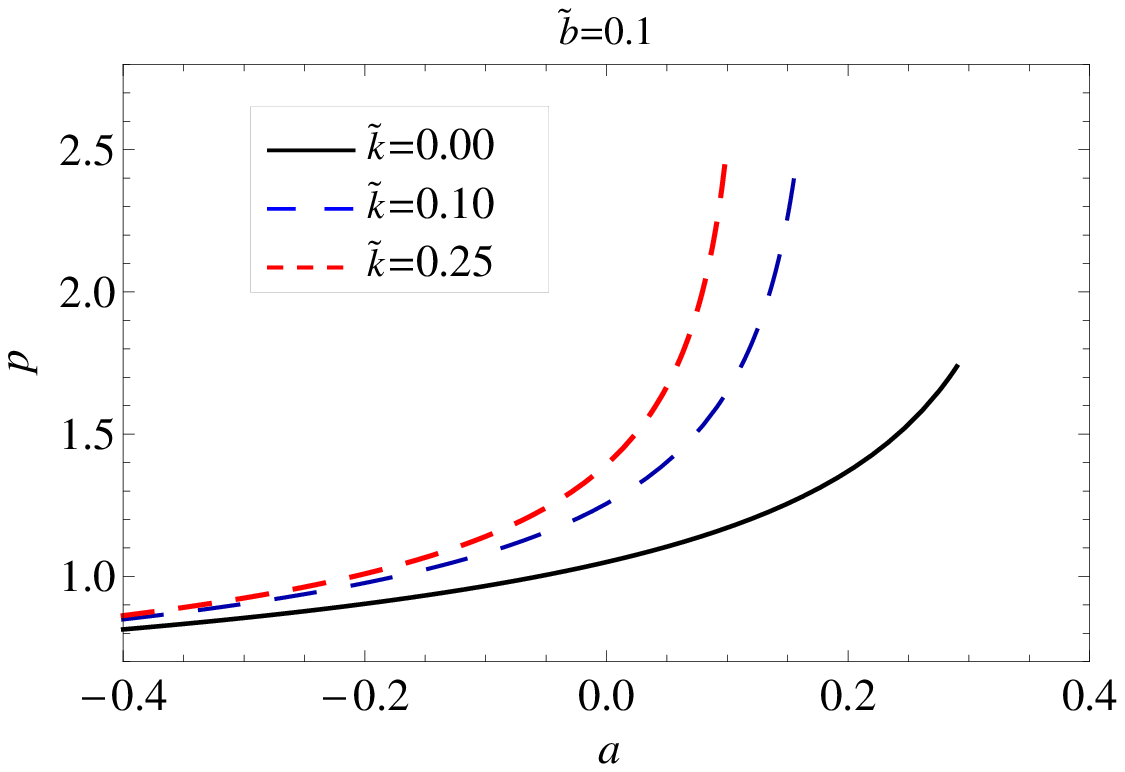}\hspace*{-0.9cm}&
		    \includegraphics[scale=0.77]{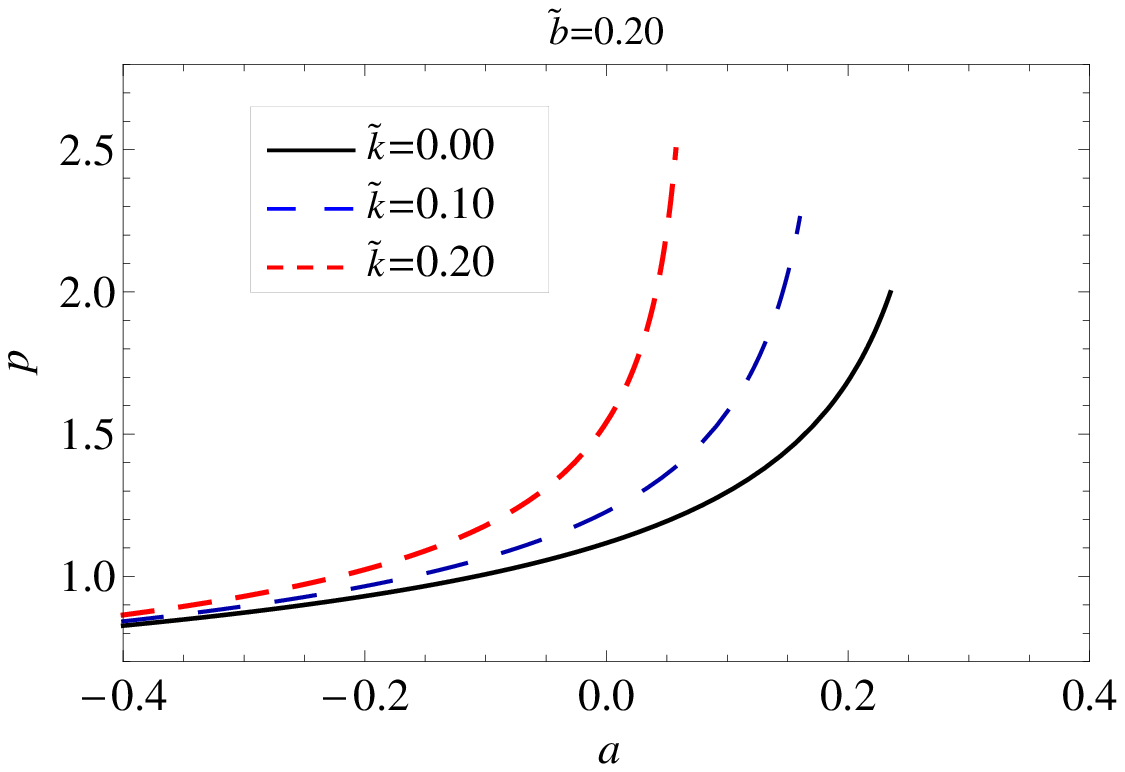}\\
			\includegraphics[scale=0.77]{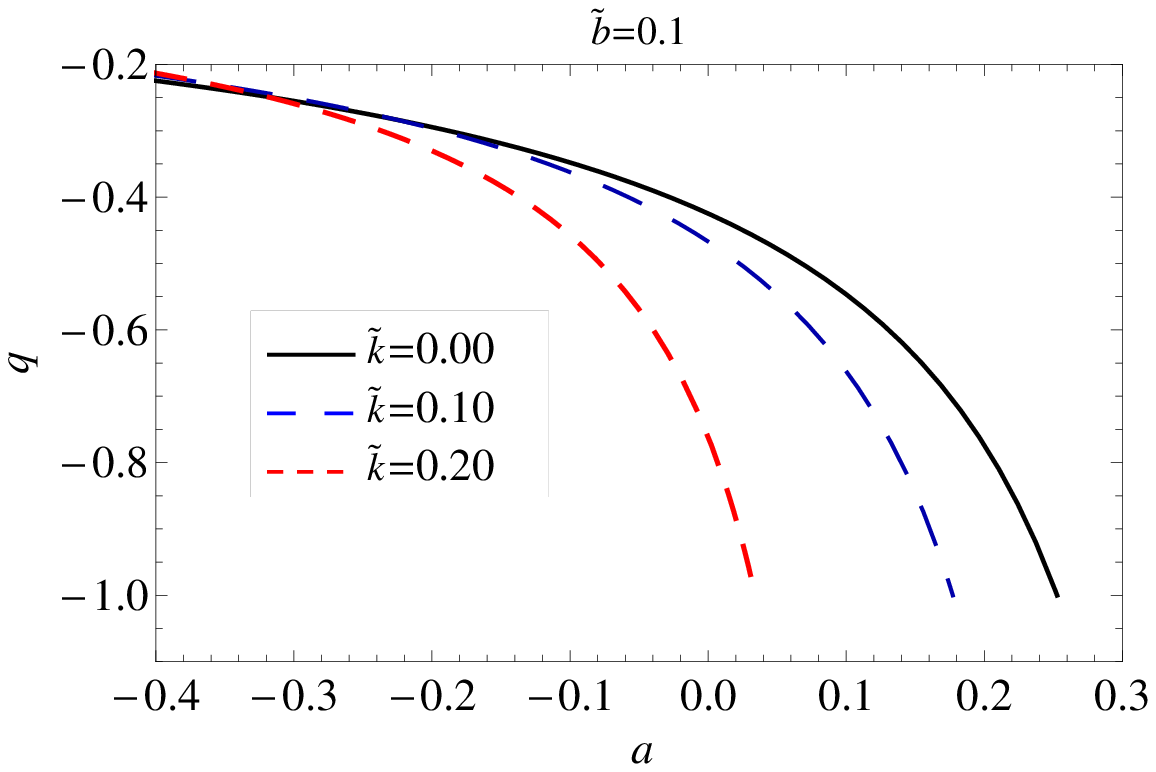}\hspace*{-0.9cm}&
		    \includegraphics[scale=0.77]{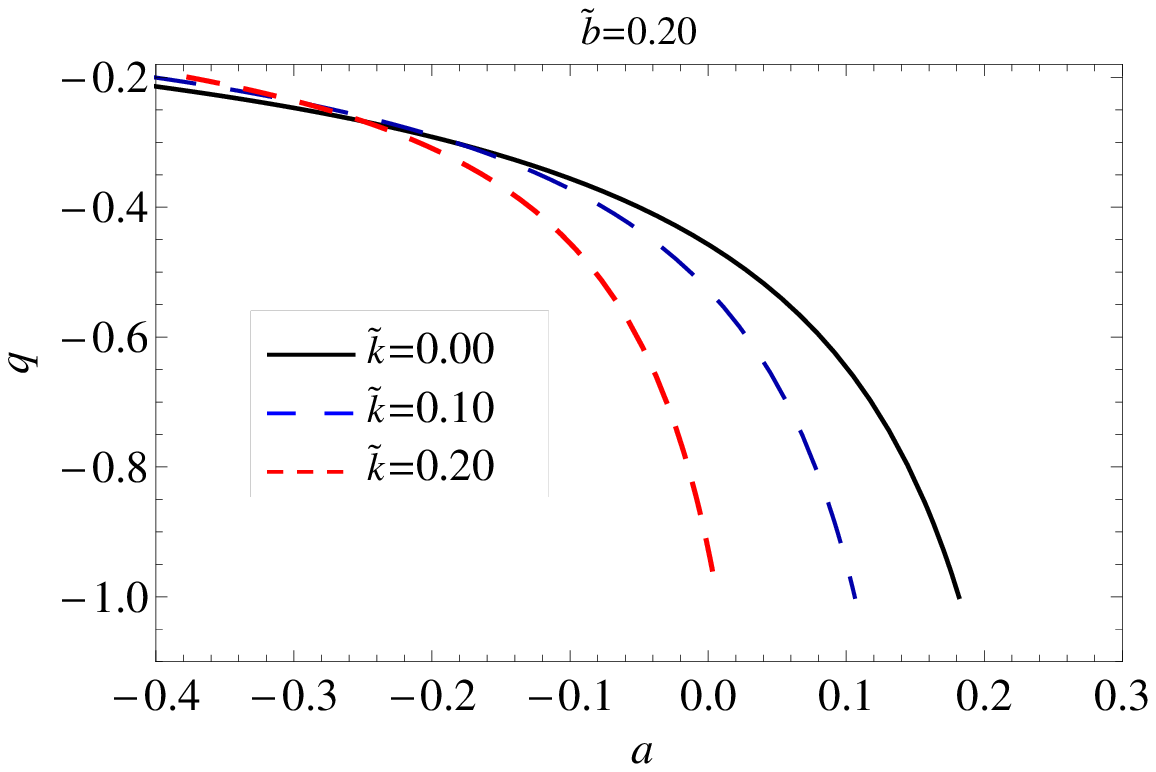}
			\end{tabular}
	\end{centering}
	\caption{Plot showing the behaviour of strong lensing coefficients $p$ and $q$ as a function of black hole spin $a$ and different values of $\tilde{b}$ and $\tilde{k}$.}\label{plot2}		
\end{figure*}
Solving Eq.~(\ref{xm})  gives photon orbit radius in terms of $a$ , $\tilde{k}$ and $\tilde{b}$. The plots in Fig.~\ref{plot3} between $x_m$ and $a$ for different values of $\tilde{k}$  and $\tilde{b}$ suggest that for higher values of $a$, photons are allowed to get closer to the black hole but stay further away from the black hole for negative $a$.   

The divergence of the integral (\ref{integral}) occurs only for the quadratic term in $z$, as for  $x_0 \to x_m $, $\alpha = 0$ and $f(z,x_0) \approx 1/z $, which diverges as $z \to 0$. We  split the Integral (\ref{integral})  into  divergent part and regular part as
\begin{eqnarray}
I(x_0)= I_D(x_0) + I_R(x_0),
\end{eqnarray}
with
\begin{eqnarray}\label{div}
I_D(x_0)&=&\int_{0}^{1} R(0,x_m)f_0(z,x_0)dz,\\
I_R(x_0)&=&\int_{0}^{1} [R(z,x_0)f(z,x_0)-R(0,x_0)f_0(z,x_0)]dz. \label{reg}
\end{eqnarray}
The integral in Eq.~(\ref{div}) has logarithmic divergence at $x_0=x_m$ and  can be analytically  solved to
\begin{eqnarray}\label{phiR}
I_D(x_0)&=& \frac{2 R(0,x_m)}{\sqrt{\gamma}} \log\Bigg(\frac{\sqrt{\gamma + \alpha}+\sqrt{\gamma}}{\sqrt{\alpha}}\Bigg),
\end{eqnarray}
and the integral in Eq.~(\ref{reg}) has divergence subtracted and in order to solve it we expand it in the powers of $(x_0-x_m)$ upto $\mathcal{O}(x_0-x_m)$ as follows
\begin{eqnarray}\label{phiT}
I_R(x_m) = \int_{0}^{1} [R(z,x_m)f(z,x_m)-R(0,x_m)f_0(z,x_m)]dz,
\end{eqnarray} 
which can only be solved numerically. Using Eqs.~(\ref{phiR}) and (\ref{phiT}), one can write the deflection angle as
\begin{eqnarray}\label{def}
\alpha_{D}(\theta)=-p \log\Big(\frac{\theta D_{OL}}{u_m}-1\Big)+q + \mathcal{O}\left(u-u_m\right),
\end{eqnarray}
where $D_{OL}$ is the observer to lens distance. The deflection angle coefficients $p$ and $q$ in the strong field limit are given by 
\begin{figure*}[t]
	\begin{centering}
		\begin{tabular}{cc}
		    \includegraphics[scale=0.75]{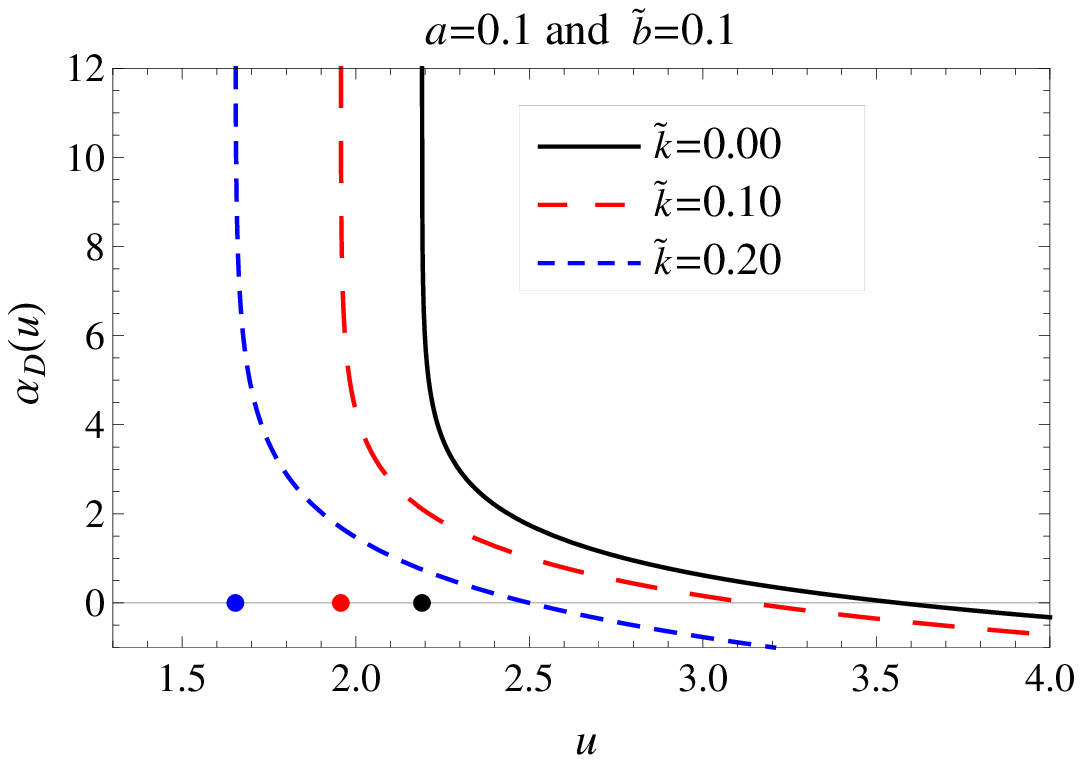}\hspace*{-0.9cm}&
		    \includegraphics[scale=0.75]{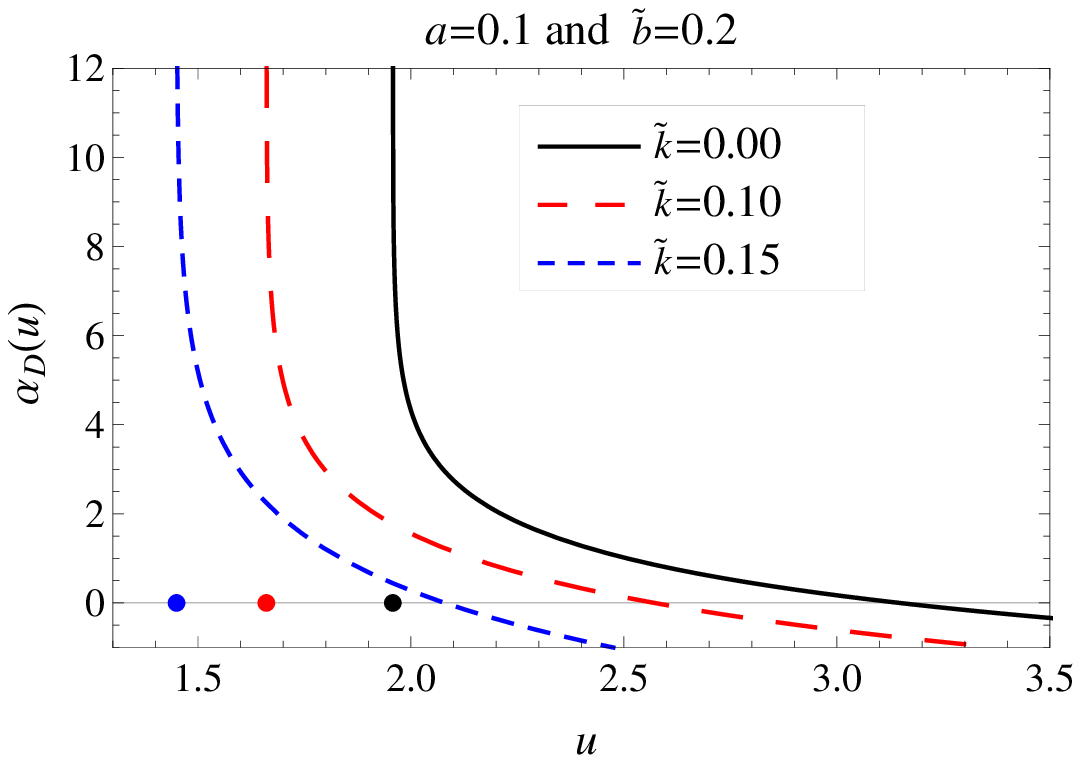}\\
		     \includegraphics[scale=0.75]{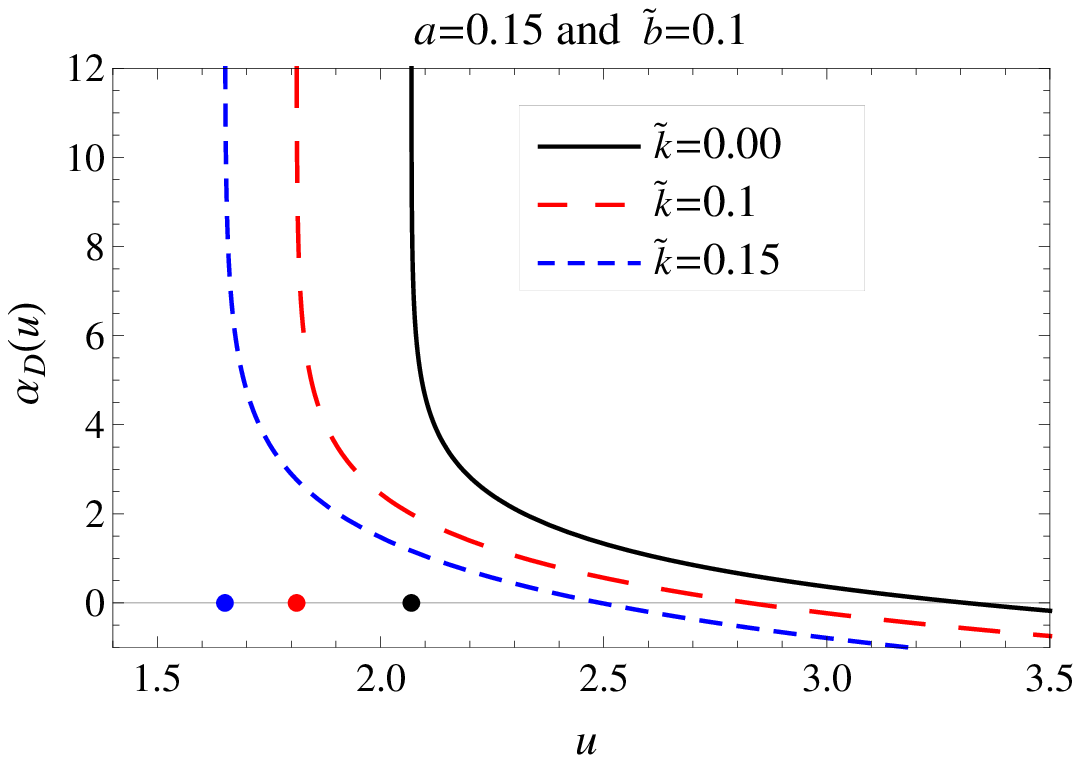}\hspace*{-0.9cm}&
		    \includegraphics[scale=0.75]{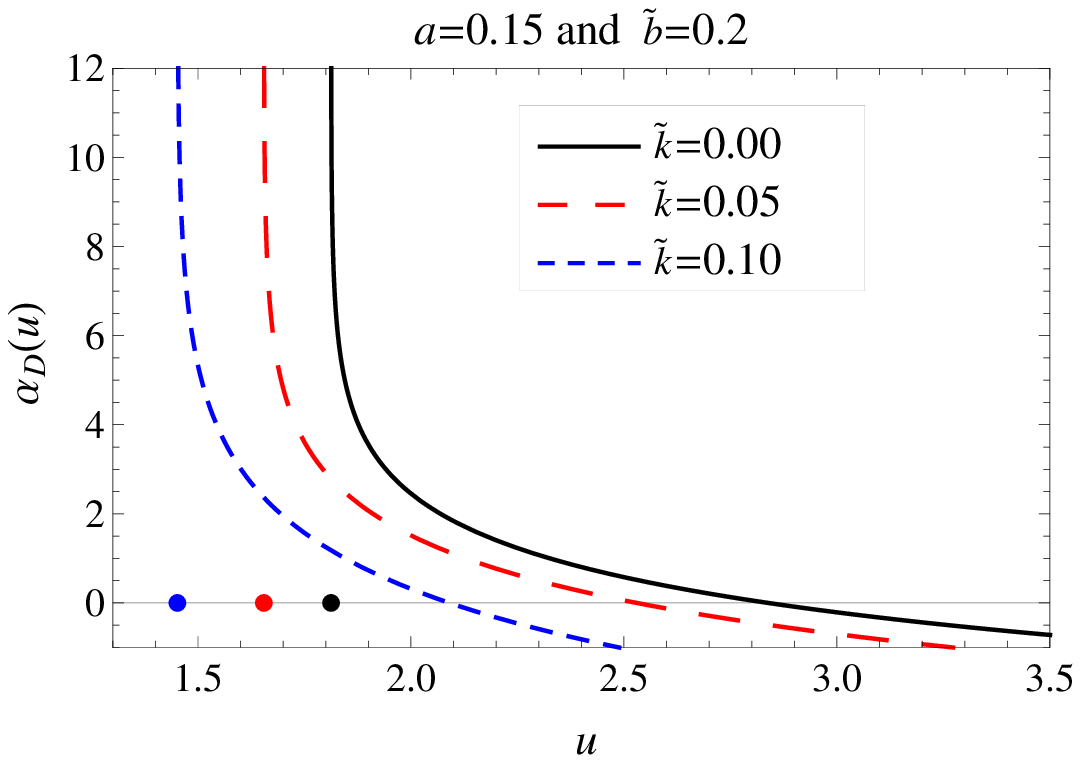}\\
			\end{tabular}
	\end{centering}
	\caption{Plot showing the variation of light deflection angle as a function of impact parameter $u$ for different values of $a$, $\tilde{b}$ and $\tilde{k}$. Points on the horizontal axis correspond to the impact parameter $u = u_m$ at which deflection angle diverges.}\label{plot4}		
\end{figure*}
\begin{eqnarray}
p = \frac{R(0,x_m)}{2\sqrt{\gamma _m}}, ~~~ \textrm{and}~~~ q = -\pi +I_R(x_m) + p\log\frac{c x_m^2 }{u_m^2}
\end{eqnarray}
which in terms of the metric elements can be written as
\begin{eqnarray}\label{abar}
p = \frac{x_m^{3/2}}{\sqrt{\left(a^2 \frac{e^{\tilde{k}/x_m}}{(c_1-1)} + x_m \right)\left(a^2 (1+1/c_1)+x_m c_1 e^{\tilde{k}/x_m}\right)\gamma_m}},
\end{eqnarray}
where
\begin{eqnarray}
c_1 &=& 2 b + x_m e^{\tilde{k}/x_m}.
\end{eqnarray}
Expanding Eq.~(\ref{angmom}) about $x_0=x_m$, we get 
\begin{eqnarray}
u-u_m &=& c (x_0-x_m)^2.
\end{eqnarray}
All the functions with subscript $m$ are evaluated at $x_0=x_m$. The $p$ and $q$, in Eq.~(\ref{abar}), are the strong-field deflection coefficients and are plotted in Fig.~\ref{plot2}, which shows that  $p$ and $q$, respectively, increase and decrease with increasing $a$. Both the coefficients diverge for higher values of $a$, suggesting that strong-field limit is invalid at higher values. In the limit $a\to 0$ , $\tilde{k}\to 0$, and $\tilde{b}\to 0$, the coefficients retain the values for  Schwarzschild black hole as $p$ = 1 and $q=-0.4002$ \cite{Kumar:2020sag,Islam:2020xmy}, whereas in the limit $\tilde{b}\to 0$ and  $\tilde{k}\to 0$, the values of $p = 1.090$ and $q=-0.4887$ correspond to the Kerr black hole with $a=0.1$. If only  
$\tilde{b}\to 0$, the values of $p$ and $q$ for the nonsingular black hole \cite{Islam:2020sag}, respectively, are 1.1753 and -0.55412. The resulting light deflection angle $\alpha_{D}(u)$ in the strong deflection limit is plotted in Fig~\ref{plot4} and this is evident that it diverges at $u=u_m$. For a fixed value of $u$, the deflection angle decreases with increasing $\tilde{b}$ and $\tilde{k}$, suggesting that deflection angle is highest for the Schwarzschild black hole than the Kerr and Kerr-sen black holes.  Also the presented study of deflection angle are valid only in the vicinity of $u_m$. For $u \gg u_m$, the strong deflection is not a valid approximation and a different expansion should be performed.

\section{Observables and relativistic images}\label{Sec4}
The geometrical configuration for gravitational lensing constitutes a light source, an observer and a black hole that act as a lens and lies in between the source and observer. The photons emitted by the source (S) is deviated by the black hole (L) and reach the observer (O). The observer perceives the image of the source at an angle $\theta$ concerning the optical axis OL, whereas,  the source is at an angular position $\beta$ with respect to OL. The emission direction and the detection direction make an angle $\alpha_{D}(\theta)$, which is called the deflection angle. \\

We introduce a coordinate independent Ohanian lens equation \citep{Ohanian:1987pc} connecting the source and observer positions as
\begin{eqnarray}\label{oho}
\xi &=& \frac{D_{OL}+D_{LS}}{D_{LS}}\theta-\alpha_{D}(\theta),   
\end{eqnarray} 
$\xi$ is the angle between the direction of source and optical axis. $D_{LS}$ is the lens-source distance. The angle $\xi$ and $\beta$ are related by 
\begin{eqnarray}\label{rel}
\frac{D_{OL}}{\sin(\xi-\beta)} &=& \frac{D_{LS}}{\sin \beta}
\end{eqnarray}
The lensing effects are most evident when all the objects are almost aligned. In fact, this is the case when the relativistic images are most prominent. So we study the case when the angles $ \beta $, $\xi$ and $\theta$ are very small. However, if a ray of light emitted by the source S follow multiple loops around the black hole before reaching the observer, $\alpha_D$ must be very close to a multiple of 2$\pi$. Replacing  $\alpha_{D}(\theta)$  by  $2n\pi + \Delta\alpha _n$ ,with  $n \in N $ and $ 0 < \Delta\alpha _n \ll 1 $, we can rewrite Eq.~(\ref{oho}) for small values of  $\beta$, $\xi$ and $\theta$ using Eq.~(\ref{rel}) as,
\begin{eqnarray}\label{lensequation}
\beta &=& \theta -\frac{D_{LS}}{D_{OL}+D_{LS}} \Delta\alpha _n.
\end{eqnarray}  
Given the angular position of source $\beta $ and the distances of observer and source from the black hole, one can calculate the image positions by using Eq.~(\ref{lensequation}). The deflection angle increases as the light ray trajectory gets closer to the event horizon or photon sphere, such that for a particular value of $u$, the light form loops around the black hole resulting in $ \alpha_{D}(\theta)> 2\pi$. With further decreasing impact parameter, the light ray winds several times around the black hole before escaping to the observer. Finally, for critical impact parameter, corresponding to the closest distance $x_m$, the deflection angle diverges. For each loop of the light geodesic, there is a particular value of impact parameter at which light reaches the observer from the source. So infinite sequence of images will be formed on each side of the lens. Equation~(\ref{def}) with $\alpha_{D}(\theta_n{^0}) = 2n\pi $ leads to 
\begin{eqnarray}\label{theta}
\theta_n{^0} &=& \frac{u_m}{D_{OL}}(1+e_n),
\end{eqnarray}  
where 
\begin{eqnarray}
e_n &=& e^{\frac{q-2n\pi}{p}}.
\end{eqnarray}

Here, $n$ is the number of loops followed by the photons around the black hole. The deflection angle is expanded around $\theta_n{^0}$ as:
\begin{eqnarray}
\alpha_{D}(\theta) &=& \alpha_{D}(\theta_n {^0}) +\frac{\partial \alpha_{D}(\theta)}{\partial \theta } \Bigg |_{\theta_n{^0}}(\theta-\theta_n{^0})+\mathcal{O}(\theta-\theta_n{^0}). 
\end{eqnarray}
On using (\ref{theta}) and setting $\Delta\theta_n= (\theta-\theta_n{^0}) $ we obtain 
\begin{eqnarray}
\Delta\alpha_n &=& -\frac{pD_{OL}}{u_m e_n}\Delta\theta_n.
\end{eqnarray}
Finally the lens equation (\ref{lensequation})  becomes
\begin{eqnarray}
\beta &=& ( \theta_n{^0} + \Delta\theta_n )+\frac{D_{LS}}{D_{OL}+D_{LS}}\Bigg(\frac{p D_{OL}}{u_m e_n}\Delta\theta_n\Bigg).
\end{eqnarray}
The second term in the above equation is very small as compared to the third term, there by neglecting it, we get 
\begin{eqnarray}\label{angpos}
\theta_n &=& \theta_n{^0} + \frac{D_{OL}+D_{LS}}{D_{LS}}\frac{u_me_n}{D_{OL}p}(\beta-\theta_n{^0}).
\end{eqnarray}
\begin{figure*}[t]
	\begin{centering}
		\begin{tabular}{cc}
		    \includegraphics[scale=0.75]{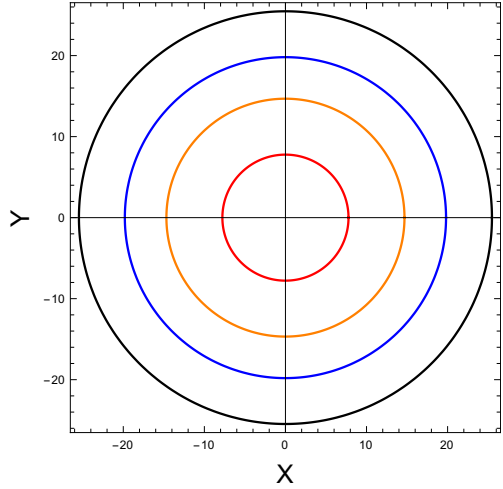}&
		    \end{tabular}
	\end{centering}
\caption{Plot showing the outermost Einstein rings for supermassive black holes at the center of nearby galaxies in case of Schwarzschild geometry. Black, red, blue and green circles, respectively, correspond to Sgr A*, M87*, NGC 4649 and NGC 1332 black holes.}
\label{plotER}	 
\end{figure*}	
The above equation, for $\beta>0$, determines images only on the same side of the source ($\theta>0$). To obtain the images on the opposite side, one can solve the same equation with the source placed at $-\beta$. Finally using Eq.~(\ref{angpos}), we obtain full set of primary ($+\beta$) and secondary images ($-\beta$). 
\subsection{Einstein Ring}
Einstein ring is the most spectacular effect of gravitational lensing, which arises when the lens is in the line of sight of source, i.e., the source lies on the optical axis connecting the observer and lens \cite{Einstein}. Multiple Einstein rings may arise in case of complex lens systems \cite{Ohanian:1987pc,Virbhadra:1999nm,Virbhadra:2002ju} whereas a partial double Einstein ring was also discovered in Ref.~\cite{Gavazzi:2008} in which the authors suggested that these rings are formed from two sources situated at different distances from the lens. Location of the Einstein rings can be obtained by solving the lens equation with $\beta=0$. The Einstein rings are said to be relativistic if the deflection angle is larger that $2\pi$. For the particular orientation of source, lens and the observer being perfectly aligned ($\beta=0$), the equation (\ref{angpos}) reduces to
	\begin{eqnarray}\label{Ering}
		\theta_n^{E} &=& \left(1-\frac{D_{OL}+D_{LS}}{D_{LS}}\frac{u_me_n}{D_{OL}p} \right) \theta_n^{0},
	\end{eqnarray} 
and its solution gives radii of the Einstein rings. For a particular case when  the lens is exactly midway between the source and observer, Eq.~(\ref{Ering})  yields
	\begin{eqnarray}\label{Ering2}
		\theta_n^{E} &=& \left(1-\frac{2 u_me_n}{D_{OL}p} \right) \left(\frac{u_m}{D_{OL}}(1+e_n)\right).
	\end{eqnarray} 
	Since $D_{OL} \gg u_m$, the Eq.~(\ref{Ering2}) yields 
	\begin{eqnarray}\label{Ering3}
		\theta_n^{E} &=& \frac{u_m}{D_{OL}}\left(1+e_n \right),
	\end{eqnarray}

\begin{figure*}[t]
	\begin{centering}
		\begin{tabular}{c c}
		    \includegraphics[scale=0.77]{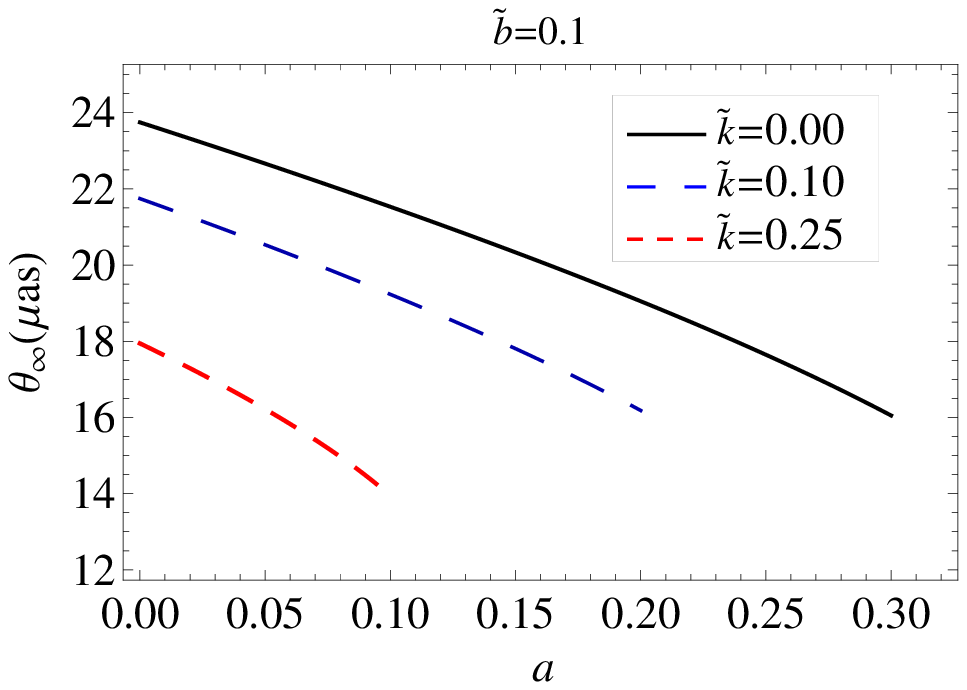}\hspace*{-0.7cm}
		    \includegraphics[scale=0.77]{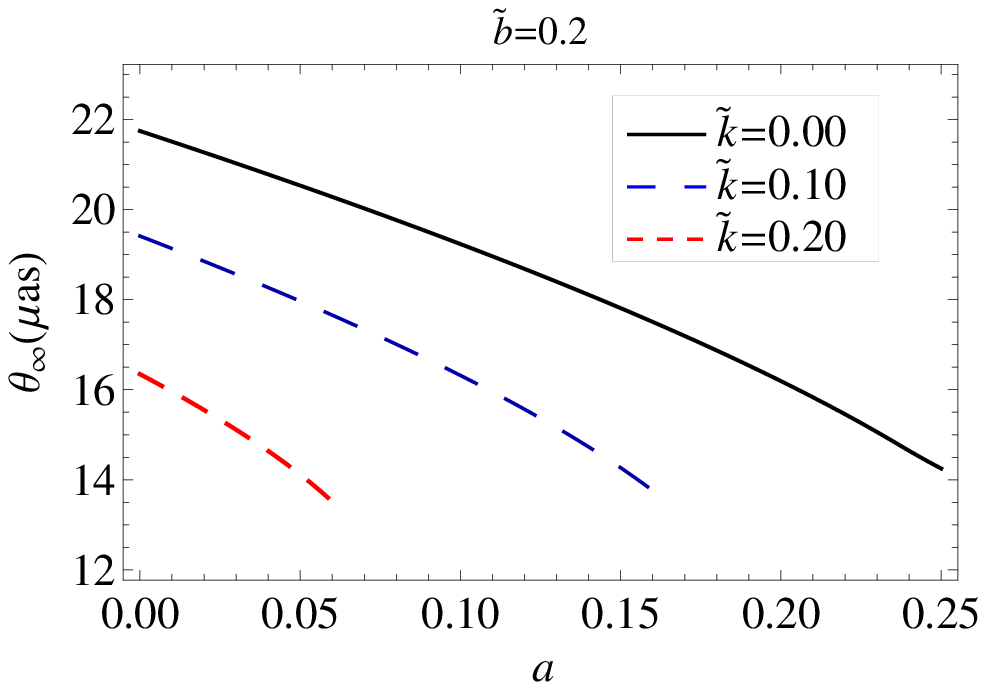}\\
			\includegraphics[scale=0.77]{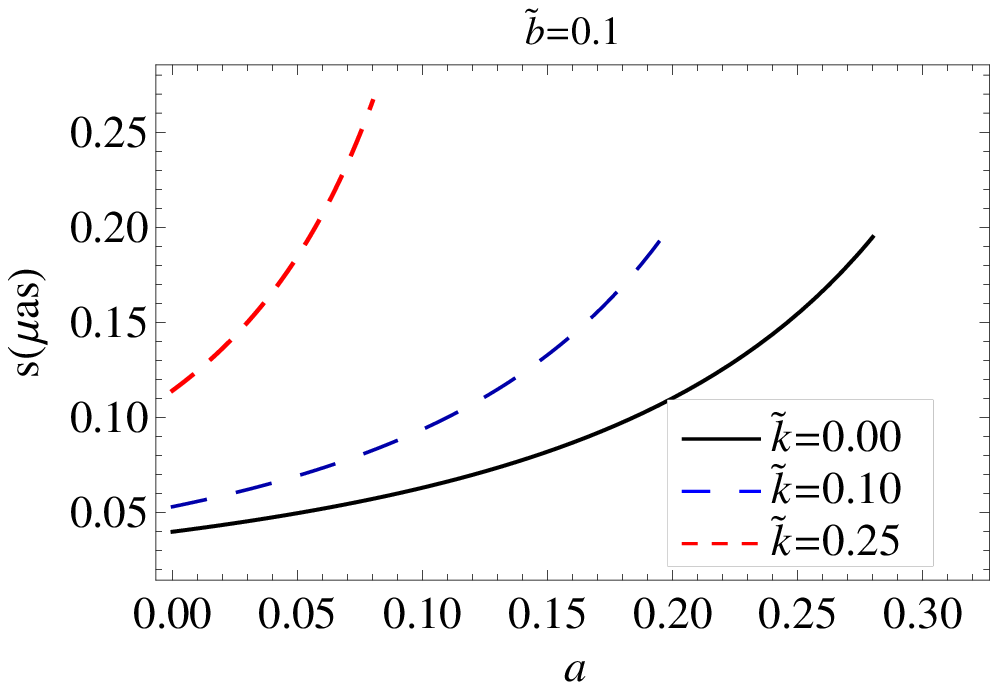}\hspace*{-0.7cm}
		    \includegraphics[scale=0.77]{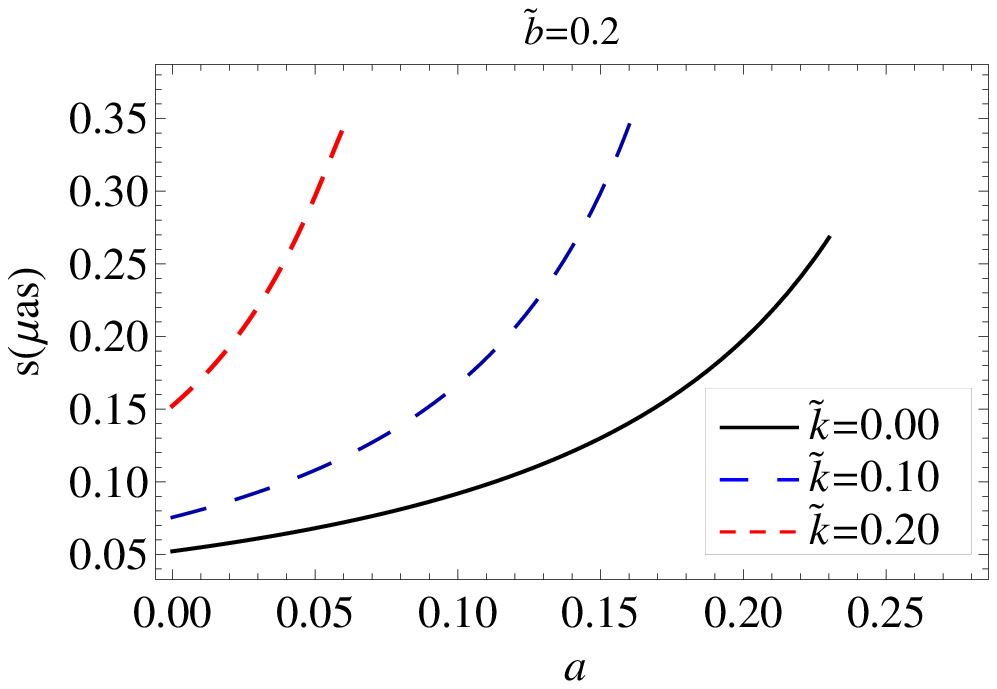}\\
			\includegraphics[scale=0.78]{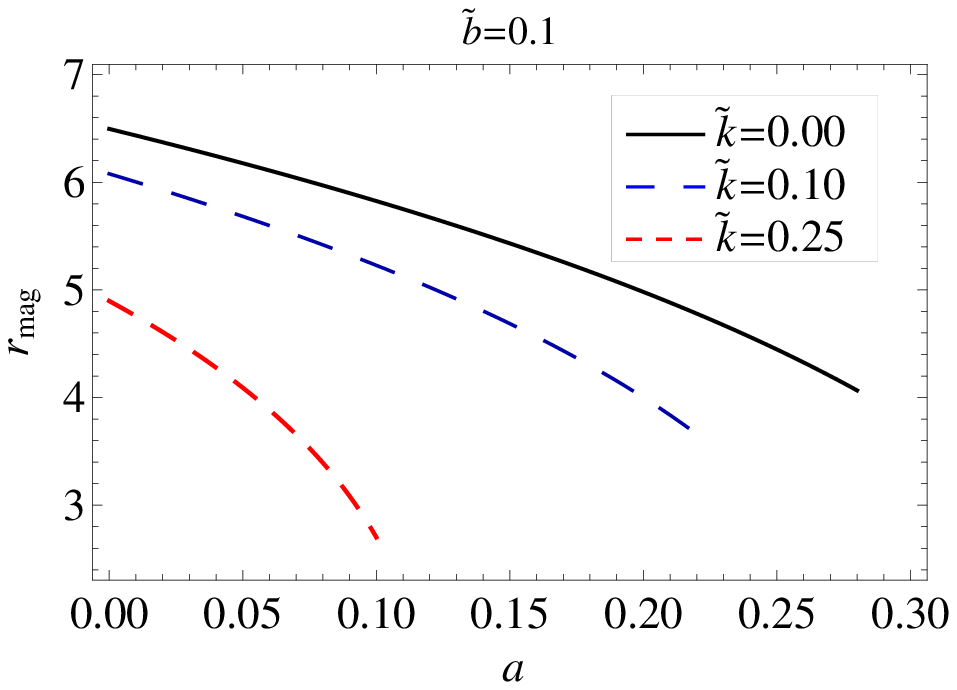}\hspace*{-0.7cm}
		    \includegraphics[scale=0.78]{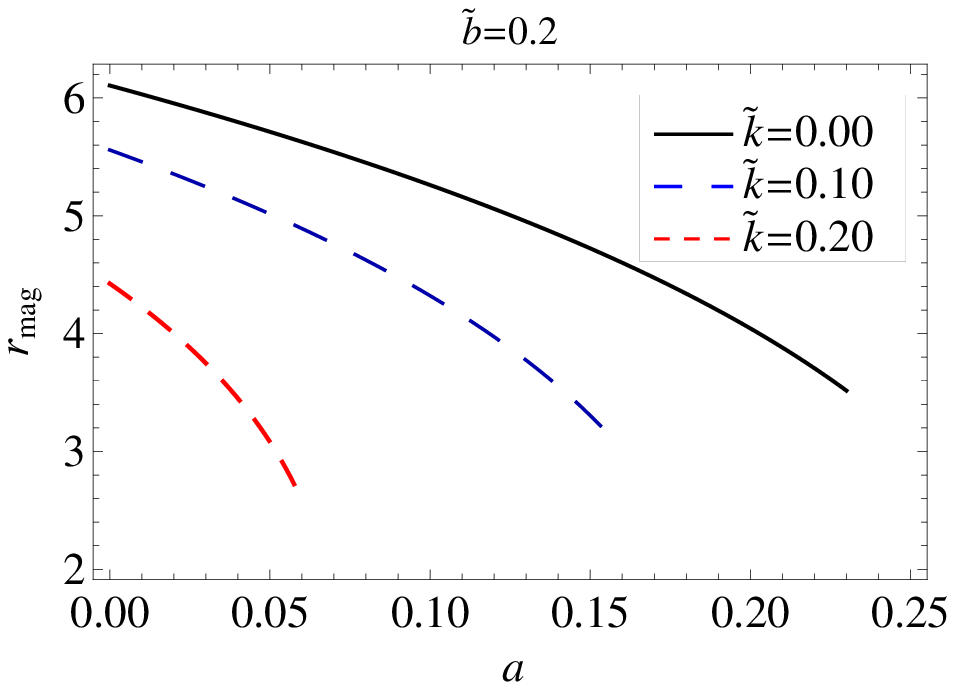}
			\end{tabular}
	\end{centering}
	\caption{Plot showing the behavior of strong lensing observables  $\theta_{\infty}$, $s$, and $r_{mag}$ as a function of $a$ and different values  of $\tilde{b}$ and $\tilde{k}$ for the Sgr A* black hole.}\label{plot1a}		
\end{figure*}
\begin{figure*}[t]
	\begin{centering}
		\begin{tabular}{c c}
		    \includegraphics[scale=0.77]{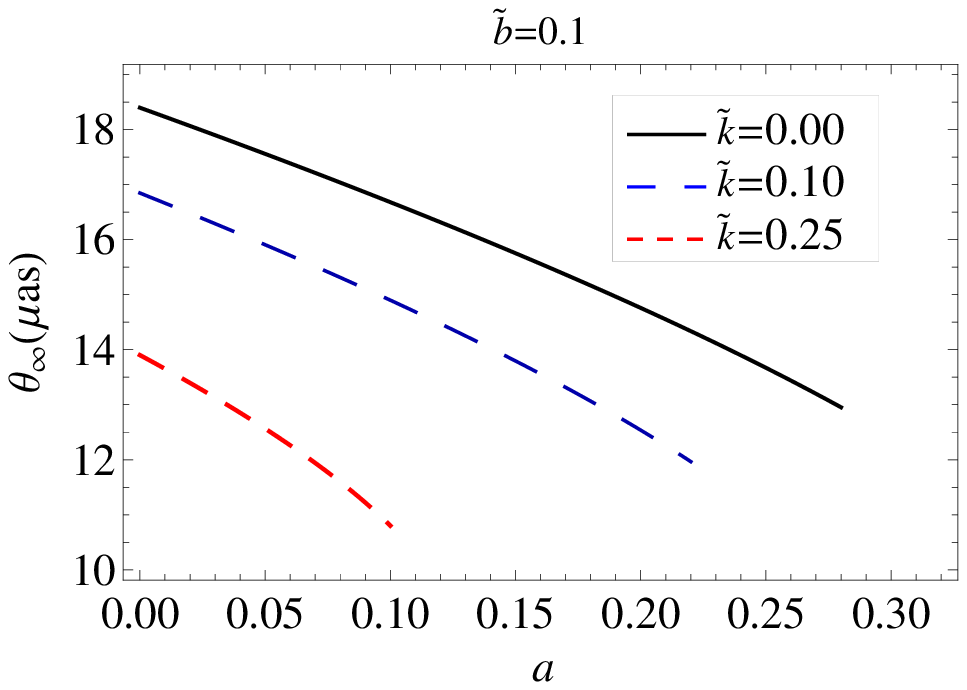}\hspace*{-0.7cm}
		    \includegraphics[scale=0.77]{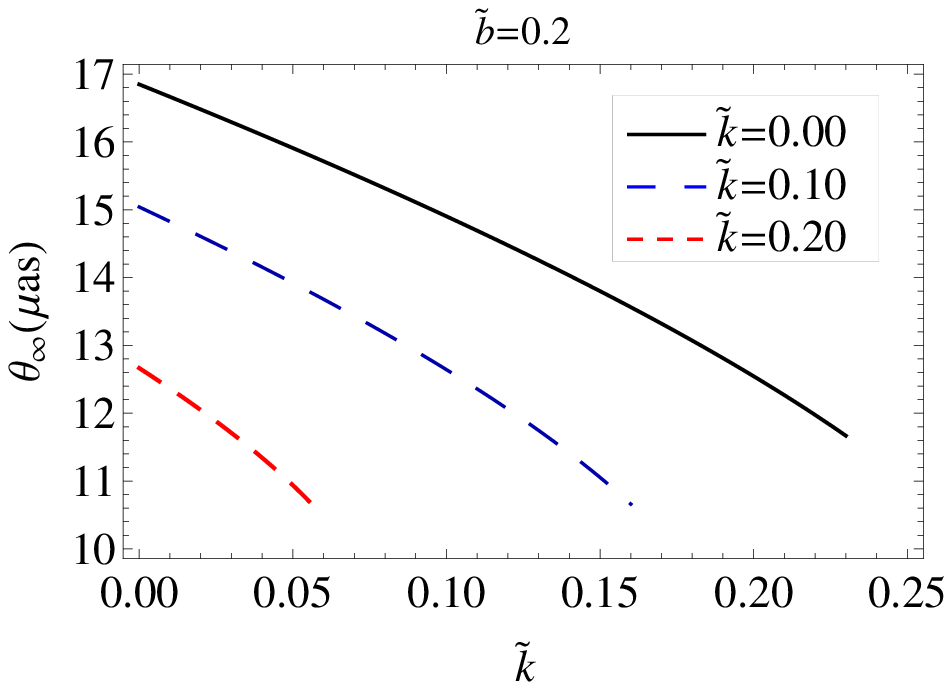}\\
			
			\includegraphics[scale=0.77]{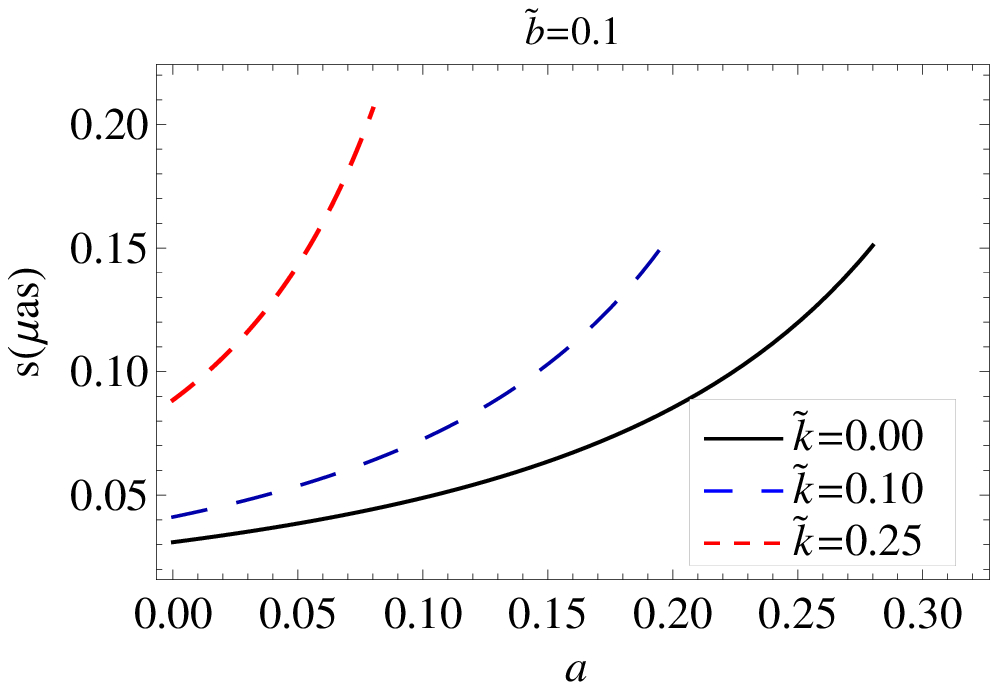}\hspace*{-0.7cm}
		    \includegraphics[scale=0.77]{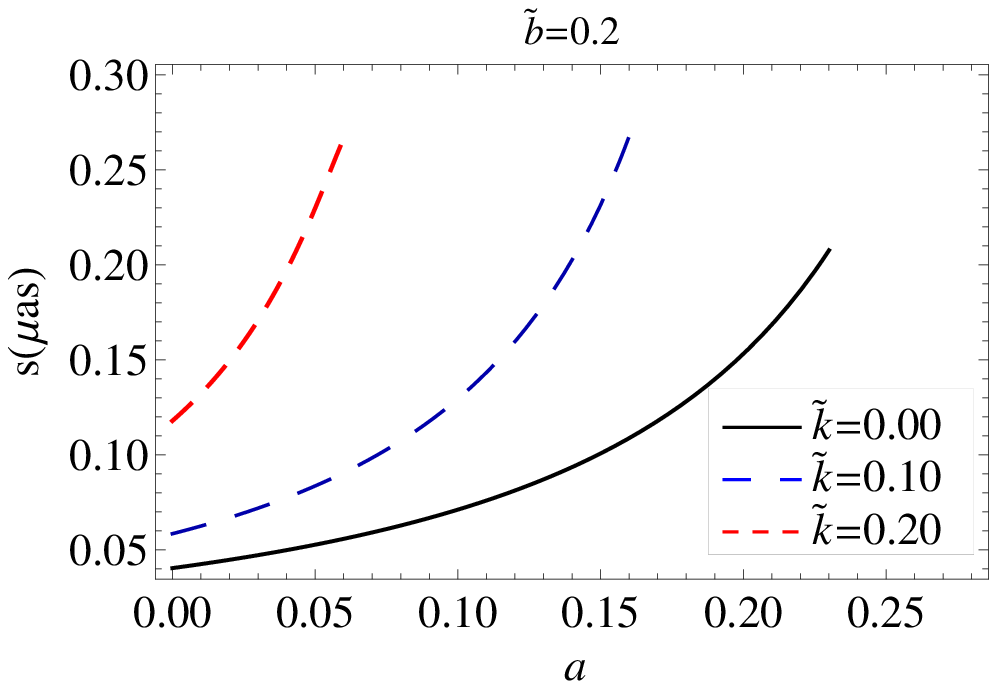}\\
			
			\includegraphics[scale=0.79]{mag1.eps}\hspace*{-0.7cm}
		    \includegraphics[scale=0.79]{mag2.eps}
			\end{tabular}
	\end{centering}
	\caption{Plot showing the behavior of strong lensing observables $\theta_{\infty}$, $s$, and $r_{mag}$ as a function of $a$ and different values  of $\tilde{b}$ and $\tilde{k}$  for the M87* black hole.}\label{plot1b}		
\end{figure*}  
which gives the radius of the $n$th relativistic Einstein ring. Note that $n=1$ represents the outermost ring, and as $n$ increases, the  radius of the ring decreases. Also, it can be conveniently determined from Eq.~(\ref{Ering3}) that radius of the Einstein ring increases with the mass of the black hole and decreases as the distance between the observer and lens increases. In  the Fig.~\ref{plotER},  we plot the outermost relativistic Einstein rings of Sgr A*, M87* and other nearby black holes.

\begin{table*}[t]
\centering

	\begin{tabular}{p{1cm} p{1cm} p{1cm} p{1.6cm} p{1.6cm} p{1cm}}
		
\hline\hline
\multicolumn{3}{c}{}&
\multicolumn{2}{c}{Lensing Coefficients}&
\multicolumn{1}{c}{}\\
{$a$ } & {$\tilde{b}$} & {$\tilde{k}$} & {$p$}&{$q$} & {$u_m/R_s$}\\ \hline
\hline
\\
\multirow{7}{*}{0.0} & 0.0 & 0.00 & 1.00 & -0.400230 & 2.59808 \\                                            
                      
                    & 0.1 & 0.00 & 1.05025 &-0.424710 & 2.41633\\

                     & 0.1 & 0.10 & 1.12235& -0.466846 &2.21261\\      

                     & 0.1 & 0.25 &1.39307 & -0.761242& 1.82579\\                       

                     & 0.2 &0.00 &1.11731 & -0.457238 & 2.2129\\

                     & 0.2 & 0.10  &1.22731 &-0.527185& 1.97543\\              
                     
                     & 0.2 & 0.25  &1.54174& -0.926537& 1.66338\\               
\hline 
\\
\multirow{7}{*}{0.1} & 0.0 &0.00  & 1.09030& -0.488790& 2.39162\\      
                     
                      & 0.1 &0.00  &1.17103& -0.546462& 2.19087\\                    

                     & 0.1 &0.10 &1.30478 & -0.662210& 1.95691\\                                  

                     & 0.1 &0.20  &1.65085& -1.104970& 1.65322\\                                 

                     & 0.2 &0.00 &1.29664& -0.645978& 1.95737\\      

                     & 0.2 &0.10 &1.57844& -0.947654& 1.66042\\                       

                     & 0.2 & 0.15 &2.05301& -1.711780& 1.44961\\                                   
\hline
\\
\multirow{7}{*}{0.15}  &0.00 &0.00  &1.14883& -0.549965& 2.2826\\
                       
                      & 0.10 &0.00 &1.25587& -0.639491& 2.06901\\                                           
                      & 0.10 &0.10 &1.45559& -0.845230& 1.81185\\                         
                      & 0.10 &0.15  &1.66897& -1.124820& 1.65135\\

                     & 0.20 &0.00  &1.44421& -0.821852& 1.81247\\
    
                      &0.20 & 0.05 &1.62771& -1.036550& 1.65454\\
  
                     & 0.20 & 0.10  &2.06371& -1.694630& 1.45167\\
		\hline\hline
	\end{tabular}
	
\caption{Estimates for the strong lensing coefficients for different values of $a$, $\tilde{b}$ and $\tilde{k}$. The cases $a=\tilde{b}=\tilde{k}=0$, $\tilde{b}=\tilde{k}=0$ and $\tilde{k}=0$, respectively, correspond to Schwarzschild, Kerr and Kerr--Sen black holes.  
\label{table1}  
	}
\end{table*}   

\begin{table*}[t]
\resizebox{1\textwidth}{!}{
 \begin{centering}	
	\begin{tabular}{p{0.8cm} p{0.8cm} p{1cm} p{1.5cm} p{2cm} p{1.5cm} p{2cm} p{1.5cm} p{2cm} p{1.5cm} p{2cm} p{1.5cm} p{1.3cm} p{1.5cm}p{1.3cm}}
		
\hline\hline
\multicolumn{3}{c}{}&
\multicolumn{2}{c}{Sgr A*}&
\multicolumn{2}{c}{M87*}& 
\multicolumn{2}{c}{NGC 4649}&
\multicolumn{2}{c}{NGC 1332}\\
{$a$ } & {$\tilde{b}$} & {$\tilde{k}$} & {$\theta_\infty $ ($\mu$as)} & {$s$ ($\mu$as) }  & {$\theta_\infty $ ($\mu$as)} & {$s$ ($\mu$as) }  & {$\theta_\infty $ ($\mu$as)} & {$s$ ($\mu$as) } &{$\theta_\infty $ ($\mu$as)} & {$s$ ($\mu$as) } &  {$r_m$} \\ \hline
\hline
\\
\multirow{7}{*}{0.0}  & 0.0 & 0.00 & 25.5324& 0.031953 &   19.7820 & 0.0247571   & 14.6615 & 0.0183488 & 3.31684 & 0.00415101 & 6.82188 \\                                            
                       
                        & 0.1 & 0.00 & 23.7462 & 0.039973 &  18.3982 & 0.0309709 & 13.6359 & 0.0229542 & 3.0848 & 0.00519287  & 6.49549\\
                     
                         & 0.1 & 0.10 & 21.7442& 0.053137  &      16.8470 &0.041170 & 12.4862 & 0.0305132 & 2.82473 & 0.00690293 & 6.07824 \\      
                      
                        & 0.1 & 0.25 & 17.9428 & 0.114225 & 13.9017 & 0.0884992 & 10.3033 & 0.0655915 & 2.33089 & 0.0148386 & 4.89702\\                       
                     
                        & 0.2 & 0.00 & 21.7471& 0.052170 &    16.8493 & 0.0404207 & 12.4879 & 0.0299579 & 2.8251 & 0.0067773  &  6.10563\\
                      
                       & 0.2 & 0.10 & 19.4134& 0.075541 &  15.04110 & 0.0585281 & 11.1478 & 0.0433783  &2.52193 & 0.00981335 & 5.55842\\              
                     
                        & 0.2 & 0.20 & 16.3467 & 0.152237 & 12.6651 & 0.117951 & 9.38681 & 0.0874196 & 2.12355 & 0.0197767 & 4.42478\\               
\hline 
\\
\multirow{7}{*}{0.1}    & 0.0 & 0.00 & 23.5035 & 0.047172 &  18.210 & 0.0365481 & 13.4965 & 0.0270878 & 3.05327 & 0.006128 & 6.25687\\      
                       & 0.1 & 0.00 & 21.5306 & 0.063121 & 16.6815 & 0.048905 & 12.3635 & 0.0362461 & 2.79697 & 0.0081998 & 5.82554\\                    
                        & 0.1 & 0.10 & 19.2314 & 0.093807 & 14.9001 & 0.0726804 & 11.0433 & 0.0538673 & 2.49829 & 0.0121863 &  5.22839\\                                  
                      & 0.1 & 0.20 & 16.2469 & 0.184991  &      12.5878 & 0.1433270 & 9.32947 & 0.106228 & 2.11058 & 0.0240316 & 4.13233\\                                 
                       & 0.2 & 0.00 &  19.2359 & 0.091888 & 14.9036 & 0.0711934  & 11.0459 & 0.0527652& 2.49888 & 0.0119369 &  5.26122\\      
                       & 0.2 & 0.10 & 16.3176 & 0.167168 &      12.6426 & 0.1295190 & 9.37011 & 0.0959935 & 2.11978 & 0.0217164  &4.32191\\                       
                        & 0.2 & 0.15 & 14.2459 & 0.290025 & 11.0375 & 0.224706 & 8.18048 & 0.166542  & 1.85065 & 0.0376763 & 3.32286\\                                   
\hline
\\
\multirow{7}{*}{0.15}    & 0.0 & 0.00 &  22.4321& 0.058575 
& 17.3800 & 0.045383 &12.8812 & 0.0270878 & 2.91409 & 0.006128 & 5.93812\\
                        & 0.1 & 0.00 &  20.333 & 0.082083 & 15.7536 & 0.063596 & 11.6759 & 0.0471347  & 2.6414 & 0.0106632& 5.43201 \\                                                           
                        & 0.1 & 0.10 &  17.8058 & 0.132954 & 13.7956 & 0.103010 & 10.2247 & 0.0763465 & 2.3131 & 0.0172717 & 4.68668\\                         
                      & 0.1 & 0.15 & 16.2285 & 0.191688   & 12.5735 & 0.148516 & 9.31892 & 0.110073 &  2.10819 & 0.0249016 & 4.08747 \\
                        
                         & 0.2 & 0.00 & 17.8119 & 0.130053   & 13.8004 & 0.100763 & 10.2282 & 0.0746805 & 2.3139 & 0.0168948 & 4.72362\\
                        
                        & 0.2 & 0.05 &  16.2599 & 0.181180   & 12.5979 & 0.140375 & 9.33694 & 0.10404 & 2.11227 & 0.0235366  &  4.19110\\
                        
                        & 0.2 & 0.10 &  14.2662 & 0.298834 & 11.0532 & 0.231531 & 8.19209 & 0.1716 & 1.85328 & 0.0388206  &  3.30564\\
		\hline\hline
	\end{tabular}
\end{centering}
}	
	\caption{Estimates for the lensing observables for supermassive black holes at the center of nearby galaxies for different values of $a$, $\tilde{b}$ and $\tilde{k}$. The cases $a=\tilde{b}=\tilde{k}=0$, $\tilde{b}=\tilde{k}=0$ and $\tilde{k}=0$, respectively, correspond to Schwarzschild, Kerr and Kerr--Sen black holes.
\label{table2}  
	}
\end{table*}   

Image magnification is another good source of information. It is defined as the ratio between solid angles of the image and the source. Magnification of the $n$th image is given by \cite{Bozza:2002zj,Bozza:2002af}
\begin{eqnarray}
\mu_n &=& \frac{1}{\beta} \Bigg[\frac{u_m}{D_{OL}}(1+e_n) \Bigg(\frac{D_{OS}}{D_{LS}}\frac{u_me_n}{D_{OL}p}  \Bigg)\Bigg].
\end{eqnarray}
Thus magnification decreases exponentially  with $n$ and the images become fainter as $n$ increases.\\
Following  \cite{Bozza:2002zj} the three observables for nonsingular Kerr--Sen black hole are defined as 
\begin{eqnarray}
\theta_\infty &=& \frac{u_m}{D_{OL}}\\
s &=& \theta_1-\theta_\infty \approx \theta_\infty (e^{\frac{q-2\pi}{p}})\\
r_{\text{mag}} &=& \frac{\mu_1}{\sum{_{n=2}^\infty}\mu_n } \approx 
e^{\frac{2 \pi}{p}}
\end{eqnarray} 
Here, $s$ is the angular separation between the first image ($n=1$) and the rest images which are supposedly packed at $\theta_\infty$. $r_{\text{mag}}$ is the difference in magnitude of  flux of the first image and flux from all the other images. For numerical estimation of these observables we are considering a realistic case of black holes, namely, Sgr A* in our galactic center, M87* in Messier 87 galaxy, NGC 4649 and NGC 1332 \cite{Kormendy:2013}. For Sgr A*, we consider $M=4.3\times 10^6 M_{\odot}$ and $d=8.35$~Kpc \cite{Do:2019vob}, whereas for M87* the mass $M=6.5 \times 10^9 M_{\odot} $ and distance $d=16.8$~Mpc \cite{Akiyama:2019eap}. \\
We have tabulated the strong lensing coefficients and the observables in Table~\ref{table1} and Table~\ref{table2}. For comparisons we have estimated these values for the Schwarzschild black hole and the Kerr black hole and found that the angular separation for nonsingular Kerr--Sen black hole is larger. From the plots in Figs.~\ref{plot1a} and ~\ref{plot1b} it can be immediately followed that angular separation $s$ increases but angular position ($\theta_{\infty}$) and flux magnitude ($r_{\text{mag}}$) decrease  with $\tilde{b}$ and $\tilde{k}$. 

\section{Black hole shadow and parameter estimation}\label{Sec5}
The optical appearance of a black hole, embedded in a bright background or surrounded by a geometrically thick and optically thin emission region due to accretion flows, is known as \textit{shadow}, which is a dark region on the observer's celestial sky fringed by the bright and sharp photon ring. Indeed, a shadow is the gravitationally lensed image of the surrounded photon region around the black hole's event horizon. The study of black hole shadow was led by the pioneering work of Synge \cite{Synge:1966} and Luminet \cite{Luminet:1979}. Later, Bardeen \cite{Bardeen} studied the shadow of the Kerr black hole.  Potential applications of shadow in determining the black hole parameters and unraveling the valuable information regarding the near-horizon field features of gravity have aroused flurry of activities in the analytical investigations, observational studies and numerical simulation of shadows for large varieties of black holes in general relativity as well in modified theories of gravity over the past decades \cite{Falcke:1999pj,De,Shen:2005cw,Amarilla:2010zq,Amarilla:2011fx,Yumoto:2012kz,Amarilla:2013sj,Atamurotov:2013sca,Abdujabbarov:2016hnw,Amir:2017slq,Abdujabbarov:2015xqa,Cunha:2018acu,Mizuno:2018lxz,Shaikh:2019fpu,Mishra:2019trb,Ghosh:2020ece,Singh:2017vfr,Amir:2016cen,Atamurotov:2015xfa,Papnoi:2014aaa}. 

In this section, we investigate the propagation of freely moving photons in rotating nonsingular Kerr--Sen black hole spacetimes whose metric is given by~(\ref{metric1}). Such photons follow the spacetime geodesics, whose equations can be greatly simplified due to the presence of integrals of motion associated with spacetime symmetries. The projection of photon four-momentum $p^{\mu}$ along the two obvious commuting Killing vectors $\xi_{(t)}^{\mu}=\delta^{\mu}_t$ and $\xi_{(\phi)}^{\mu}=\delta^{\mu}_{\phi}$ entails the existence of two linearly independent conserved canonical momentum components, namely, energy $E$ and axial angular momentum component $L$ \cite{Frolov:1998wf,Chandrasekhar:1992}
\begin{equation}
p_t=g_{tt}p^{t}+g_{t\phi}p^{\phi}=-E,\qquad p_{\phi}=g_{t\phi}p^{t}+g_{\phi\phi}p^{\phi}=L.
\end{equation}
Solving these two coupled equations lead to the following equations of motion for $t$ and $\phi$ coordinates \cite{Chandrasekhar:1992}
\begin{align}
\Sigma \frac{dt}{d\tau}&=\frac{r^2+2bre^{-k/r}+a^2}{\Delta}\Big({ E}(r^2+a^2+2bre^{-k/r})-a{ L}\Big)-a(aE\sin^2\theta-L),\label{teq} \\
\Sigma \frac{d\phi}{d\tau}&=\frac{a}{\Delta}\Big({ E}(r^2+a^2+2bre^{-k/r})-a{ L}\Big)-\left(a{ E}-\frac{{ L}}{\sin^2\theta}\right).\label{phieq}
\end{align}

\begin{figure*}
	\centering
\begin{tabular}{c c}
	\includegraphics[scale=0.77]{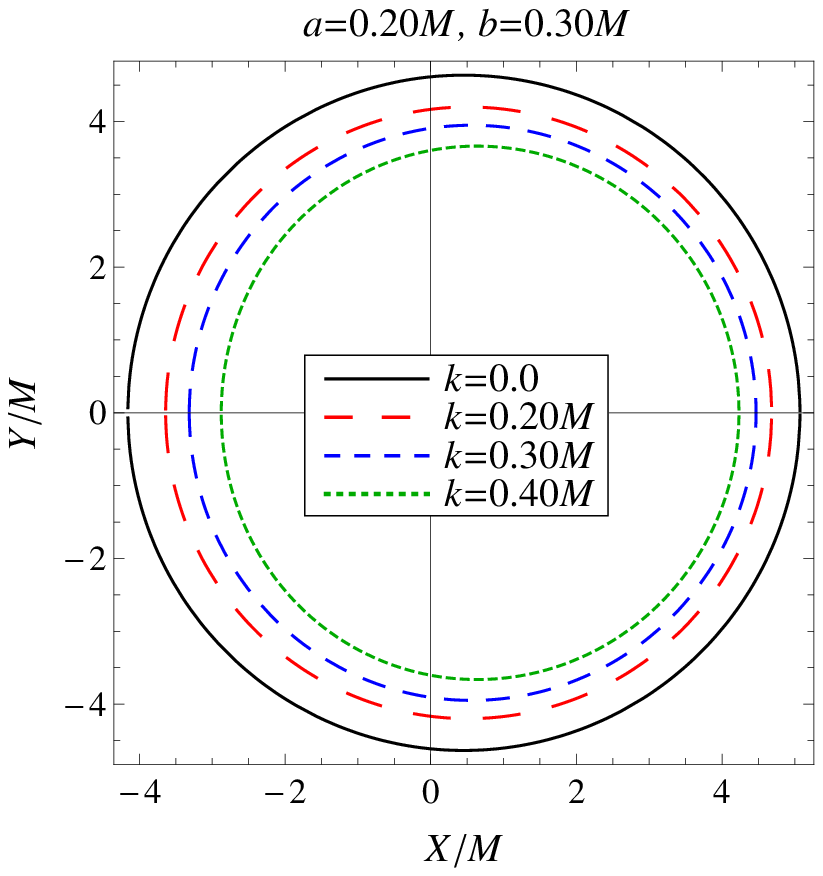}&
	\includegraphics[scale=0.77]{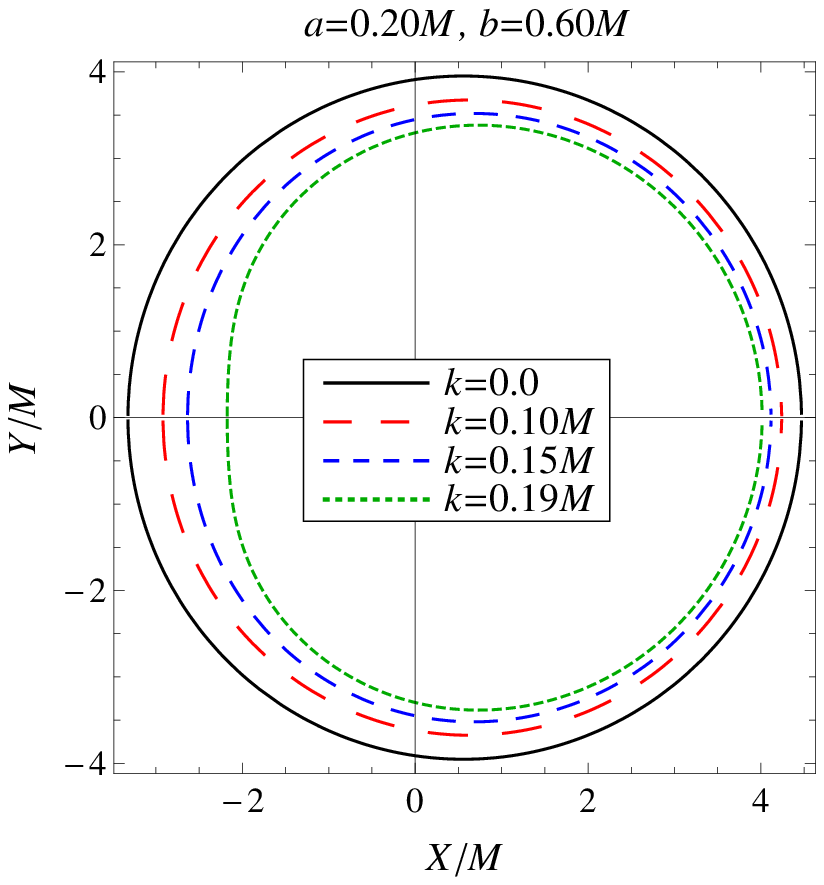}\\
	\includegraphics[scale=0.77]{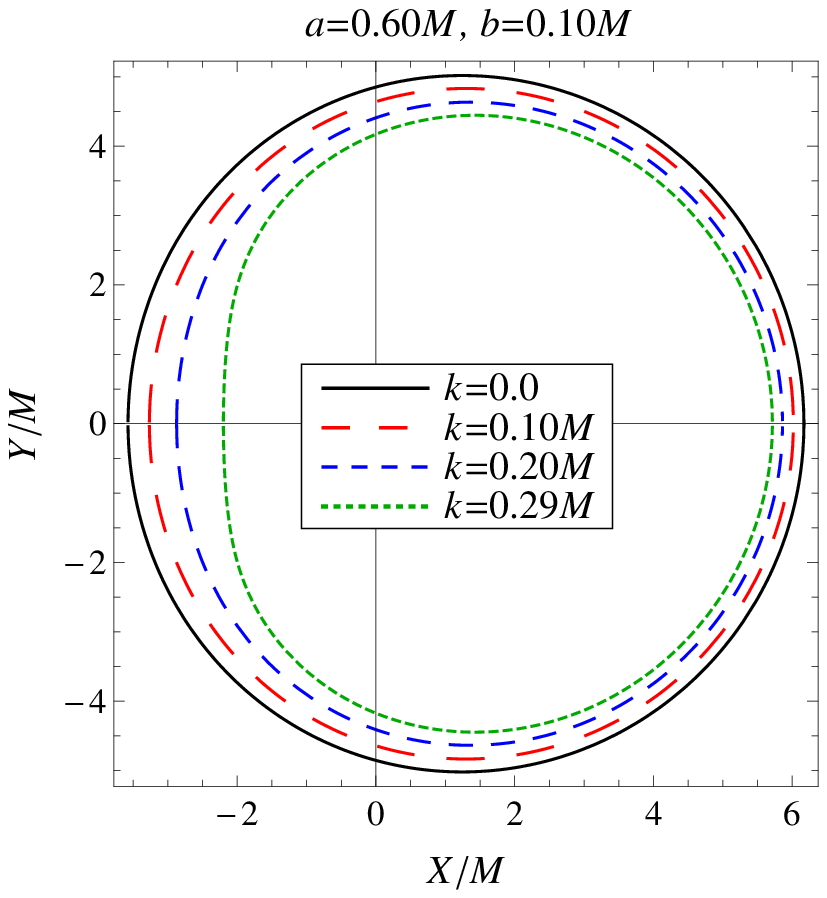}&
	\includegraphics[scale=0.77]{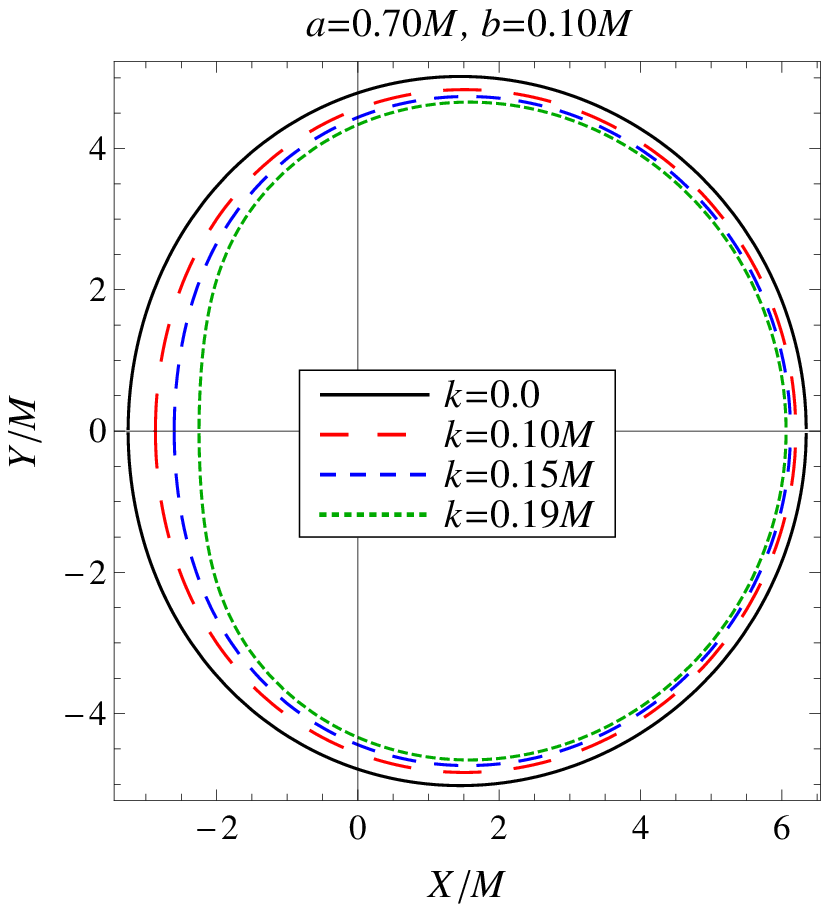}\\
	\includegraphics[scale=0.77]{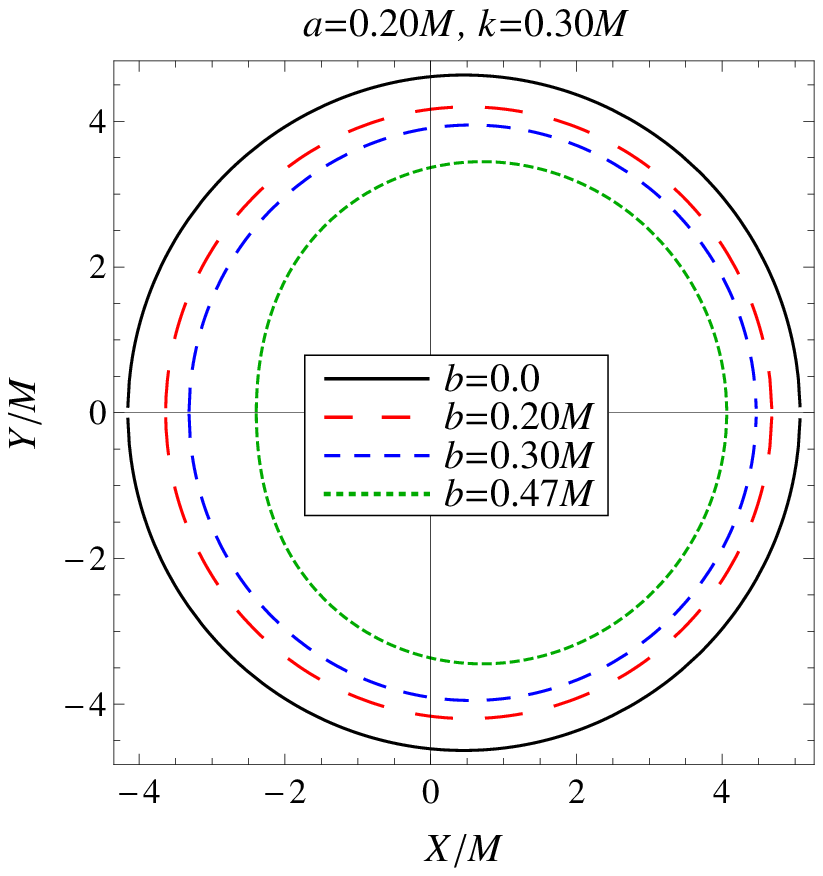}&
	\includegraphics[scale=0.77]{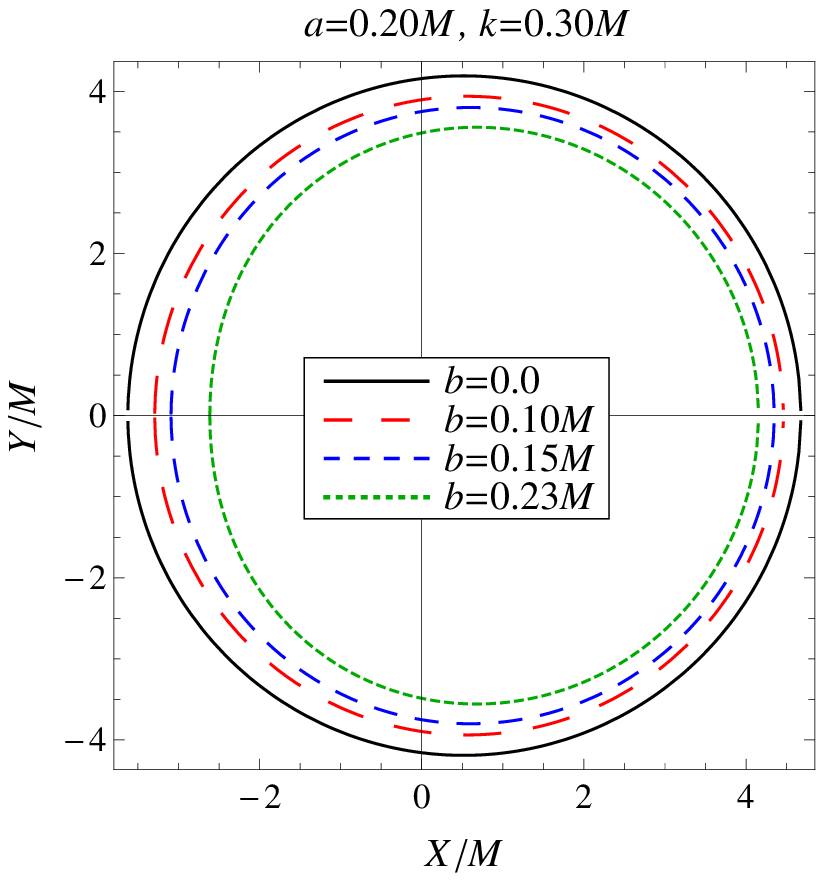}
\end{tabular}
\caption{Plot showing the rotating nonsingular Kerr--Sen black hole shadows with varying $b$ and $k$. The black curves are for the Kerr--Sen black holes in the upper and middle panels, whereas they are for the rotating nonsingular black holes in the lower panel.}\label{shadow}	
\end{figure*}

\begin{figure*}
\begin{tabular}{c c}
	\includegraphics[scale=0.77]{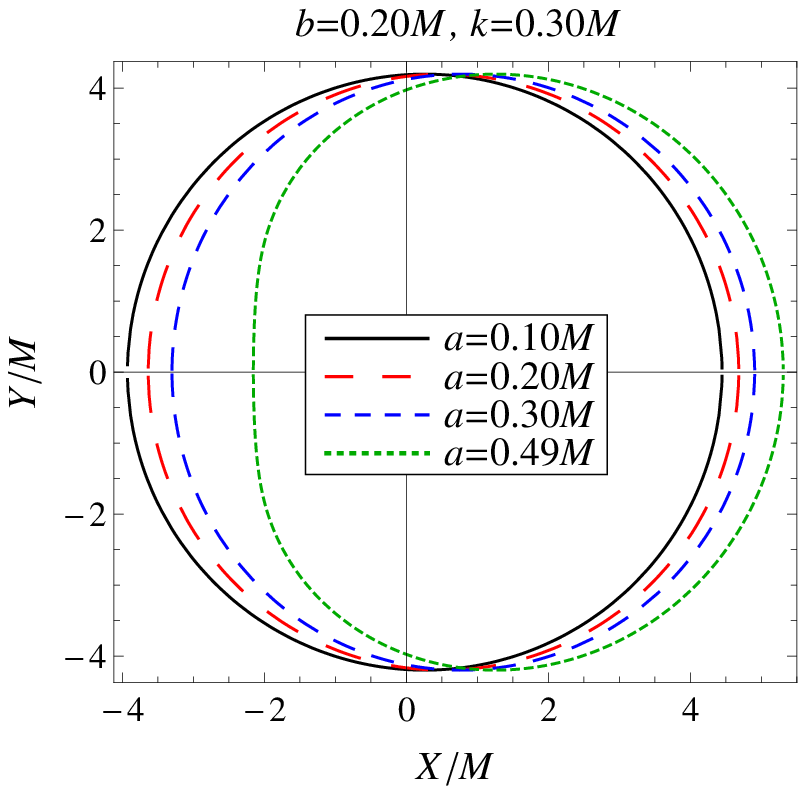}&
	\includegraphics[scale=0.77]{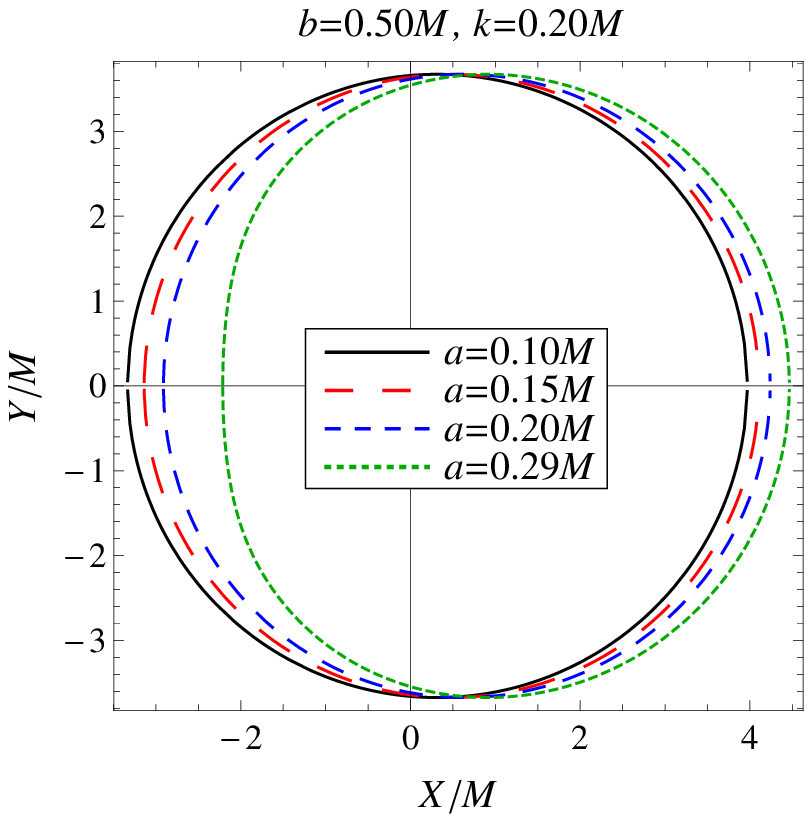}
\end{tabular}
\caption{Plot showing the rotating nonsingular Kerr--Sen black hole shadows with varying $a$.}\label{shadow1}
\end{figure*}

\begin{figure*}[ht!]
	\begin{center}
		\begin{tabular}{c c}
			\includegraphics[scale=0.63]{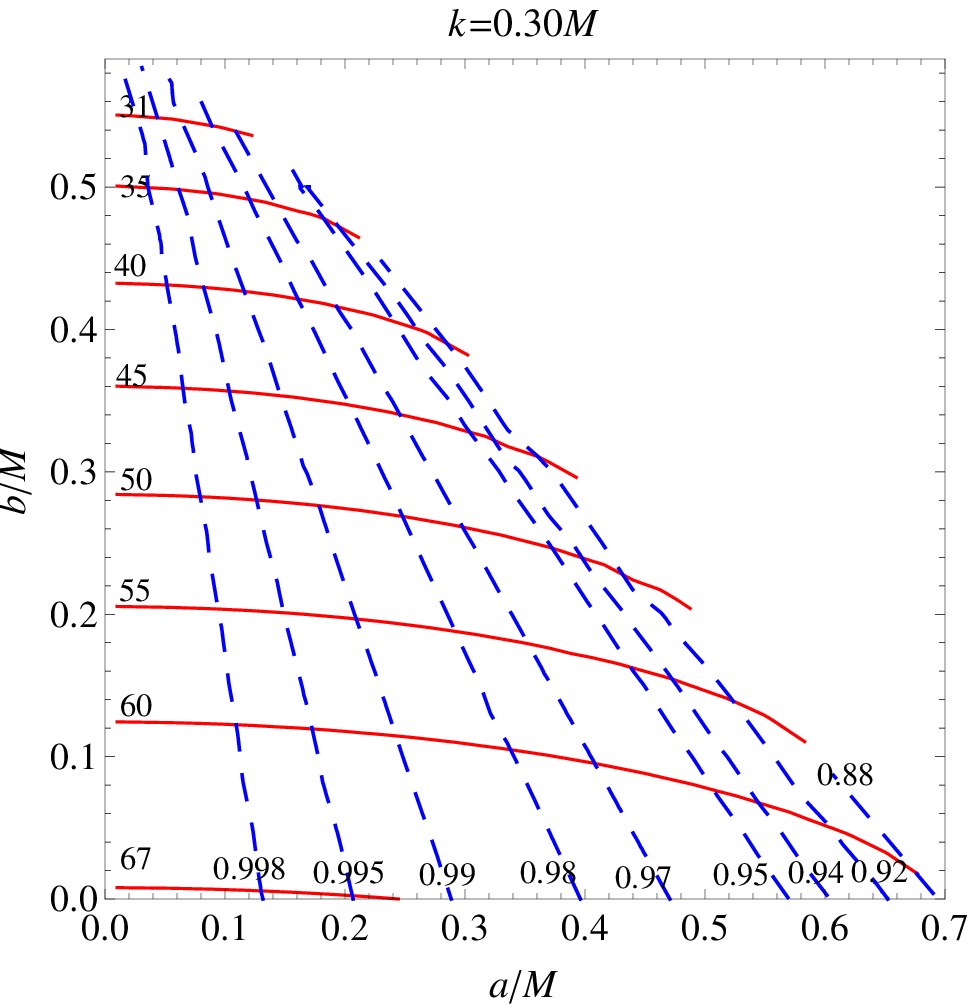}&
			\includegraphics[scale=0.63]{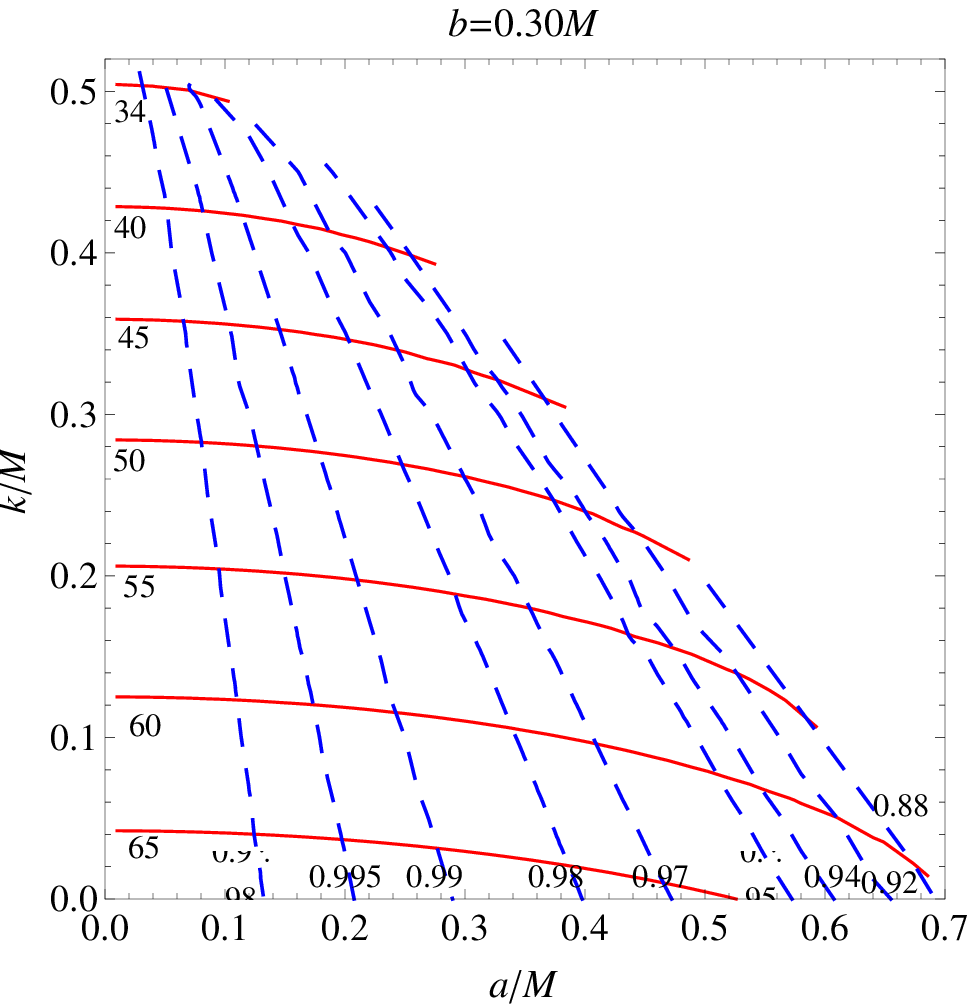}
		\end{tabular}	
	\end{center}
	\caption{Contour plots of the observables $A$ and $D$ in the nonsingular Kerr--Sen black hole parameter space. Solid red curves correspond to the constant $A$, and dashed blue curves are for the constant oblateness parameter $D$. }
	\label{parameterestimation}
	\begin{tabular}{c c}
		\includegraphics[scale=0.65]{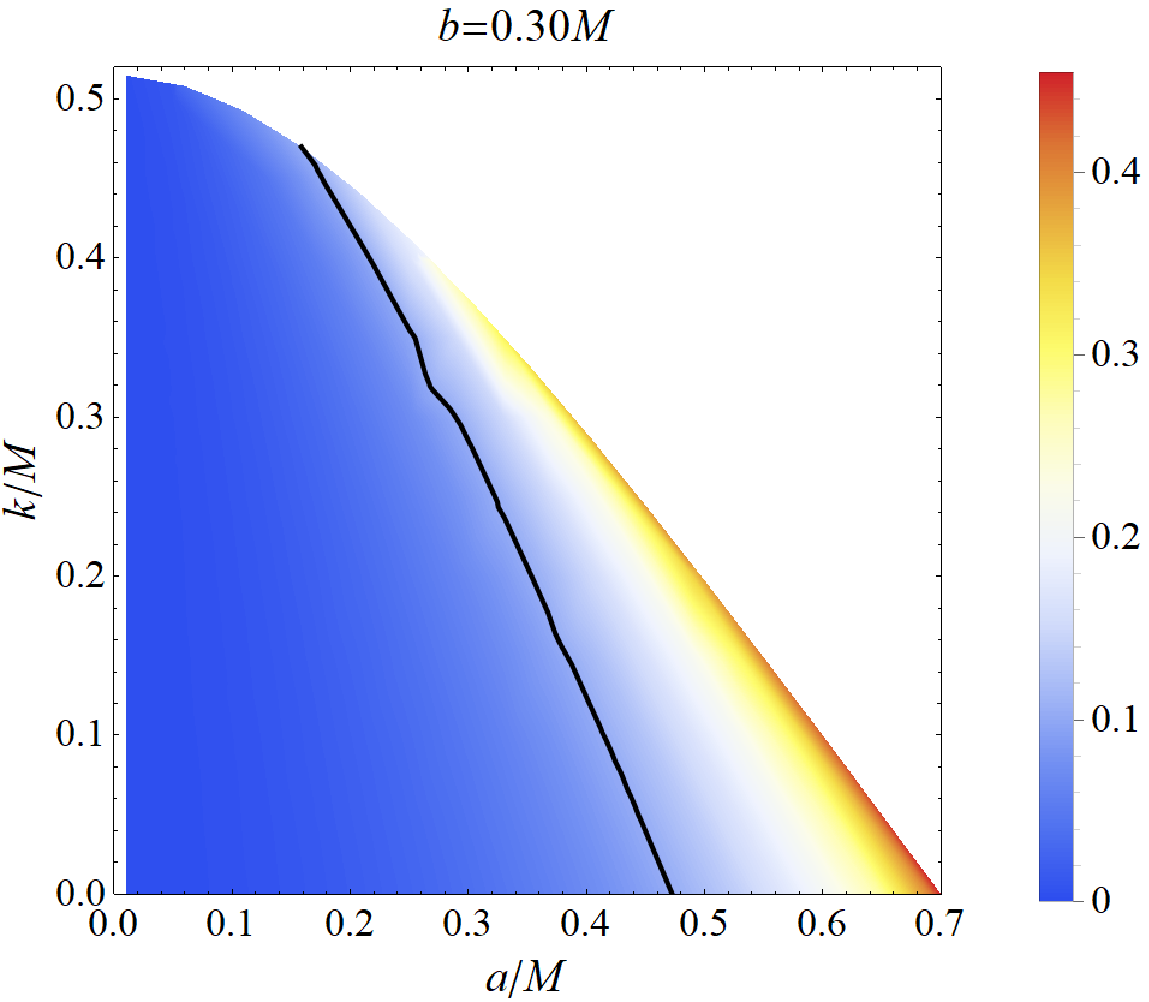}&
		\includegraphics[scale=0.65]{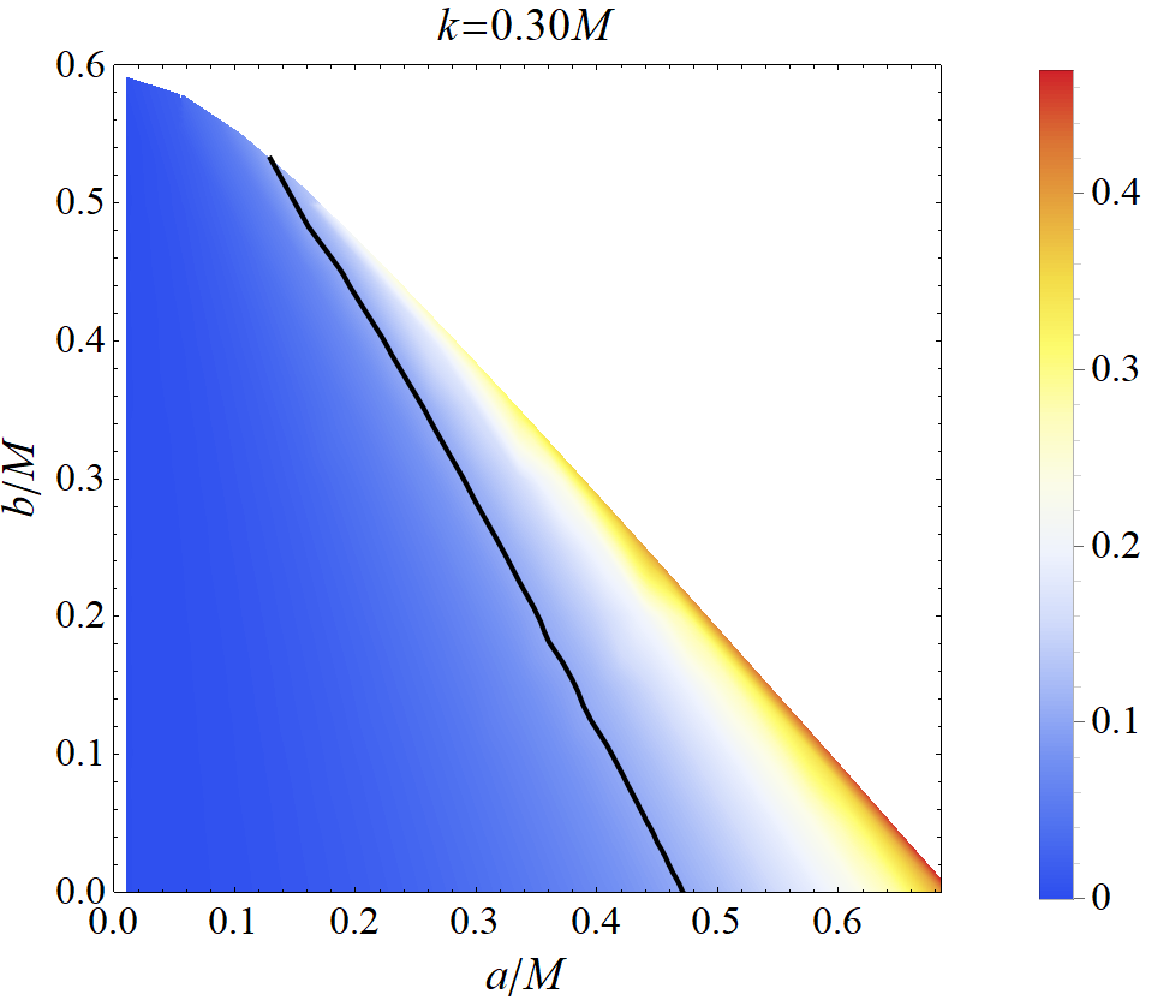}\\
		\includegraphics[scale=0.65]{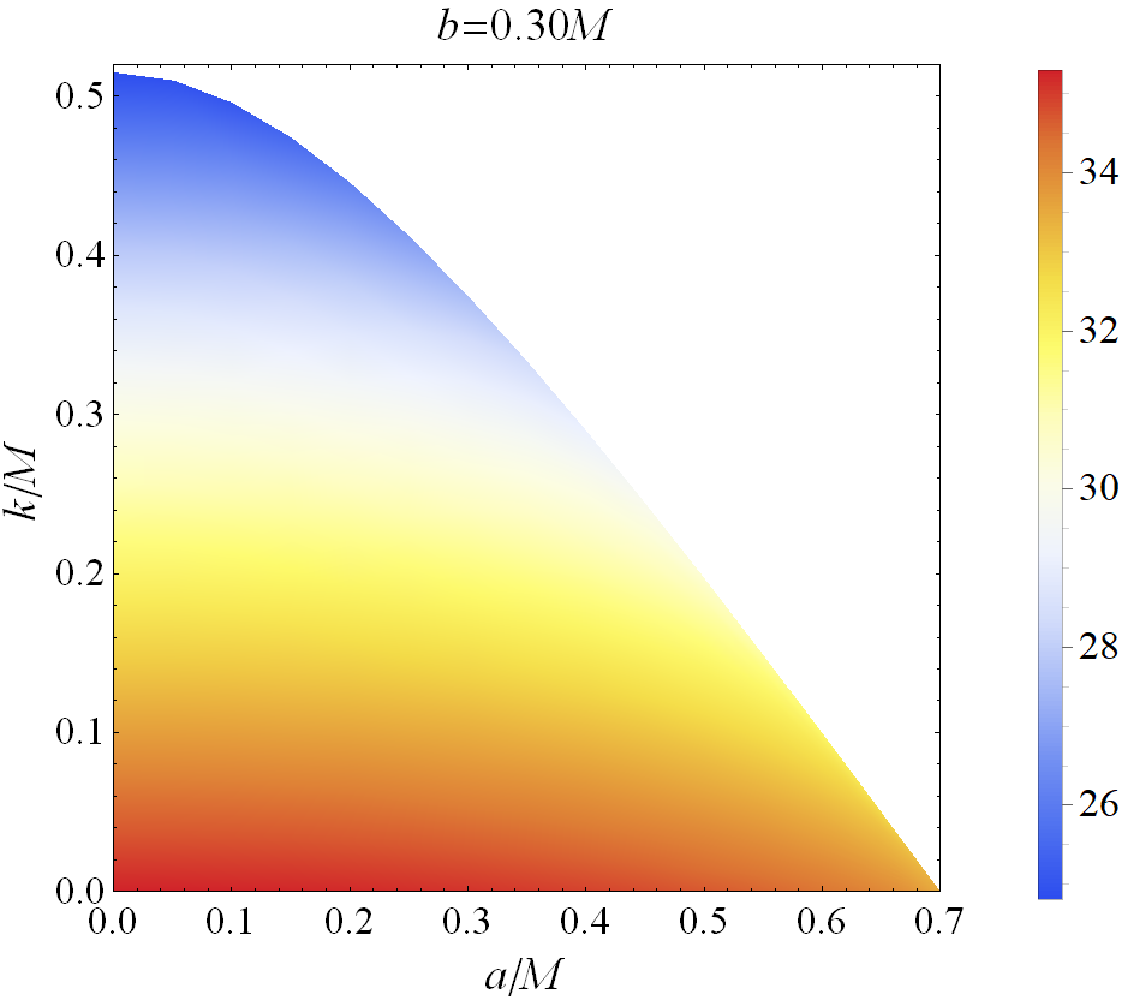}&
		\includegraphics[scale=0.65]{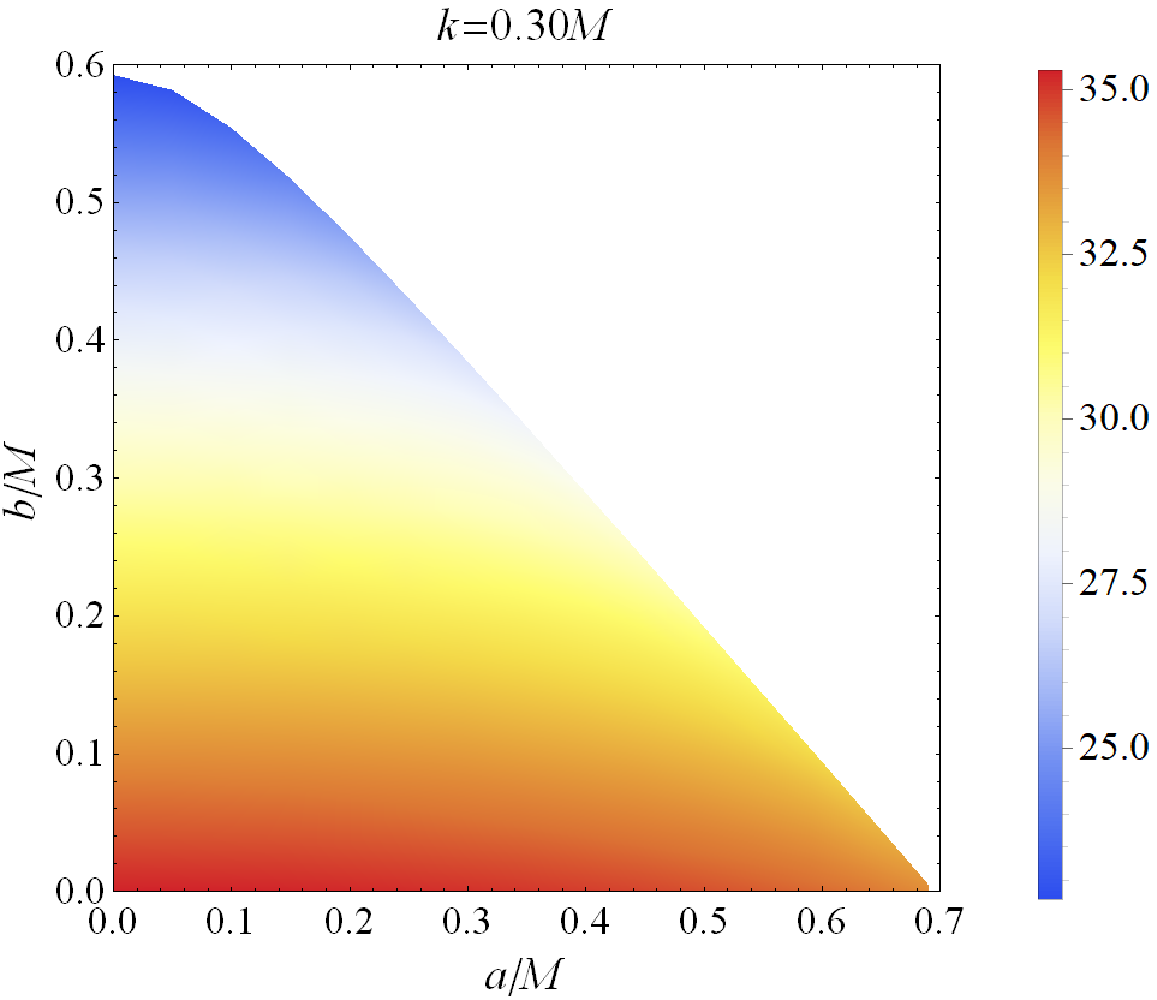}\\
	\end{tabular}
	\caption{Circularity deviation observable $\Delta C$ (upper panel) and the angular diameter $\theta_d$ (lower panel) as a function of black holes parameters for $\theta_o=\pi/2$. Black solid lines correspond to the M87* black hole shadow bounds, $\Delta C=0.10$. For these parameters values, the $\theta_d$ is smaller than the M87* shadow angular diameter within $1\sigma$ region.}	\label{M87}
\end{figure*}

Due to the broken spherical symmetry, the equations of motion are reasonably complicated; the Euler-Lagrange equation for the rest of coordinates gives second-order differential equations. The Abelian isometry group of $\xi_{(t)}^{\mu}$ and $\xi_{(\phi)}^{\mu}$ is insufficient for the complete integrability of the geodesics equations of motion in the Kerr--Sen spacetimes. It is for this reason we follow the Hamilton-Jacobi equation and the integral approach pioneered by Carter \cite{Carter:1968rr} to obtain the equations of motion in the first-order differential form
\begin{eqnarray}
\label{HmaJam}
\frac{\partial S}{\partial \tau} = -\frac{1}{2}g^{\mu\nu}\frac{\partial S}{\partial x^\mu}\frac{\partial S}{\partial x^\nu}.
\end{eqnarray}
Using the conserved $p_t$ and $p_{\phi}$, the Jacobi action $S$ admits a separable solution in the first integral form as follows
\cite{Carter:1968rr}
\begin{eqnarray}
S=-Et + L \phi +S_r(r)+S_\theta(\theta) \label{Jaction}.
\end{eqnarray}
Here, $S_r(r)$ and $S_{\theta}(\theta)$, respectively, are functions only of the $r$ and $\theta$ coordinates.
On using the four-momentum relations 
\begin{align}
&\frac{dS_r}{dr}=p_r= g_{rr}p^r=\frac{\Sigma}{\Delta}\frac{dr}{d\tau},\label{dS1}\\
&\frac{dS_{\theta}}{d\theta}= p_{\theta}=g_{\theta\theta}p^{\theta}=\Sigma\frac{d\theta}{d\tau},\label{dS2}
\end{align}
to the Eq.~(\ref{HmaJam}), the following separable equations of motion in the first-order differential form are obtained
\begin{eqnarray}
\Sigma \frac{dr}{d\tau}&=&\pm\sqrt{\mathcal{R}(r)}\ ,\label{req}\\
\Sigma \frac{d\theta}{d\tau}&=&\pm\sqrt{\Theta(\theta)}\ ,\label{theq}
\end{eqnarray}
where the sign "$\pm$" in $r$ and $\theta$ equations are mutually independent and, respectively, corresponds to two different possible radially incoming or outgoing and the zenith value increasing or decreasing geodesics, such that the sign changes at turning points of geodesic motion. The effective potentials $\mathcal{R} (r)$ and ${\Theta}(\theta)$ for radial and polar motion are given by 
\begin{align}
\mathcal{R}(r)&=\Big((r^2+2bre^{-k/r}+a^2){ E}-aL\Big)^2-\Delta \Big({\cal K}+(a{ E}-{ L})^2\Big),\label{Rpot}\\ 
\Theta(\theta)&={\cal K}-\left(\frac{{ L}^2}{\sin^2\theta}-a^2 {E}^2\right)\cos^2\theta.\label{theta0}
\end{align}
where $\mathcal{K}\geq 0$ is the separability constant related to the Carter constant of motion associated with the second-rank irreducible tensor field of \textit{hidden symmetry}. Indeed, this hidden symmetry plays the pivotal role for the separability of geodesics equation of motion for rotating spacetimes. It is worth-mentioning over here that any four-dimensional vacuum solution of Petrov type-D admits Killing tensor field associated with the hidden symmetries. Although, the Kerr--Sen black hole is not a vacuum solution and belongs to the Petrov type-I, however, it still possess a irreducible second-rank Killing tensor apart from the metric tensor $g_{\mu\nu}$. This tensor will yield a fourth constant of motion when contracted twice with the four-momenta and thus the geodesics equations are fully integrable \cite{Hioki:2008zw}. Therefore, the Hamilton-Jacobi equation is also separable for the rotating nonsingular Kerr--Sen black holes. The allowed region for photon motion is constrained by $\mathcal{R}\geq 0$ and $\Theta(\theta)\geq 0$ and the motion is symmetric around the equatorial plane (cf. Eq.~(\ref{theta0})). Depending on the values of $E, L$ and $\mathcal{K}$, the photon geodesics can be classified into capturing, scattering and bound orbits at constant Boyer-Lindquist coordinate. These bound orbits can be determined by the solutions of $\dot{r}=0$ and $\ddot{r}=0$ and corresponds to the unique extremum of the radial potential in the exterior region to the event horizon $r_+$, i.e., $r>r_+$, such that \cite{Frolov:1998wf,Chandrasekhar:1992}
\begin{equation}
\left.\mathcal{R}\right|_{(r=r_p)}=\left.\frac{\partial \mathcal{R}}{\partial r}\right|_{(r=r_p)}=0,\,\, \text{and}\,\, \left.\frac{\partial^2 \mathcal{R}}{\partial r^2}\right|_{(r=r_p)}> 0,\label{vr} 
\end{equation}
where $r_p$ is the unstable photon orbit radius. We define two dimensionless impact parameters $\xi=L/E, \eta=\mathcal{K}/E^2$ as energy-rescaled angular momentum and energy-rescaled energy. Solving Eq.~(\ref{vr}) for Eq.~(\ref{Rpot}) results into the critical values of impact parameters ($\xi_c, \eta_c$) for the unstable orbits 
\begin{align}
\xi_c=&\frac{e^{-k/r_p}}{a r_p \Delta'}\Big(2 M r_p^2 \Delta' - \Delta (4 M (k + r_p) + e^{k/r_p} r_p \Delta')\Big),\nonumber\\
\eta_c=&\frac{e^{-2 k/r_p}}{a^2 r_p^2 \Delta'^2}\Bigg(-r_p^2 (a^2 e^{k/r_p} - 2 M r_p)^2 \Delta'^2 - \Delta^2 \left(4 M (k + r_p) + e^{k/r_p} r_p \Delta'\right)^2 \nonumber\\
&+ 2 \Delta \Big(8 a^2 M^2 (k + r_p)^2 + r_p (a^2 e^{k/r_p} + 2 M r_p) \Delta' \left(4 M (k + r_p) + e^{k/r_p} r_p \Delta'\right)\Big)\Bigg),\label{CriImpPara}
\end{align}
\begin{figure*}
\begin{tabular}{c c}
	\includegraphics[scale=0.77]{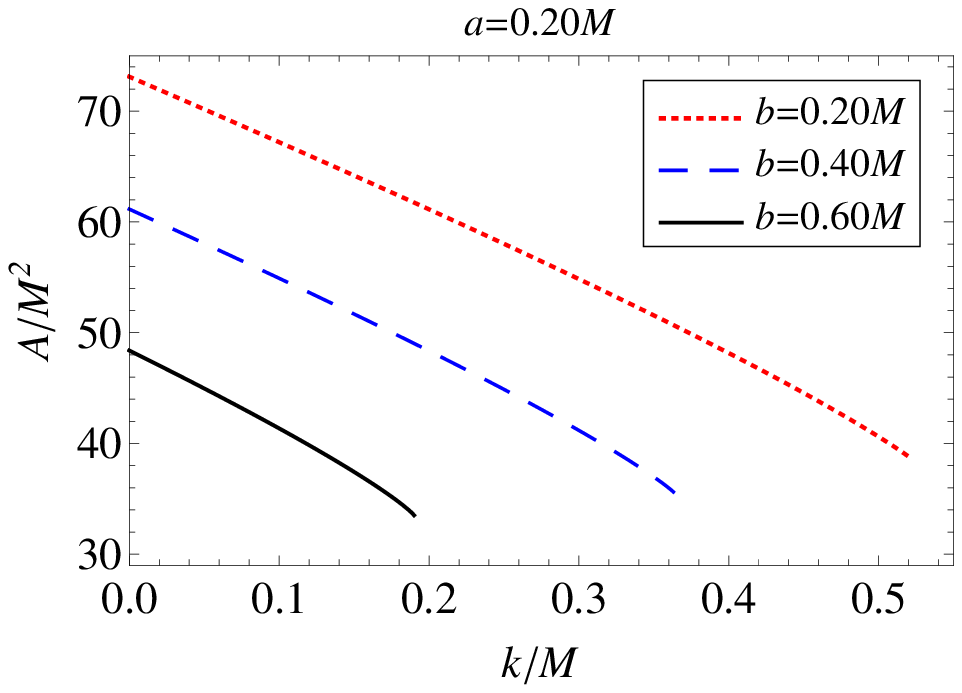}\hspace*{-1cm}&
	\includegraphics[scale=0.77]{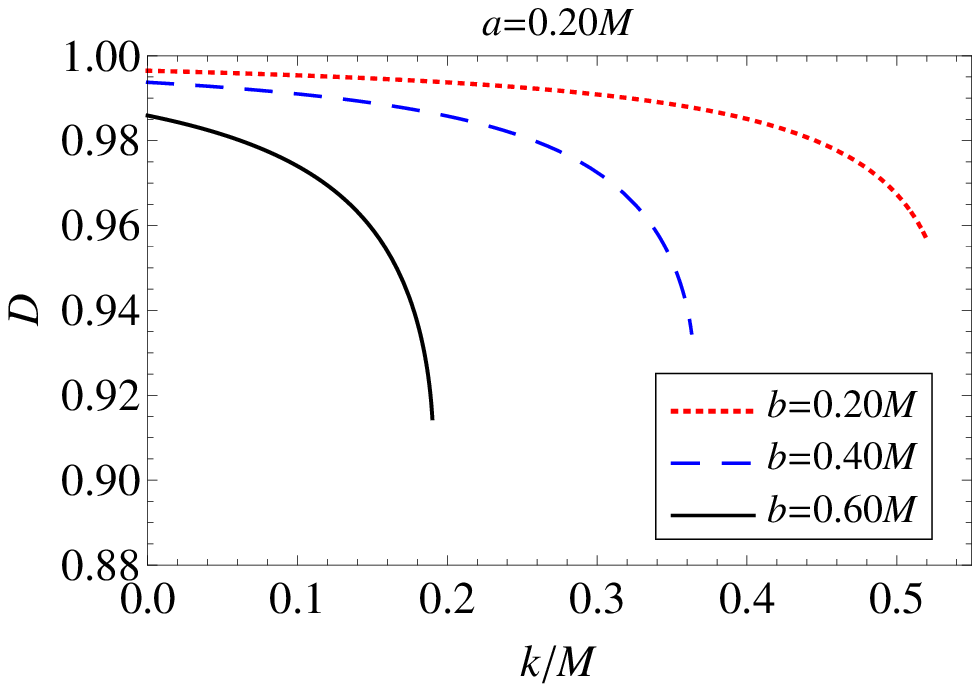}\\
	\includegraphics[scale=0.77]{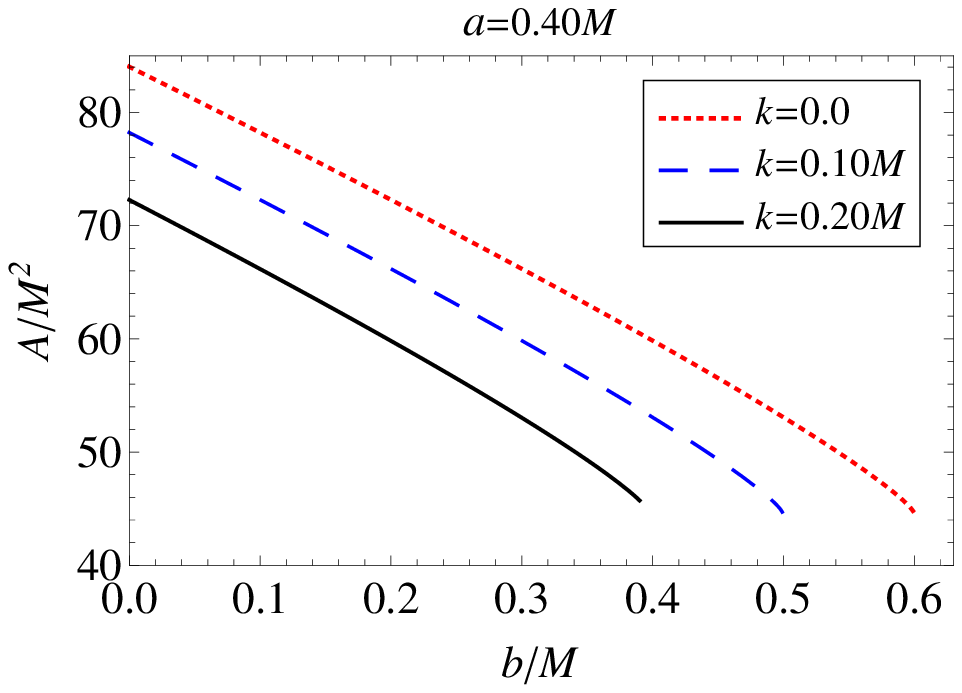}\hspace*{-1cm}&
	\includegraphics[scale=0.77]{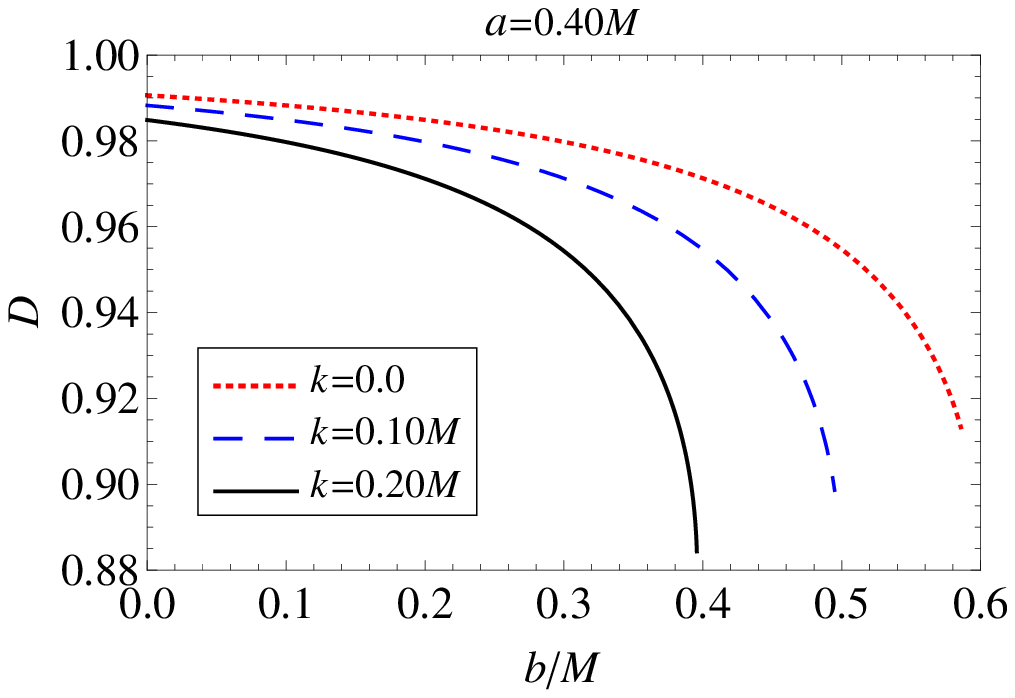}\\
	\includegraphics[scale=0.77]{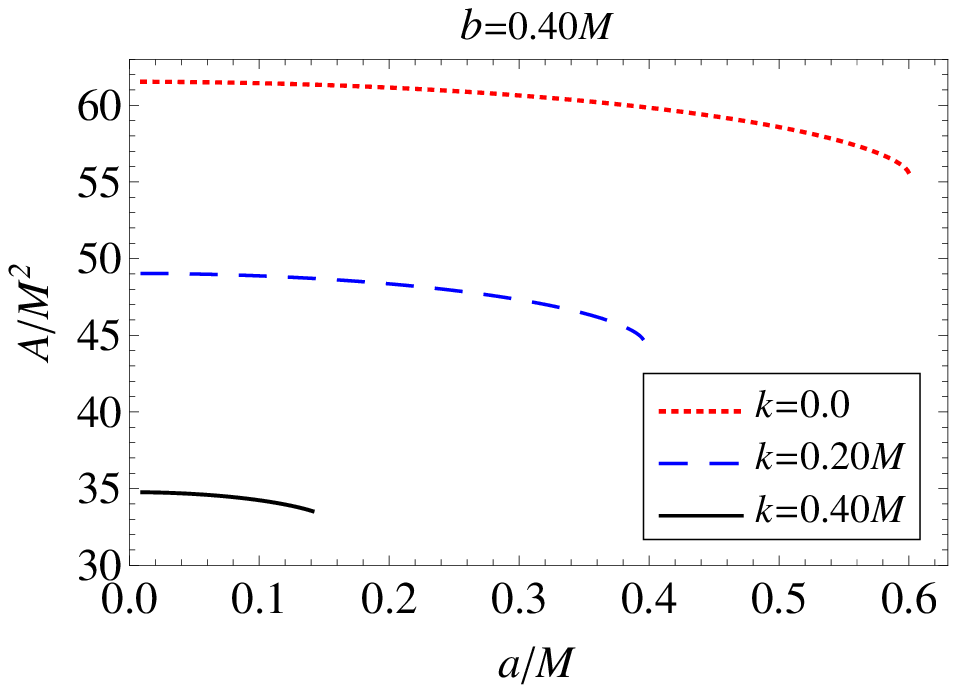}\hspace*{-1cm}&
	\includegraphics[scale=0.77]{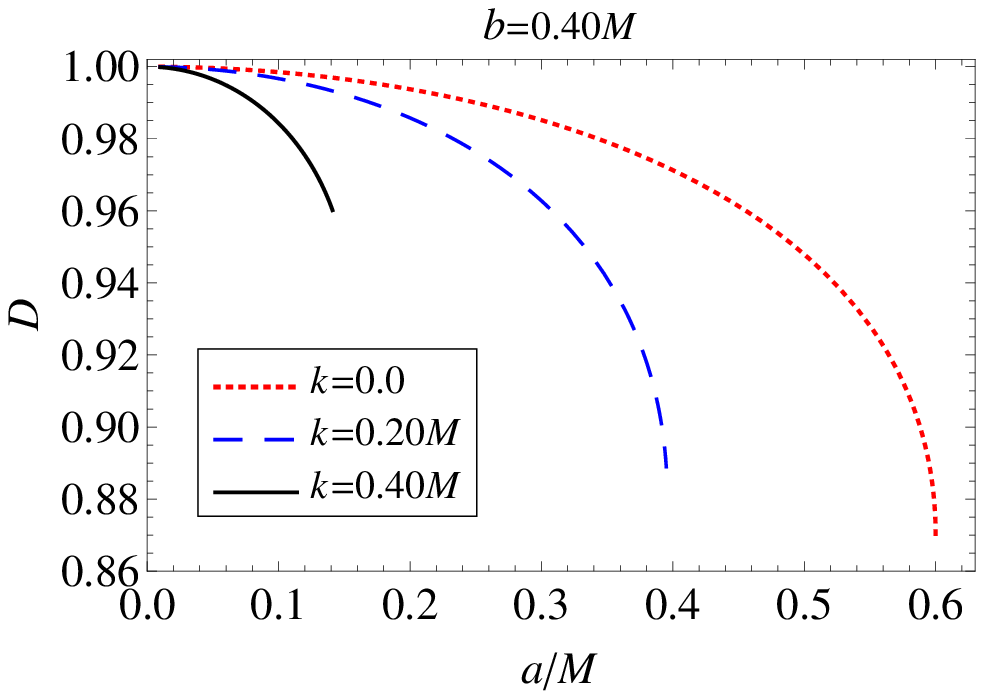}	
\end{tabular}
\caption{The shadow area $A$ and oblateness observables $D$ with varying parameters for the nonsingular Kerr--Sen black holes.}\label{obs}
\end{figure*}
where $'$ stands for the derivative with respect to the radial coordinate. Equation (\ref{CriImpPara}) in the limit $k\to0$ reduces to the critical impact parameter for the Kerr--Sen black holes \cite{Lan:2018lyj,Xavier:2020egv}, whereas for $b\to 0$ it describes the critical impact parameter for the rotating nonsingular black hole \cite{Amir:2016cen}.  A small outward perturbation to unstable bound photon orbits forces the photons to leave the orbit and direct them to infinity on a hyperbolic trajectory, whereas an inward perturbation guides them to fall into the black hole \cite{Chandrasekhar:1992}. In particular, photons with $\mathcal{\eta}_c=0$ form the planar circular orbits confined only to the equatorial plane, whereas $\mathcal{\eta}_c>0$ lead to three-dimensional spherical (or non-planar) orbits \cite{Chandrasekhar:1992}. Contrary to the non-rotating black holes, the axial-symmetry and frame-dragging effects in the rotating black holes also lead to nonplanar orbits at constant radii. In rotating spacetimes, photons can either have prograde motion or retrograde motion relative to the black hole rotation, whose respective radii at the equatorial plane, $r_p^{-}$ and $r_p^{+}$, can be identified as the real positive roots of $\eta_c=0$ for $r_p\geq r_+$, and all other spherical photon orbits have radii $r_p^-< r_p<r_p^+$. Furthermore, generic orbits at constant radius $r$ at a plane other than $\theta=\pi/2$ are nonplanar ($\dot{\theta}\neq 0$) and repeatedly cross the equatorial plane ($\eta>0$) while oscillating symmetrically about it. These spherical photon orbits construct a photon region around the rotating black hole. Indeed, these photon essentially go closet to the black hole, yet outside the event horizon, and can still escape to a distant observer with high redshift. The resulting optical appearance of a black hole, embedded in a bright background or surrounded by a geometrically thick and optically thin emission region due to accretion flows, is known as "shadow", which is a dark region on the observer's celestial sky fringed by the bright and sharp closed photon ring. A shadow is the gravitationally lensed image of the surrounded photon region around the black hole's event horizon. In principle, multiple nearly-circular photon rings are expected depending on the number of revolution around the black holes, such that for infinitely large number of loops they are described by the critical impact parameters defined in Eq.~(\ref{CriImpPara}). By making a stereographic projection of black hole shadow on the observer's celestial sky to the image plane, the shadow boundary can be described by the following coordinates
\begin{eqnarray}
X&=&-r_o\frac{p^{(\phi)}}{p^{(t)}}=\lim_{r_o\rightarrow\infty}\left(-r_o^2 \sin{\theta_o}\frac{d\phi}{d{r}}\right),\nonumber\\ Y&=&r_o\frac{p^{(\theta)}}{p^{(t)}}=\lim_{r_o\rightarrow\infty}\left(r_o^2\frac{d\theta}{dr}\right),\label{Celestial}
\end{eqnarray} 
where $p^{(\mu)}$ are the tetrad components of photon four-momentum and the $Y$ axis is chosen along the black hole's rotational axis. Using the geodesics equations, this can be simplified to
\begin{eqnarray}
X&=&-\xi_c\csc\theta_o\xrightarrow{\theta_o=\pi/2} -\xi_c,\nonumber\\
Y&=&\pm\sqrt{\eta_c+a^2\cos^2\theta_o-\xi_c^2\cot^2\theta_o}\xrightarrow{\theta_o=\pi/2}\pm\sqrt{\eta_c}.\label{pt}
\end{eqnarray} 
The parametric curve $Y$ vs $X$ delineates the black hole shadow. Besides the black hole parameters, the shadow also depends on the observer's inclination angle $\theta_o$, such that the black hole rotation effect is most prominent for $\theta_o=\pi/2$. This is because for $\theta_o=\pi/2$ a full set of spherical photon orbits $r_p^-\leq r\leq r_p^+$ mapped onto the shadow boundary, whereas for $\theta_o\neq \pi/2$ only a subset of orbits contribute. Considering $\theta_o=\pi/2$ the shadows of nonsingular Kerr--Sen black holes are shown for varying $b$ and $k$ in  Fig.~\ref{shadow}. Whereas, the effect of rotation parameter $a$ on the shadows are depicted in Fig.~\ref{shadow1}. 
Clearly the shadows are distorted from a perfect circle, and to characterize their size and shape, we use the observables, namely, area $A$ and oblateness $D$ \cite{Kumar:2018ple}
\begin{equation}
A=2\int{Y(r_p) dX(r_p)}=2\int_{r_p^{-}}^{r_p^+}\left( Y(r_p) \frac{dX(r_p)}{dr_p}\right)dr_p,\label{Area}
\end{equation} 
\begin{equation}
D=\frac{X_r-X_l}{Y_t-Y_b},\label{Oblateness}
\end{equation}
where subscripts $r, l, t$ and $b$, respectively, stand for the right, left, top and bottom of the shadow boundary. We plotted the observables $A$ and $D$ in Fig.~\ref{obs} and it is evident that, in contrast to the parameter $a$, parameters $b$ and $k$ have profound impact on the shadow area. The shadow size decrease with independently increasing $b$, $k$ and $a$, however, $A$ falls sharply with $b$ and $k$ in comparing with $a$. Similarly, the oblateness increase with independently increasing $b$, $k$ and $a$. Moreover, for fixed values of parameters $a$ and $k$, the shadows of nonsingular Kerr--Sen black holes are smaller and more distorted than those for the rotating nonsingular black holes and Kerr black hole. The oblateness is more prominent for the near-extremal black holes. To estimate the black hole parameters, we depicted these observables ($A,D$) in the black hole parameter space ($a,b$) and ($a,k$) in Fig.~\ref{parameterestimation}. 
It is evident from figure that the lines of constant area and oblateness intersects each other at a unique point, which determines the precise values of black hole parameters. Therefore, with the given values of observables $A$ and $D$, one can uniquely determines upto two black hole parameters.
\begin{figure*}
	\begin{tabular}{c c}
		\includegraphics[scale=0.65]{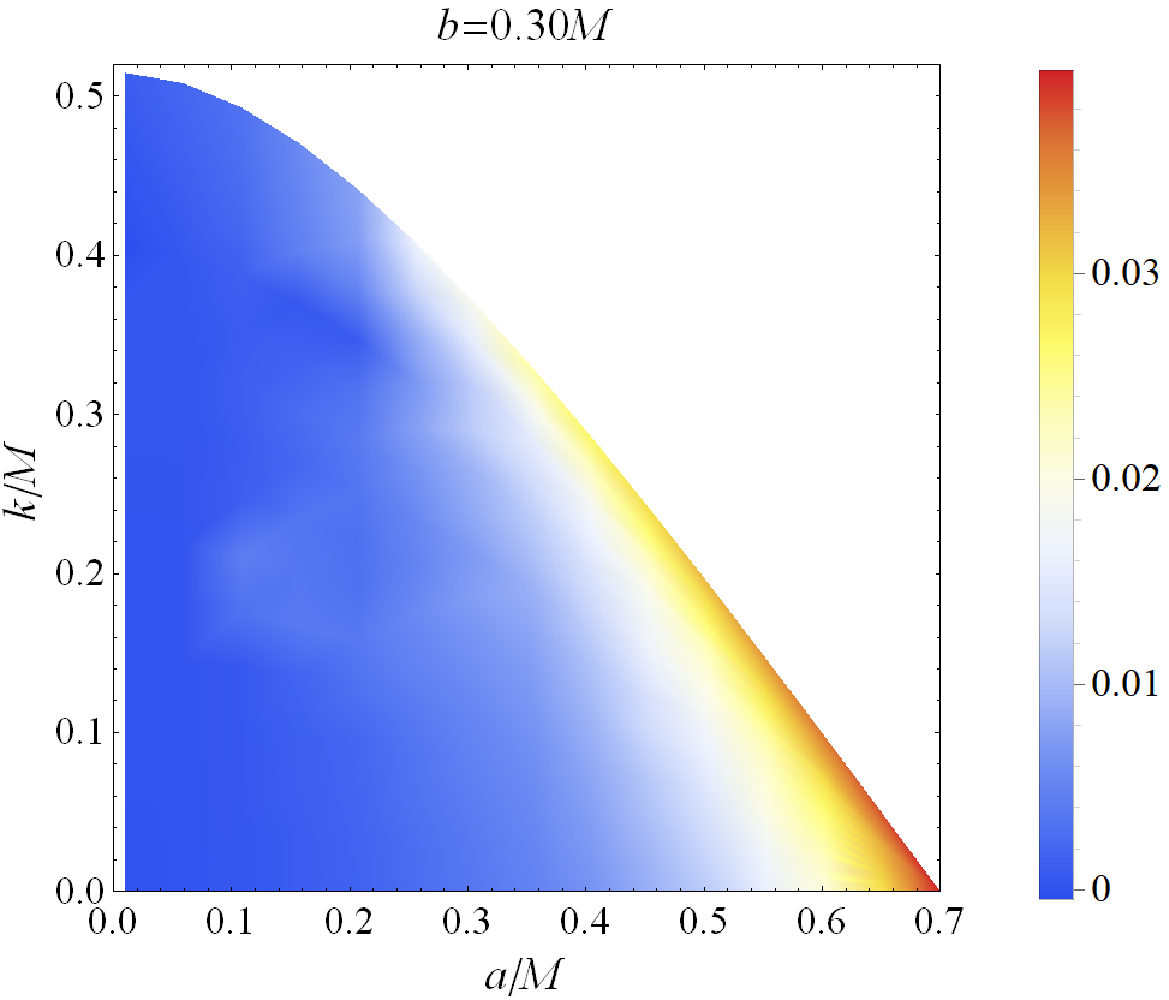}&
		\includegraphics[scale=0.65]{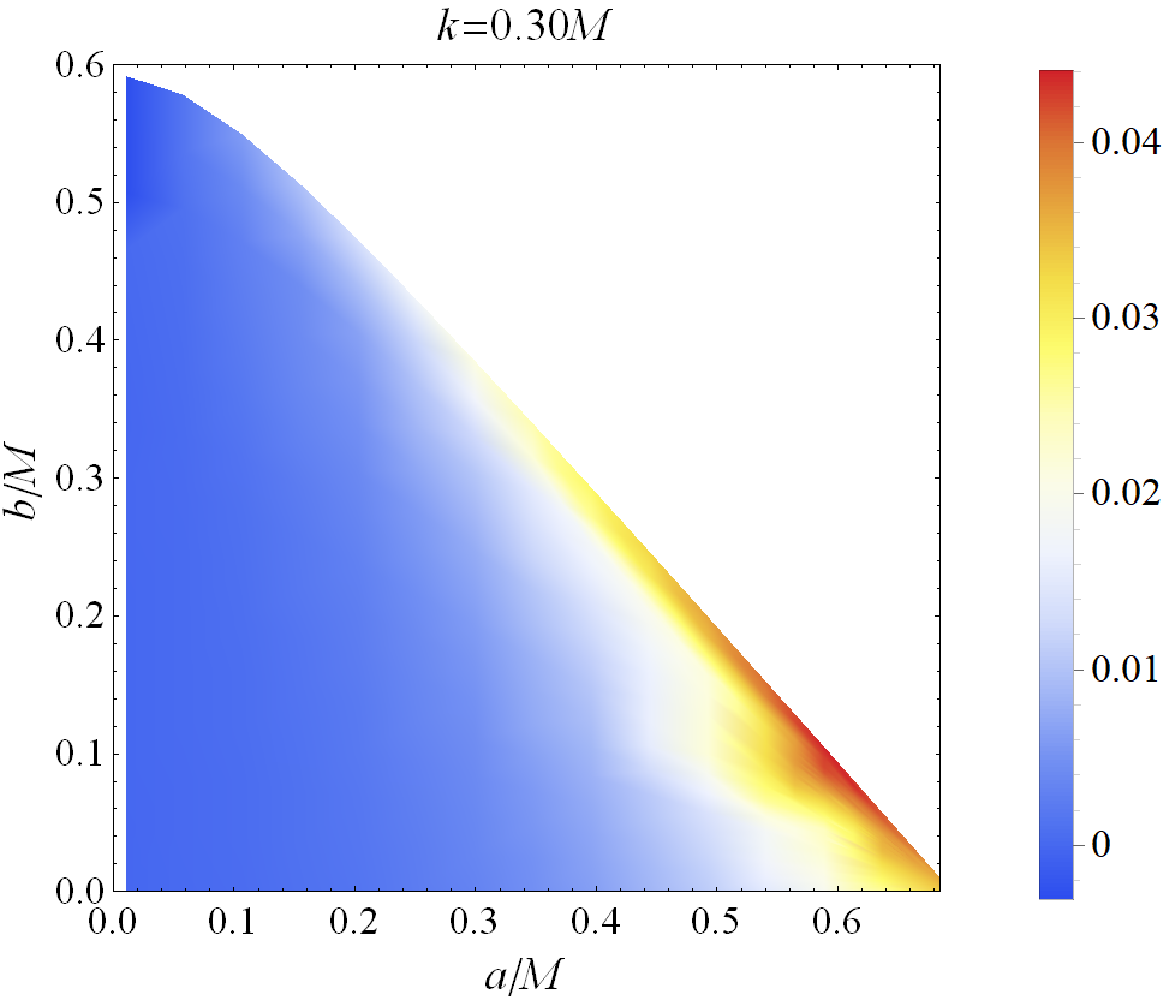}\\
		\includegraphics[scale=0.65]{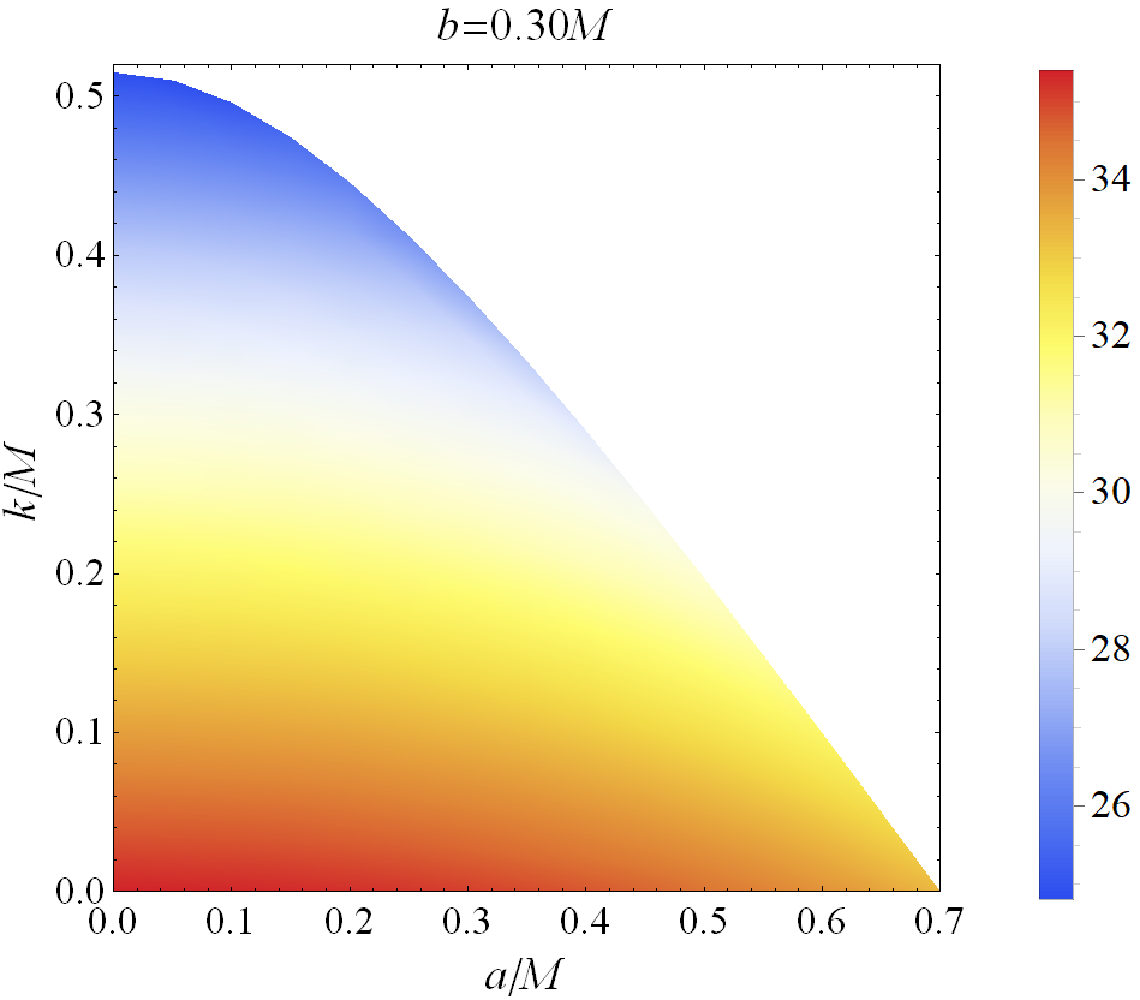}&
		\includegraphics[scale=0.65]{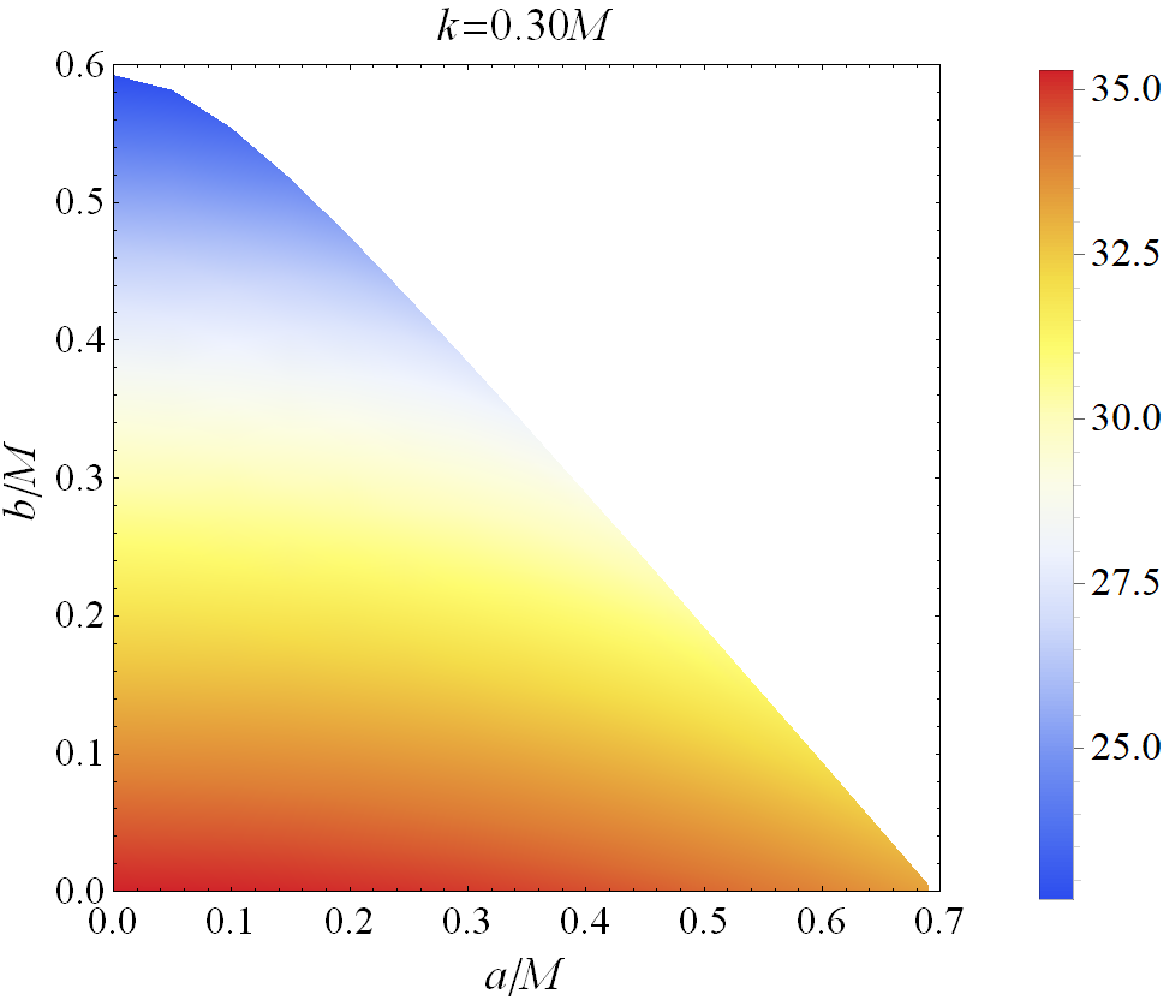}\\
	\end{tabular}
	\caption{Circularity deviation observable $\Delta C$ (upper panel) and the angular diameter $\theta_d$ (lower panel) as a function of black holes parameters for inclination angle $\theta_o=17^{o}$. For these parameters values, the shadow observables are smaller than those deduced for the M87* shadow.}	\label{M87_1}
\end{figure*}

We now compare the shadows produced by the nonsingular Kerr--Sen black holes with that for the M87* black hole. For this, we consider the two astronomical observables, the circularity deviation $\Delta C\leq 0.10$ and angular diameter $\theta_d=42\pm 3\, \mu$as, deduced for the M87* black hole by the EHT \cite{Akiyama:2019cqa,Akiyama:2019eap}. On the observer's screen the shadow boundary is described by the coordinates ($R(\varphi),\varphi$) in a polar coordinate system with the origin at the shadow center ($X_C,Y_C$). Due to the black hole rotation, the shadow shift only in the direction perpendicular to the black hole rotation. Furthermore, considering black hole center at $(0,0)$, the shadow center can be defined as  \cite{Johannsen:2015qca,Johannsen:2010ru}
\begin{equation}
Y_C=0,\qquad X_C=\frac{(X_r - X_l)}{2}.
\end{equation}
The radial coordinate of shadow from its center, $R(\varphi)$, and the angular coordinate $\varphi$ are defined as follows
\begin{equation}
R(\varphi)= \sqrt{(X-X_C)^2+(Y-Y_C)^2},\;\ \varphi\equiv \tan^{-1}\left(\frac{Y}{X-X_C}\right)\nonumber.
\end{equation}
The shadow average radius $\bar{R}$ is calculated as \cite{Johannsen:2010ru}
\begin{equation}
\bar{R}=\frac{1}{2\pi}\int_{0}^{2\pi} R(\varphi) d\varphi.
\end{equation}
\begin{figure}
	\begin{tabular}{c c}
	\includegraphics[scale=0.73]{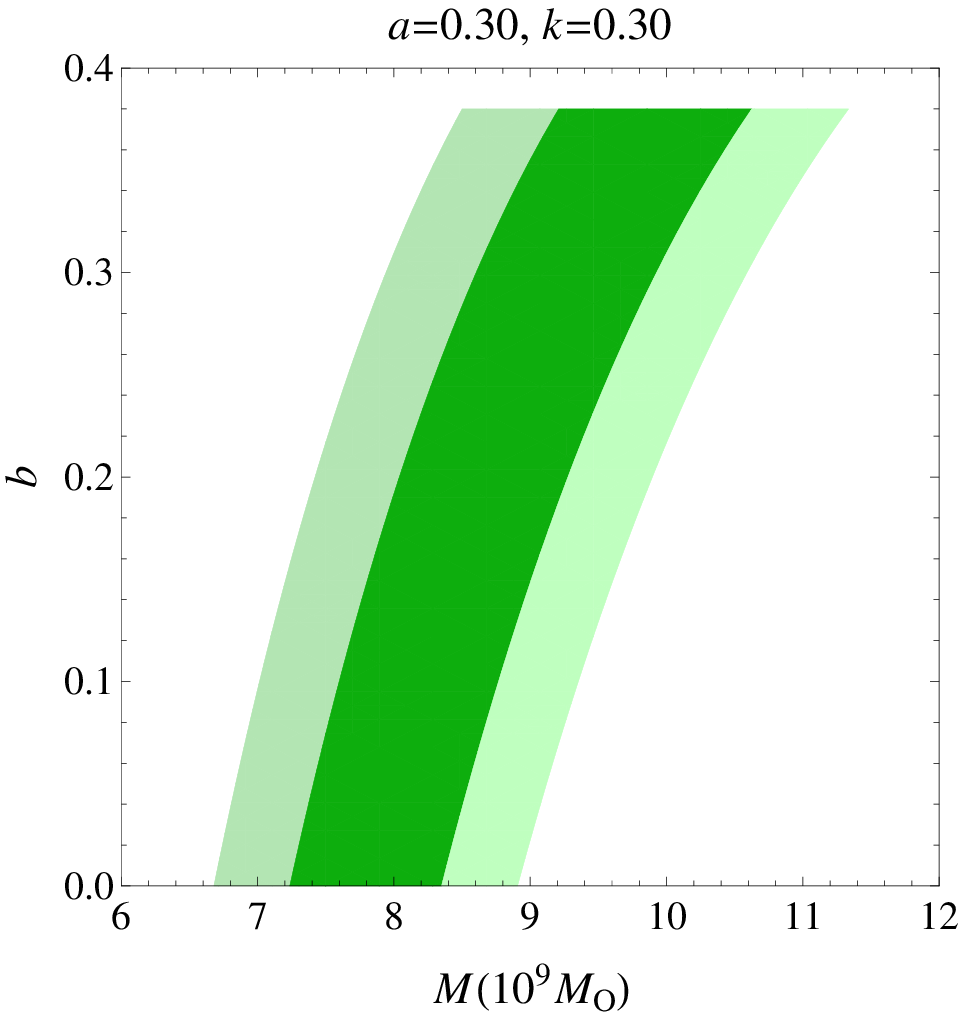}&
	\includegraphics[scale=0.73]{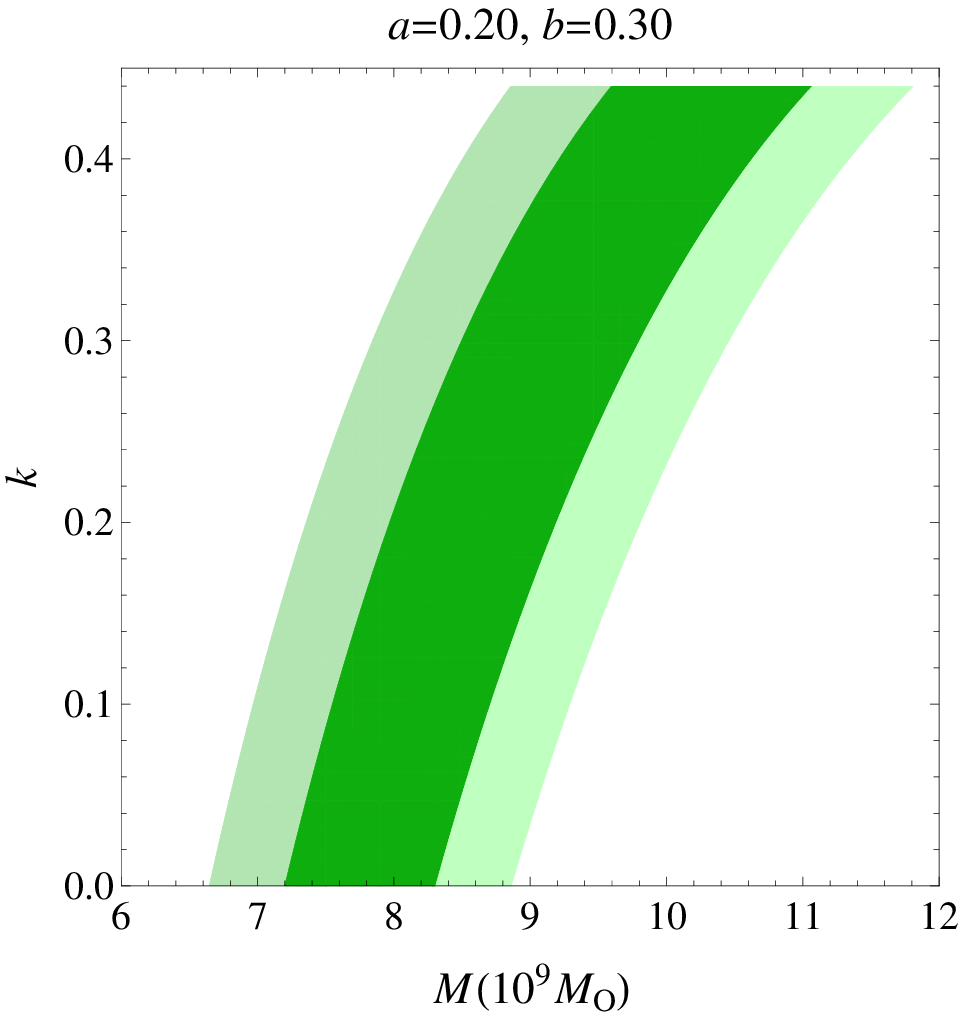}
	\end{tabular}
\caption{Constraints on the parameter $b$, $k$ and estimated M87* black hole mass $M$ using M87* shadow angular size within $1\sigma$ (dark green region) and $2\sigma$ (light green region).}\label{mass_M87}
\end{figure}
We describe the circularity deviation $\Delta C$ as a measures of the root-mean-square deviation of $R(\varphi)$ from the shadow average radius $\bar{R}$ \cite{Johannsen:2010ru,Johannsen:2015qca}
\begin{equation}
\Delta C=2\sqrt{\frac{1}{2\pi}\int_0^{2\pi}\left(R(\varphi)-\bar{R}\right)^2d\varphi}.
\end{equation}
The shadow angular diameter is also defined as follow
\begin{equation}
\theta_d=2\frac{R_s}{d},\;\;\;\; R_s=\sqrt{A/\pi},
\end{equation}    
where $d$ is the distance of the M87* black hole from the earth. In Figs.~\ref{M87} and \ref{M87_1}, the circularity deviation $\Delta C$ and angular diameter $\theta_d$ are shown for nonsingular Kerr--Sen black holes shadows for inclination angles $\theta_o=90^o$ and $\theta_o=17^o$. Considering $M=6.5 \times 10^9 M_{\odot}$ and $d=16.8\,$Mpc for the M87* black hole, we use the observables $\Delta C\leq 0.10$ and  $\theta_d=42\pm 3\,\mu$as deduced by the EHT to constrain the nonsingular Kerr--Sen black hole parameters. We place the constraints on ($a,k$) and ($a,b$) by, respectively, fixing the $b$ and $k$ values. From Fig.~\ref{M87}, it is clear that over a finite parameter space, black hole shadows satisfy the bound $\Delta C\leq 0.10$. However, the angular diameter are smaller than that for the M87* black hole within $1\sigma$ region, i.e., $\theta_d=39\,\mu$as. The maximum angular diameter for $b=0.30M$ and $k=0.30M$ are, respectively, $\theta_d=35.3469\,\mu$as and $\theta_d=35.3363\,\mu$as. For fixed values of black hole parameters the apparent shadow circularity deviation and angular diameter decrease as the observer move away from the equatorial plane toward the polar axis. For $\theta_o=17^o$ the $\Delta C$ and $\theta_d$ for the nonsingular Kerr--Sen black holes shadows are smaller than the bounds deduced for the M87* black hole. Furthermore, with the aid of M87* black hole shadow angular size we depicted the constraints on the black hole mass and parameters $b$ and $k$ in Fig.~\ref{mass_M87}.

\section{Conclusions}\label{Sec6}
General relativity is an elegant theory of gravity which agrees with most of the observations; one can expect that at sufficiently small length scales,  in particular when quantum effects become relevant, general relativity is likely to breakdown. The presence of spacetime singularities and other issues have motivated the study of viable modified theories of gravity. These theories aim to reproduce general relativity in the weak gravitational field regime, but they can differ substantially from it in the strong-field regime. Among the various attempts to quantize gravity, string theory holds as one of the most promising candidates, with most of the analysis therein focused on the low-energy limit. In the context of the low-energy limit of heterotic string theory, a rotating black hole solution is the famous Kerr--Sen black hole. While in the absence of a well-defined quantum gravity theory, researchers turned attention to explore nonsingular models of black hole solutions. Therefore, it is instructive to explore the gravitational lensing to find the dependence of observables on the free parameter $k$ and charge $b$ and also obtain the shadow of nonsingular Kerr--Sen black holes and compare the results with those for the Kerr black holes.

With this motivation, we have analysed the strong gravitational lensing of light due to nonsingular Kerr--Sen black holes and compare the results with those for the Kerr black holes which besides the mass $M$ and angular momentum $a$, also depends on two parameters $k$ and $b$. We have examined the effects of black hole parameters $k$ and $b$ on the light deflection angle and lensing observables, in the strong-field observation, due to nonsingular Kerr--Sen black holes and compared with those due to Kerr black holes of general relativity and Kerr-Sen black hole.  We have used asymptotic lens equation to describe the geometrical relation between the black hole, observer and the source and expanded the deflection angle near the photon sphere and used its leading order terms in the lens equation to express the lensing observables in terms of strong lensing coefficients. We have numerically calculated the strong lensing coefficients $p$ and $q$, and lensing observables $ \theta_{\infty} $, $s$, $r_{\text{mag}}$,  $u_{m}$ as functions of $k$ and $b$ for the source relativistic images. These observables show that the separation between the two relativistic images for nonsingular  Kerr--Sen black hole is greater than that for the Kerr black holes.  Interestingly, we find that $p$ increases whereas $q$ and deflection angle $\alpha_D$ decrease with increasing $k$ and $b$ and observe the diverging behaviour of deflection angle.  We have calculated position and relative magnification of these images for supermassive black holes, namely, Sgr A*, M87, NGC 4649 and NGC 1332.  The minimum impact parameter $u_m$ decreases with $a$ similar to photon sphere radius $x_m$. At higher values of $k$ and $b$, photon comes closer to the black hole.  Similarly $p$ grows and $q$ decreases with $a$, both diverging at critical value of $a$. Analytical treatment of the lens equation gives two infinite sets of images on either side black hole.

In turn, we have analyzed the shadow of the nonsingular Kerr-Sen black hole to investigate how shadows get modified due to the parameters $b$ and $k$.  Interestingly, we have found that the size of the black hole shadow for a fixed value of the rotation parameter $a$, decreases and become more distorted with an increase in both parameters $b$ and $k$, i.e., resulting in a smaller and distorted shadow than in Kerr--Sen and  Kerr geometries.
Shadow observables $A$ and $D$ are used to characterize the shadows and in turn to back estimate the unique and precise value of the black hole parameters.  The departure produced by the parameters $b$ and $k$ for the M87* black hole shadow angular diameter is of $\mathcal{O}(\mu$as) and significant enough to be measured observationally with the advance EHT soon. Furthermore, we placed the stringent constraints on the parameters for which nonsingular Kerr--Sen black holes shadows satisfy the M87* black hole shadow observables deduced by the EHT. Our study empirically recommends that the M87* shadow observations do not wholly rule out nonsingular Kerr-Sen black holes; however, it severely bound the deviations.

Many interesting avenues are amenable for future work from nonsingular Kerr-sen black holes. For example, it will be intriguing to apply these solutions to study the effects of the parameters $b$ and $k$ in the light of EHT results to determine if they correspond to Kerr black hole or Kerr-Sen black hole of modified theories of gravity is admissible.  The results presented here are the generalization of previous discussions, on the Kerr and Kerr--Sen black holes, to a more general setting, and the possibility of a different conception of these result to the weak-field gravitational lensing is an interesting problem for future research.

\section{Acknowledgments} S.G.G. and  S.U.I  would like to thank SERB-DST for the ASEAN project IMRC/AISTDF/CRD/2018/000042. R.K. would like to thanks UGC for providing SRF.

\end{document}